\documentclass[%
 reprint,
 amsmath,amssymb,
 aps,
]{revtex4-2}

\usepackage{graphicx}
\usepackage{dcolumn}
\usepackage{bm}
\usepackage[utf8]{inputenc}
\usepackage{amsmath}
\usepackage{subcaption}

\usepackage{xcolor}
\usepackage{microtype}
\usepackage{siunitx}
\usepackage{booktabs}
\usepackage{soul}

\usepackage{ragged2e}

\newcommand{\jcaption}[1]{\caption{\justifying #1}}

\begin{document}

\preprint{APS/123-QED}

\title{Towards metrology with highly charged isomeric ions from antiproton annihilation}

\author{Sara Alfaro}
\author{Lorenz Panzl}%
\email{lorenz.panzl@uibk.ac.at}
\author{Tommaso Faorlin}%
\author{Yannick Weiser}%
\author{Thomas Lafenthaler}%
\author{Thomas Monz}

\affiliation{%
 Universität Innsbruck, Institut für Experimentalphysik, Technikerstraße 25/4, A-6020 Innsbruck, Austria\\
}%

\author{Fredrik Parnefjord Gustafsson}
\author{Matthias Germann}
\author{Michael Doser}
\affiliation{%
 Experimental Physics Department, CERN, 1211 Geneva 23, Switzerland
}%

\author{Jakub Zieliński}
\author{Georgy Kornakov}
\affiliation{%
 Warsaw University of Technology, Faculty of Physics, ul. Koszykowa 75, 00-662 Warsaw, Poland
}%

\author{Sankarshan Choudapurkar}
\author{Giovanni Cerchiari}
\email{giovanni.cerchiari@uni-siegen.de}
\affiliation{
 University of Siegen, Department of Physics, Walter-Flex-Straße 3, 57072 Siegen, Germany
}%

\date{\today}

\begin{abstract}
We describe how the annihilation of antiprotons can be utilized to generate highly charged isomeric ions in an ion-trap setup. We identify optical transitions in the hyperfine splitting of Hydrogen-like atoms composed of an isomer and a single electron in the ground state. We identify promising candidates in the isomers of Y, Nb, Rh, In, and Sb, for which the hyperfine transition lies in the infrared and whose excited state level lifetime is in the hundreds of milliseconds, which is suitable for metrology applications.
\end{abstract}

\maketitle


\section{\label{sec:intro}Introduction}

The study of baryonic antimatter at low energy has substantially progressed over the past decades thanks to the deceleration of antiprotons ($\bar{\mathrm{p}}$) in charged particle traps~\cite{ELENAring}. Studies on low-energy antimatter in ion traps aim at testing the boundaries of the Standard Model (SM) of particle physics, explain the asymmetry between matter and antimatter~\cite{Smorra2017}, and measure how antimatter motion is affected by earth's gravitational pull~\cite{Anderson2023}. One of the major obstacles that hinders this field from expanding further is the lack of an area of application in which antimatter can find future use beyond fundamental studies. This perspective may change with the advent of new experiments that aim to use antiprotons for nuclear research~\cite{Teixeira2022, TRZCINSKA2004157, Zhang2025} and with the development of transport devices that can carry antiprotons outside the synthesis facilities~\cite{Aumann2022, Gibney2024, Leonhardt2025}.

\begin{figure}[h!]
    \centering
    \includegraphics[width=0.95\linewidth]{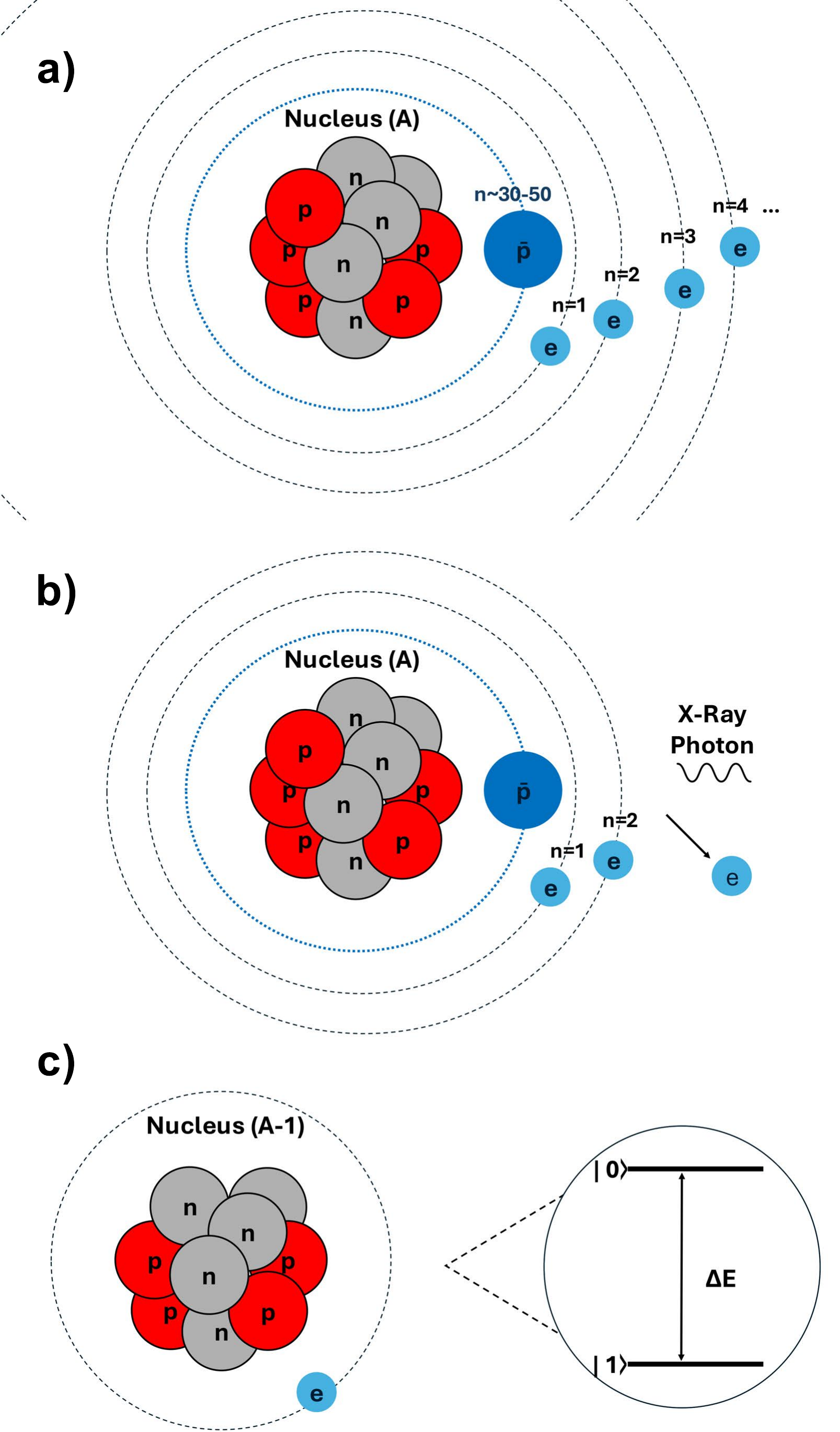}
    \jcaption{Schematic depiction of the process for HCI formation via antiproton annihilation. a) Initially, the antiproton ``$\bar{\text{p}}$'' is captured by the atom (mass number ``A'') at a  Rydberg orbit, which has a large principal quantum number (``n''), typically 30-50, but is found deep inside the electron cloud. b) The antiproton then initiates the Auger emission of electrons via bound-bound Coulomb collision energy transfer, ejecting or exciting the orbiting electrons ``e'' around the nucleus and generating X-ray photons in the process. c) Finally, the atom is left fully or partially stripped, with a proton ``p'' (or a neutron ``n'') from the nucleus undergoing annihilation with the antiproton in the process. This results in an HCI two-level system with energy difference $\Delta E$  and a single electron either left over after the stripping process or introduced during experimentation.}
    \label{fig:HCIsynthesis}
\end{figure}

In this article, we propose to utilize antiproton annihilation to produce isomeric ions that may find application in metrology and quantum technology.  The isomers could be used in these applications by measuring the hyperfine (HF) transition of hydrogen-like (H-like) highly charged ions (HCIs). The idea to utilize HF transitions in HCI as metrology reference is well established in literature as HCI are considered an interesting case for precision spectroscopy for fundamental tests of physics ~\cite{Shabaev2018, Indelicato2019, Morgner2023, Kozlov2018}, study of nuclear structure~\cite{Sun2024, Campbell2016}, and metrology in general ~\cite{LopezUrrutia2016}. The synthesis of HCI via $\bar{\mathrm{p}}$ annihilation differs from state-of-the-art techniques. To date, HCIs have been produced either by accelerating electrons onto atoms in Electron Beam Ion Traps (EBITs)~\cite{Levine1989} or by accelerating ions in particle accelerators onto targets~\cite{Steck2020}. The main difference is that a single antiproton can react with an atom and produce HCI via reaction that involves controlling the ions at $\unit{\electronvolt}$ activation energies or lower, which is an energy closer to a chemical reaction and it is several orders of magnitude lower than what is required in EBITs and particle accelerators. During the reaction, antiprotons most likely annihilate with one of the surface nucleons, releasing approximately $500$~MeV~\cite{PDG}. The reaction strips away the external electrons in an Auger cascade and produces highly charged nuclei that typically have one to a few nucleons less compared to the reagent~\cite{Kornakov2022, Gotta2008, Bacher1988}. The annihilation can be engineered in an ion trap either by mixing $\bar{\mathrm{p}}$ with positive ions, as is done for the formation of anti-hydrogen~\cite{Amoretti2002} or by charge exchange starting from negative ions as precursors~\cite{Gerber2019, Kornakov2022, Aumann2022}. The mechanism for HCI ion production via capture of an antiproton is depicted in Fig.~\ref{fig:HCIsynthesis}. Therefore, it is worth exploring whether annihilation can be controlled for the synthesis of HCI for fundamental studies~\cite{Morgner2023} and technological interest. In this article, we identify a few nuclei for a first experimental exploration starting from criteria of technological relevance where the use of antiprotons can be preferred to other existing techniques.

\section{\label{sec:qubit}HCI from annihilation}

We search for possible candidates considering the following restrictions. We select elements heavier than Kr because the resulting products are sufficiently heavy and highly charged to remain confined in an ion trap in most of the annihilation events~\cite{Kornakov2022}. We restrict elements to be lighter than Xe because it has been shown experimentally that Kr is fully stripped by annihilation and Xe is only partially stripped with one or two electrons remaining~\cite{Bacher1988, Gotta2008}. Annihilation of elements heavier than Kr results in HCI products that can have a transition wavelength of the HF splitting in the infrared (IR) part of the optical spectrum that can, therefore, be addressed with laser radiation fields. We also consider nuclear products whose lifetime is longer than 60~min, which is a practical observation time for metrology experiments that allow performing several measurement cycles, which are typically in the order of 1~s~\cite{Ludlow2015}. The limit of 60~min may seem small for applications. However, antiprotons can be preserved indefinitely in an ion trap~\cite{Sellner2017} and thus annihilated on demand when new HCI are needed. Finally, the annihilation process not only strips the electrons, but also potentially excites the nucleus at energies that are typically inaccessible in compact EBITs~\cite{Micke2018}, which can achieve stripping with electron beams at tens of keV kinetic energy. This peculiar property of the annihilation allows us to select isomeric nuclear products, which, in general, tend to have a stronger magnetic dipole moment than ground state nuclei, and often have larger HF splitting.

Table~\ref{tab:candidates} presents a list of candidates that meet the aforementioned criteria. In the table, we calculate the wavelengths of the H-like hyperfine transition using the model described in Ref.~\cite{Shabaev1994}. The calculations are conducted using the nuclear data provided in Ref.~\cite{Mertzimekis2016}. A detailed description of the calculations can be found in Appendix~\ref{app:calculations}.

\begin{table}[]
    \jcaption{Potential isomeric candidates that will result from antiproton annihilation with masses between Kr and Xe, and with lifetime longer than 60 minutes. The candidates are sorted by increasing proton number and neutron number as second criterion. The nuclear spin $I$, magnetic moment $\mu$/$\mu_N$, excitement energy $E$, and half-life $t_{1/2}$, and calculated ground state hyperfine splitting wavelength $\lambda_{HF}$ for each of the respective hydrogen-like ion is shown. The time units are ``m`` minutes, ``h'' hours, ``d'' days, and ``y'' years. The calculations are conducted using the nuclear data provided in Ref.~\cite{Mertzimekis2016}.}
    \centering
    \begin{tabular}{cccccc}
    \toprule
                Ion &     $I$ &     $\mu$/$\mu_N$ &  $E$[keV] &  $t_{1/2}$ & $\lambda_\text{HF}$~(\si{\micro\meter}) \\
    \midrule
     $^{83}$Kr$^ {35+}$ &  1/2- & 0.5910 &           42 &    1.83 h &       19.42455 \\
 $^{85}$Kr$^ {35+}$ &  1/2- & 0.6320 &          305 &    4.48 h &       18.16445 \\
 $^{82}$Rb$^ {36+}$ &    5- & 1.5096 &           69 &    6.47 h &       12.51460 \\
 $^{85}$Sr$^ {37+}$ &  1/2- & 0.5990 &          239 &    67.6 m &       16.10615 \\
 $^{87}$Sr$^ {37+}$ &  1/2- & 0.6240 &          388 &    2.82 h &       15.46087 \\
  $^{85}$Y$^ {38+}$ &  9/2+ & 6.2000 &           20 &     4.9 h &        2.567061 \\
  $^{87}$Y$^ {38+}$ &  9/2+ & 6.2400 &          381 &    13.4 h &        2.550630 \\
  $^{90}$Y$^ {38+}$ &    7+ & 5.2800 &          682 &    3.19 h &        3.125522 \\
 $^{92}$Nb$^ {40+}$ &    2+ & 6.1370 &          135 &   10.15 d &        1.968983 \\
 $^{93}$Mo$^ {41+}$ & 21/2+ & 9.9300 &         2425 &    6.85 h &        1.336406 \\
 $^{99}$Rh$^ {44+}$ &  9/2+ & 5.6200 &           65 &     4.7 h &        1.769000 \\
$^{101}$Rh$^ {44+}$ &  9/2+ & 5.4300 &          157 &    4.34 d &        1.830757 \\
$^{102}$Rh$^ {44+}$ &    6+ & 4.0100 &          141 &    3.74 y &        2.540022 \\
$^{106}$Ag$^ {46+}$ &    6+ & 3.7040 &           90 &     8.3 d &        2.374345 \\
$^{108}$Ag$^ {46+}$ &    6+ & 3.5800 &          109 &     438 y &        2.455607 \\
$^{110}$Ag$^ {46+}$ &    6+ & 3.6020 &          118 &     250 d &        2.440786 \\
$^{110}$In$^ {48+}$ &    2+ & 4.3650 &           62 &    69.1 m &        1.529835 \\
$^{114}$In$^ {48+}$ &    5+ & 4.6460 &          190 &    49.5 d &        1.625155 \\
$^{116}$Sb$^ {50+}$ &    8- & 2.5900 &          383 &    60.3 m &        2.587845 \\
$^{118}$Sb$^ {50+}$ &    8- & 2.3200 &          250 &     5.0 h &        2.880392 \\
$^{120}$Sb$^ {50+}$ &    8- & 2.3400 &        0 + x &    5.76 d &        2.856473 \\
$^{119}$Te$^ {51+}$ & 11/2- & 0.8940 &          261 &    4.70 d &        6.946341 \\
$^{121}$Te$^ {51+}$ & 11/2- & 0.8950 &          294 &     164 d &        6.938580 \\
    \bottomrule
\end{tabular}
    
    \label{tab:candidates}
\end{table}

For most of the candidates in Tab.~\ref{tab:candidates} production via beta-decay is unfeasible. The nuclei $^{87\textrm{m}1}$Y, $^{93\textrm{m}1}$Mo, $^{101\textrm{m}1}$Rh, $^{110\textrm{m}1}$In, $^{119\textrm{m}1}$Te, $^{118\textrm{m}1}$Sb are daughter by beta-decay of nuclei whose lifetimes are in the order of an hour and for $^{92\textrm{m}1}$Nb, $^{102\textrm{m}1}$Rh, $^{106\textrm{m}1}$Ag, $^{108\textrm{m}1}$Ag, $^{114\textrm{m}1}$In, and $^{120\textrm{m}1}$Sb there is no beta-decay path to these isotopes. Only $^{90\textrm{m}1}$Y can be produced from $^{90}$Sr, which has a half-life of 28.9 years. Therefore, annihilation is favorable for producing the listed nuclei in a trap on demand.

We acknowledge that the selection of a minimum of 60~min lifetime and the requirement of the nuclear product to be an isomer make a compelling case to explore the annihilation technique for HCI synthesis. However, this criterion might exclude other viable candidates and we present an extension of Tab.~\ref{tab:extra_1min} in Appendix~\ref{app:extended table}, including nuclear products with masses beyond the aforementioned range of interest.

\begin{table*}[]
    \jcaption{Shortlisted candidates from Tab.~\ref{tab:candidates} that can be produced and trapped in more than 5\% of the annihilation events (``Isotope yield''). Each isomeric ``product'' emerging from the annihilation is reported together with the rest excess energy (``isomeric state''). For each product, we indicate a suggested list of precursors (``Reagent'') with corresponding ``natural abundance''. Finally, we report the transition wavelength of Hyperfine splitting for the Hydrogen-like atom calculated according to Ref.~\cite{Shabaev1994} (``$\lambda_a$'') and with the GRASP~\cite{Jonsson2013,Jonnsson2007,Bieron2023} software (``$\lambda_\textrm{GRASP}$''). The corresponding lifetimes of the Hyperfine excited state (``Lifetime'') are calculated with eq. 22 of Ref.~\cite{shabaev1998}.}
    \centering
    \begin{tabular}{cccccccc}
        \hline
         Product & \begin{tabular}{@{}c@{}}Isomeric  \\ state (keV) \end{tabular} & Reagent & \begin{tabular}{@{}c@{}}Natural  \\ abundance (\%) \end{tabular} & \begin{tabular}{@{}c@{}}Isotope  \\ yield (\%) \end{tabular}  & \begin{tabular}{@{}c@{}}$\lambda_a$  \\ (\si{\micro\meter}) \end{tabular}   & \begin{tabular}{@{}c@{}}$\lambda_\textrm{GRASP}$  \\ (\si{\micro\meter}) \end{tabular}  & \begin{tabular}{@{}c@{}}Lifetime  \\ (\si{\milli\second}) \end{tabular}  \\ 
         \hline
         $^{87}$Y & 381
         & \begin{tabular}{@{}c@{}}$^{90}$Zr  \\ $^{89}$Y \end{tabular}
         & \begin{tabular}{@{}c@{}}51.45  \\ 100 \end{tabular}
         & \begin{tabular}{@{}c@{}}8.6  \\ 8.9  \end{tabular}
         & 2.551 & 2.544 & 358.8 \\
         \hline
         $^{90}$Y & 682
         & $^{92}$Zr 
         & 17.15 
         & 7.1
         & 3.126 & 3.119 & 637.1 \\
         \hline
         $^{92}$Nb & 135
         & \begin{tabular}{@{}c@{}}$^{93}$Nb  \\ $^{94}$Mo \end{tabular}
         & \begin{tabular}{@{}c@{}}100  \\ 9.2 \end{tabular}
         & \begin{tabular}{@{}c@{}}9  \\ 5.2 \end{tabular}
         & 1.969 & 1.953 & 183.7 \\
         \hline
         $^{101}$Rh & 157
         & \begin{tabular}{@{}c@{}}$^{102}$Pd  \\ $^{103}$Rh \\ $^{104}$Pd \end{tabular}
         & \begin{tabular}{@{}c@{}}1  \\ 100 \\ 11.14 \end{tabular}
         & \begin{tabular}{@{}c@{}}6  \\ 8.5 \\ 6.6 \end{tabular}
         & 1.831 & 1.814 & 132.5 \\
         \hline
         $^{102}$Rh & 141
         & \begin{tabular}{@{}c@{}}$^{104}$Pd  \\ $^{103}$Rh \end{tabular}
         & \begin{tabular}{@{}c@{}}11.14  \\ 100 \end{tabular}
         & \begin{tabular}{@{}c@{}}6.6  \\ 7.9 \end{tabular}
         & 2.540 & 2.513 & 343.7 \\
         \hline
         $^{114}$In & 190
         & $^{115}$In 
         & 95.72 
         & 6.86 
         & 1.625 & 1.613 & 93.7 \\
         \hline
         $^{118}$Sb & 250
         & $^{121}$Sb  
         & 57.21
         & 5.2 
         & 2.880 & 2.920 & 540.7 \\
         \hline
         $^{120}$Sb & 0+x
         & $^{121}$Sb
         & 57.21
         & 5.2
         & 2.856 & 2.599 & 381.4 \\

         \hline
    \end{tabular}
    
    \label{tab:shortcandidates}
\end{table*}

We continue to narrow our selection of candidates for a first experimental investigation by identifying the nuclei that can be produced at least in 5\% of the annihilations. The simulation of the yields in the reaction products was performed as follows. First, we simulated the formation of antiprotonic atoms and the resulting nuclear fragmentation using the Geant4 simulation framework \cite{agostinelli_geant4simulation_2003,j_allison_geant4_2006,allison_recent_2016}. Other toolkits, like FLUKA \cite{fluka2015,fluka2022} were considered; however, we found that Geant4 is the only toolkit that allows for a detailed study of kinetic energy and isomer states of nuclei produced from the antiproton interacting with an atom on an event-by-event basis.
The annihilation of antiprotons and the subsequent fragmentation of the nuclei is simulated following the set-up from \cite{Kornakov2022}. A 1~keV antiproton source is positioned within a nanoscale sphere of target material, all contained in a vacuum. After reaching the sphere, the antiprotons lose energy via electromagnetic interactions inside the material until they are captured by the nucleus. The annihilation process is simulated using Geant4's FTFP\textunderscore BERT\textunderscore HP physics list, which results in nuclear fragments in various isomer states.
The atomic mass and number of the fragments, together with their kinetic and excitation energies, are stored for each antiproton annihilation event.
This is simulated using one million antiprotons for all isotopes of possible selected reagent material. Finally, the expected yields are calculated assuming the formation of antiprotonic atoms inside a Penning trap with the end-cap electrodes set to 10~kV, the trap's inner diameter of 30 mm and a magnetic field of 5~T.
The trapping parameters (i.e. the high voltage on the end-cap electrodes, the diameter of the electrodes, and the strength of the magnetic field) were chosen based on the currently available trap in the AE$\overline{\textrm{g}}$IS experiment \cite{drobychev_proposal_2007,doser_aegis_2019}. These values are representative of the realistically achievable trapping conditions. Furthermore, AE$\overline{\textrm{g}}$IS has been chosen as the best candidate for implementing the formation method explained in~\cite{Kornakov2022}.
We also assume that fragments are produced isotropically in all directions. Therefore, a random emission angle is selected for each fragment to calculate whether it can be trapped. The final yields are calculated as the average of all yields obtained by this process repeated 10 times. This simulation results in a wide variety of unstable isomers that can potentially remain inside the trap. We performed additional analysis of their subsequent decay pathways toward more stable isomeric states, following the rapid decay route that results in a stable isomer with a lifetime of more than one hour.

\begin{table}[]
    \jcaption{Simulated yields of the ``product'' isomers from the annihilation of one antiproton with the ``reagent'' nucleus. For each product, we report the total ``isotope yield'' in all energy states and the ``isomer yield'' corresponding to the isomer with excitation energy ``E''. The yields for $^{92}$Nb (denoted with ``*'') are calculated as the sum of yields of all isomer states below $\qty{1}{\mega\electronvolt}$ that decay exclusively to the $\qty{135}{\kilo\electronvolt}$.}
    \centering
    \begin{tabular}{ccccc}
        \hline
         Reagent & Product & \begin{tabular}{@{}c@{}}Isotope  \\ yield (\%) \end{tabular} & \begin{tabular}{@{}c@{}}E  \\ (keV) \end{tabular} & \begin{tabular}{@{}c@{}}Isomer  \\ yield (\%) \end{tabular} \\ 
          \hline
         $^{94}$Mo & $^{93}$Mo & 14.5 & 2425 & 1.7\\
         $^{94}$Mo & $^{92}$Nb & 8.7 & 135 & 5.2$^{*}$\\
         $^{95}$Mo & $^{93}$Mo & 13.0 & 2425 & 1.5\\
         $^{95}$Mo & $^{92}$Nb & 7.7 & 135 & 4.6$^*$\\
         $^{94}$Tc & $^{93}$Mo & 10.3 & 2425 & 1.2\\
         $^{94}$Tc & $^{92}$Nb & 0.5 & 135 & 0.3$^*$\\
         $^{93}$Nb & $^{92}$Nb & 12.7 & 135  & 9$^{*}$\\
         $^{94}$Nb & $^{92}$Nb & 11.6 & 135 & 6.8$^{*}$\\
         $^{95}$Nb & $^{92}$Nb & 9.4 & 135 & 5.5$^{*}$\\
         \hline
    \end{tabular}
    
    \label{tab:yields}
\end{table}

In the interest of clarity, we describe in more detail the results that we obtained for $^{92\textrm{m}1}$Nb and $^{93\textrm{m}1}$Mo, in which we explored the annihilation yield from reagents with atomic mass $A > 92$, the atomic mass of $^{92}$Nb. Table~\ref{tab:yields} shows the main reagents for the highest yields of $^{92}$Nb and $^{93}$Mo in the isomer states of interest. There are no direct paths of obtaining $^{92}$Nb in the 135~keV state emerging from the Geant4 simulation. However, all the states 225.8~keV, 285.7~keV, 389.8~keV, and 975.0~keV decay exclusively to the 135~keV \cite{PhysRevC.93.065804} state with a halftime of less than \SI{10}{\micro\second}. Therefore, we considered all of these isomers as practically being the isomer with 135~keV excitation energy. Interestingly, the decay process might be used to tag the initialization of the 135~keV isomeric state by time-tagging the outgoing photon. From the table, we see that $^{92}$Nb can be produced 9\% of the time if $^{93}$Nb is chosen as the reagent, while $^{93}$Mo best yield is only 1.7\% from antiprotons reacting with $^{94}$Mo.

We arrive at the final list of candidates by focusing on reagent materials near the isomer of interest that can be found naturally on Earth. Moreover, we restrict the list to those product isomers whose H-like hyperfine (HF) splitting can be excited with a wavelength $\lambda_\textrm{HF}$ shorter than \SI{5.0}{\micro\meter} which are currently technologically more accessible than higher wavelengths, for example by utilizing Optical Parametric Oscillators (OPO) that are readily available on the market. This threshold was chosen after inquiring about OPO systems with different industry partners, which can provide such with different upper wavelength limits of \SI{4.1}{\micro\meter} (Radiantis: Titan XT HP), \SI{3.5}{\micro\meter} (Hübner Photonics: C-Wave NIR-II) and \SI{4.3}{\micro\meter} (Toptica: DLC TOPO).
The final list of candidates is presented in Tab.~\ref{tab:shortcandidates}, in which we report candidates with the optimal reagents, the prospect wavelength of excitation of the HF transition, and the lifetime of the excited HF states. We note that all the HF excited states have a lifetime in the hundreds of milliseconds with excellent prospects for precision spectroscopy. Unsurprisingly, a simulation performed with the atomic calculation package Hfszeeman95~\cite{Li2020}, which is part of GRASP~\cite{Bieron2023}, confirms that being HCIs,  the level splitting induced by the magnetic field can be approximated  in the linear Zeeman regime, even at a magnetic field of several Tesla. The example of $^{92\textrm{m}1}$Nb$^{40+}$ is presented in Appendix~\ref{app:HyperGRASP}.

The isomers can be produced in an ion trap by mixing positive ions with antiprotons or through the pulsed excitation of antiprotons co-trapped with negative ions as described in previous literature~\cite{Aumann2022, Gerber2019, Kornakov2022}. For the latter formation scheme involving negative ions, we discuss in Appendix~\ref{app:NbandMo} the laser pulse sequence that can be used to trigger the reaction. Finally, the isomer must be mass selected to obtain the desired one. This is nowadays feasible since modern mass spectrometry techniques in Penning traps~\cite{Schuessler2020,Tu2023} already demonstrate resolution of hundreds of eV. This resolution is two to three orders of magnitude higher than is required to identify the proposed isomeric states.

\section{\label{sec:setup}Conclusion}

In this article, we proposed the utilization of antiprotons to produce HCIs of isomeric nuclear states for spectroscopy and metrology. We focused on the HF transition of H-like ions and shortlisted less than ten candidates for the first experimental studies (see Tab.~\ref{tab:shortcandidates}). We highlighted a few key advantages that make the annihilation of antiprotons a compelling alternative to produce these HCIs. For instance, it is unfavorable or impossible to synthesize these nuclei through radioactive $\beta$ decay, and production with antiprotons allows one to study these HCI outside of radio-protected areas and over an extended period.

Our extended list of candidates (Tab.~\ref{tab:candidates}) includes elements between Kr and Xe, for which the inner-shell electron's binding energy is $\qty{30}{\kilo\electronvolt}$, so, it would require some sort of accelerator or large high voltage platform to strip the electrons via impact ionization. Therefore, annihilation offers the perspective to realize fully stripped HCI in this region of the periodic table in more compact devices and with lower activation energies. In this regard, it is important to stress that the annihilation allows not only to strip electrons, but also to obtain excited nuclei, and, for example, in our table, we reported isomer with excitation energy above $\qty{2}{\mega\electronvolt}$ ($^{93\textrm{m}1}$Mo). 

Therefore, the proposed method has the potential to extend spectroscopy studies for nuclear research and in the search for extensions to the Standard model, for example by extending points in King plots~\cite{Flambaum2018} with isomers and in the absence of inner-shell electrons. In nuclear research, it can give the possibility to produce isomers in Penning traps to perform precision mass studies to determine the mass of low-energy excited isomers~\cite{Schuessler2020} which have recently seen growing interest in the metrology community~\cite{Tiedau2024}.

To conclude, although previous studies indicate that our candidates will be fully stripped~\cite{Bacher1988, Gotta2008}, at the time of writing there exists no direct observation of the emerging charge states because the synthesis of antiprotonic atoms in UHV has never been attempted. This fact stresses the compelling case for further experiments to study the annihilation in UHV to measure the yield of isomers, to control the reaction products, and to show if the technique can find application to elements with masses higher than Xe.

\begin{acknowledgments}
This work was supported by the Ministry of Culture and Science of North Rhine-Westphalia. This project has received funding from the European Union’s Horizon Europe research and innovation program under the Marie Skłodowska-Curie grant agreement Nr 101109574. It was supported by the Research University – Excellence Initiative of Warsaw University of Technology via the strategic funds of the Priority Research Centre of High Energy Physics and Experimental Techniques, as well as the Research University – Excellence Initiative of Warsaw University of Technology via the strategic funds of the Priority Research Centre of High Energy Physics and Experimental Techniques.
The authors would like to thank J.~Braun for the installation and maintenance of GRASP on a virtual machine. The authors thank W.~Li for instructions on how to use the GRASP software. The authors would also like to thank V.~Shabaev for helpful guidance 
on literature regarding theoretical models for H-like ions. The authors would also like to thank A. Bondarev for helpful comments on an early version of the manuscript.

\end{acknowledgments}

\bibliography{refs}

\begin{thebibliography}{54}%
\makeatletter
\providecommand \@ifxundefined [1]{%
 \@ifx{#1\undefined}
}%
\providecommand \@ifnum [1]{%
 \ifnum #1\expandafter \@firstoftwo
 \else \expandafter \@secondoftwo
 \fi
}%
\providecommand \@ifx [1]{%
 \ifx #1\expandafter \@firstoftwo
 \else \expandafter \@secondoftwo
 \fi
}%
\providecommand \natexlab [1]{#1}%
\providecommand \enquote  [1]{``#1''}%
\providecommand \bibnamefont  [1]{#1}%
\providecommand \bibfnamefont [1]{#1}%
\providecommand \citenamefont [1]{#1}%
\providecommand \href@noop [0]{\@secondoftwo}%
\providecommand \href [0]{\begingroup \@sanitize@url \@href}%
\providecommand \@href[1]{\@@startlink{#1}\@@href}%
\providecommand \@@href[1]{\endgroup#1\@@endlink}%
\providecommand \@sanitize@url [0]{\catcode `\\12\catcode `\$12\catcode `\&12\catcode `\#12\catcode `\^12\catcode `\_12\catcode `\%12\relax}%
\providecommand \@@startlink[1]{}%
\providecommand \@@endlink[0]{}%
\providecommand \url  [0]{\begingroup\@sanitize@url \@url }%
\providecommand \@url [1]{\endgroup\@href {#1}{\urlprefix }}%
\providecommand \urlprefix  [0]{URL }%
\providecommand \Eprint [0]{\href }%
\providecommand \doibase [0]{https://doi.org/}%
\providecommand \selectlanguage [0]{\@gobble}%
\providecommand \bibinfo  [0]{\@secondoftwo}%
\providecommand \bibfield  [0]{\@secondoftwo}%
\providecommand \translation [1]{[#1]}%
\providecommand \BibitemOpen [0]{}%
\providecommand \bibitemStop [0]{}%
\providecommand \bibitemNoStop [0]{.\EOS\space}%
\providecommand \EOS [0]{\spacefactor3000\relax}%
\providecommand \BibitemShut  [1]{\csname bibitem#1\endcsname}%
\let\auto@bib@innerbib\@empty
\bibitem [{\citenamefont {Alanzeau}\ \emph {et~al.}(2014)\citenamefont {Alanzeau}, \citenamefont {Angoletta}, \citenamefont {Baillie}, \citenamefont {Barna}, \citenamefont {Bartmann}, \citenamefont {Belochitskii}, \citenamefont {Borburgh}, \citenamefont {Breuker}, \citenamefont {Butin}, \citenamefont {Buzio}, \citenamefont {Capatina}, \citenamefont {Carli}, \citenamefont {Carlier}, \citenamefont {Cattin}, \citenamefont {Dobers}, \citenamefont {Chiggiato}, \citenamefont {Ducimetiere}, \citenamefont {Eriksson}, \citenamefont {Fedemann}, \citenamefont {Fowler}, \citenamefont {Froeschl}, \citenamefont {Gebel}, \citenamefont {Gilbert}, \citenamefont {Hancock}, \citenamefont {Harasimowicz}, \citenamefont {Hori}, \citenamefont {Jorgensen}, \citenamefont {Kersevan}, \citenamefont {Kuchler}, \citenamefont {Lacroix}, \citenamefont {LeGodec}, \citenamefont {Lelong}, \citenamefont {Lopez-Hernandez}, \citenamefont {Maury}, \citenamefont {Molendijk}, \citenamefont {Morand}, \citenamefont {Newborough}, \citenamefont
  {Nisbet}, \citenamefont {Nosych}, \citenamefont {Oelert}, \citenamefont {Paoluzzi}, \citenamefont {Pasinelli}, \citenamefont {Pedersen}, \citenamefont {Perini}, \citenamefont {Puccio}, \citenamefont {Sanchez-Quesada}, \citenamefont {Schoerling}, \citenamefont {Sermeus}, \citenamefont {Soby}, \citenamefont {Timmins}, \citenamefont {Tommasini}, \citenamefont {Tranquille}, \citenamefont {Vanbavinckhove}, \citenamefont {Vorozhtsov}, \citenamefont {Welsch},\ and\ \citenamefont {Zickler}}]{ELENAring}%
  \BibitemOpen
  \bibfield  {author} {\bibinfo {author} {\bibfnamefont {C.}~\bibnamefont {Alanzeau}}, \bibinfo {author} {\bibfnamefont {M.-E.}\ \bibnamefont {Angoletta}}, \bibinfo {author} {\bibfnamefont {J.}~\bibnamefont {Baillie}}, \bibinfo {author} {\bibfnamefont {D.}~\bibnamefont {Barna}}, \bibinfo {author} {\bibfnamefont {W.}~\bibnamefont {Bartmann}}, \bibinfo {author} {\bibfnamefont {P.}~\bibnamefont {Belochitskii}}, \bibinfo {author} {\bibfnamefont {J.}~\bibnamefont {Borburgh}}, \bibinfo {author} {\bibfnamefont {H.}~\bibnamefont {Breuker}}, \bibinfo {author} {\bibfnamefont {F.}~\bibnamefont {Butin}}, \bibinfo {author} {\bibfnamefont {M.}~\bibnamefont {Buzio}}, \bibinfo {author} {\bibfnamefont {O.}~\bibnamefont {Capatina}}, \bibinfo {author} {\bibfnamefont {C.}~\bibnamefont {Carli}}, \bibinfo {author} {\bibfnamefont {E.}~\bibnamefont {Carlier}}, \bibinfo {author} {\bibfnamefont {M.}~\bibnamefont {Cattin}}, \bibinfo {author} {\bibfnamefont {T.}~\bibnamefont {Dobers}}, \bibinfo {author} {\bibfnamefont {P.}~\bibnamefont
  {Chiggiato}}, \bibinfo {author} {\bibfnamefont {L.}~\bibnamefont {Ducimetiere}}, \bibinfo {author} {\bibfnamefont {T.}~\bibnamefont {Eriksson}}, \bibinfo {author} {\bibfnamefont {S.}~\bibnamefont {Fedemann}}, \bibinfo {author} {\bibfnamefont {T.}~\bibnamefont {Fowler}}, \bibinfo {author} {\bibfnamefont {R.}~\bibnamefont {Froeschl}}, \bibinfo {author} {\bibfnamefont {R.}~\bibnamefont {Gebel}}, \bibinfo {author} {\bibfnamefont {N.}~\bibnamefont {Gilbert}}, \bibinfo {author} {\bibfnamefont {S.}~\bibnamefont {Hancock}}, \bibinfo {author} {\bibfnamefont {J.}~\bibnamefont {Harasimowicz}}, \bibinfo {author} {\bibfnamefont {M.}~\bibnamefont {Hori}}, \bibinfo {author} {\bibfnamefont {L.~V.}\ \bibnamefont {Jorgensen}}, \bibinfo {author} {\bibfnamefont {R.}~\bibnamefont {Kersevan}}, \bibinfo {author} {\bibfnamefont {D.}~\bibnamefont {Kuchler}}, \bibinfo {author} {\bibfnamefont {J.-M.}\ \bibnamefont {Lacroix}}, \bibinfo {author} {\bibfnamefont {G.}~\bibnamefont {LeGodec}}, \bibinfo {author} {\bibfnamefont
  {P.}~\bibnamefont {Lelong}}, \bibinfo {author} {\bibfnamefont {L.}~\bibnamefont {Lopez-Hernandez}}, \bibinfo {author} {\bibfnamefont {S.}~\bibnamefont {Maury}}, \bibinfo {author} {\bibfnamefont {J.}~\bibnamefont {Molendijk}}, \bibinfo {author} {\bibfnamefont {B.}~\bibnamefont {Morand}}, \bibinfo {author} {\bibfnamefont {A.}~\bibnamefont {Newborough}}, \bibinfo {author} {\bibfnamefont {D.}~\bibnamefont {Nisbet}}, \bibinfo {author} {\bibfnamefont {A.}~\bibnamefont {Nosych}}, \bibinfo {author} {\bibfnamefont {W.}~\bibnamefont {Oelert}}, \bibinfo {author} {\bibfnamefont {M.}~\bibnamefont {Paoluzzi}}, \bibinfo {author} {\bibfnamefont {S.}~\bibnamefont {Pasinelli}}, \bibinfo {author} {\bibfnamefont {F.}~\bibnamefont {Pedersen}}, \bibinfo {author} {\bibfnamefont {D.}~\bibnamefont {Perini}}, \bibinfo {author} {\bibfnamefont {B.}~\bibnamefont {Puccio}}, \bibinfo {author} {\bibfnamefont {J.}~\bibnamefont {Sanchez-Quesada}}, \bibinfo {author} {\bibfnamefont {D.}~\bibnamefont {Schoerling}}, \bibinfo {author}
  {\bibfnamefont {L.}~\bibnamefont {Sermeus}}, \bibinfo {author} {\bibfnamefont {L.}~\bibnamefont {Soby}}, \bibinfo {author} {\bibfnamefont {M.}~\bibnamefont {Timmins}}, \bibinfo {author} {\bibfnamefont {D.}~\bibnamefont {Tommasini}}, \bibinfo {author} {\bibfnamefont {G.}~\bibnamefont {Tranquille}}, \bibinfo {author} {\bibfnamefont {G.}~\bibnamefont {Vanbavinckhove}}, \bibinfo {author} {\bibfnamefont {A.}~\bibnamefont {Vorozhtsov}}, \bibinfo {author} {\bibfnamefont {C.}~\bibnamefont {Welsch}},\ and\ \bibinfo {author} {\bibfnamefont {T.}~\bibnamefont {Zickler}},\ }\bibfield  {title} {\bibinfo {title} {Extra low energy antiproton (elena) ring and its transfer lines - design report},\ }\href {https://cds.cern.ch/record/1694484/files/CERN-2014-002.pdf} {\bibfield  {journal} {\bibinfo  {journal} {CERN}\ } (\bibinfo {year} {2014})},\ \bibinfo {note} {\url{https://cds.cern.ch/record/1694484/files/CERN-2014-002.pdf}}\BibitemShut {NoStop}%
\bibitem [{\citenamefont {Smorra}\ \emph {et~al.}(2017)\citenamefont {Smorra}, \citenamefont {Sellner}, \citenamefont {Borchert}, \citenamefont {Harrington}, \citenamefont {Higuchi}, \citenamefont {Nagahama}, \citenamefont {Tanaka}, \citenamefont {Mooser}, \citenamefont {Schneider}, \citenamefont {Bohman}, \citenamefont {Blaum}, \citenamefont {Matsuda}, \citenamefont {Ospelkaus}, \citenamefont {Quint}, \citenamefont {Walz}, \citenamefont {Yamazaki},\ and\ \citenamefont {Ulmer}}]{Smorra2017}%
  \BibitemOpen
  \bibfield  {author} {\bibinfo {author} {\bibfnamefont {C.}~\bibnamefont {Smorra}}, \bibinfo {author} {\bibfnamefont {S.}~\bibnamefont {Sellner}}, \bibinfo {author} {\bibfnamefont {M.~J.}\ \bibnamefont {Borchert}}, \bibinfo {author} {\bibfnamefont {J.~A.}\ \bibnamefont {Harrington}}, \bibinfo {author} {\bibfnamefont {T.}~\bibnamefont {Higuchi}}, \bibinfo {author} {\bibfnamefont {H.}~\bibnamefont {Nagahama}}, \bibinfo {author} {\bibfnamefont {T.}~\bibnamefont {Tanaka}}, \bibinfo {author} {\bibfnamefont {A.}~\bibnamefont {Mooser}}, \bibinfo {author} {\bibfnamefont {G.}~\bibnamefont {Schneider}}, \bibinfo {author} {\bibfnamefont {M.}~\bibnamefont {Bohman}}, \bibinfo {author} {\bibfnamefont {K.}~\bibnamefont {Blaum}}, \bibinfo {author} {\bibfnamefont {Y.}~\bibnamefont {Matsuda}}, \bibinfo {author} {\bibfnamefont {C.}~\bibnamefont {Ospelkaus}}, \bibinfo {author} {\bibfnamefont {W.}~\bibnamefont {Quint}}, \bibinfo {author} {\bibfnamefont {J.}~\bibnamefont {Walz}}, \bibinfo {author} {\bibfnamefont {Y.}~\bibnamefont
  {Yamazaki}},\ and\ \bibinfo {author} {\bibfnamefont {S.}~\bibnamefont {Ulmer}},\ }\bibfield  {title} {\bibinfo {title} {A parts-per-billion measurement of the antiproton magnetic moment},\ }\href {https://doi.org/10.1038/nature24048} {\bibfield  {journal} {\bibinfo  {journal} {Nature}\ }\textbf {\bibinfo {volume} {550}},\ \bibinfo {pages} {371} (\bibinfo {year} {2017})}\BibitemShut {NoStop}%
\bibitem [{\citenamefont {Anderson}\ \emph {et~al.}(2023)\citenamefont {Anderson}, \citenamefont {Baker}, \citenamefont {Bertsche}, \citenamefont {Bhatt}, \citenamefont {Bonomi}, \citenamefont {Capra}, \citenamefont {Carli}, \citenamefont {Cesar}, \citenamefont {Charlton}, \citenamefont {Christensen}, \citenamefont {Collister}, \citenamefont {Cridland~Mathad}, \citenamefont {Duque~Quiceno}, \citenamefont {Eriksson}, \citenamefont {Evans}, \citenamefont {Evetts}, \citenamefont {Fabbri}, \citenamefont {Fajans}, \citenamefont {Ferwerda}, \citenamefont {Friesen}, \citenamefont {Fujiwara}, \citenamefont {Gill}, \citenamefont {Golino}, \citenamefont {Gomes~Gonçalves}, \citenamefont {Grandemange}, \citenamefont {Granum}, \citenamefont {Hangst}, \citenamefont {Hayden}, \citenamefont {Hodgkinson}, \citenamefont {Hunter}, \citenamefont {Isaac}, \citenamefont {Jimenez}, \citenamefont {Johnson}, \citenamefont {Jones}, \citenamefont {Jones}, \citenamefont {Jonsell}, \citenamefont {Khramov}, \citenamefont {Madsen},
  \citenamefont {Martin}, \citenamefont {Massacret}, \citenamefont {Maxwell}, \citenamefont {McKenna}, \citenamefont {Menary}, \citenamefont {Momose}, \citenamefont {Mostamand}, \citenamefont {Mullan}, \citenamefont {Nauta}, \citenamefont {Olchanski}, \citenamefont {Oliveira}, \citenamefont {Peszka}, \citenamefont {Powell}, \citenamefont {Rasmussen}, \citenamefont {Robicheaux}, \citenamefont {Sacramento}, \citenamefont {Sameed}, \citenamefont {Sarid}, \citenamefont {Schoonwater}, \citenamefont {Silveira}, \citenamefont {Singh}, \citenamefont {Smith}, \citenamefont {So}, \citenamefont {Stracka}, \citenamefont {Stutter}, \citenamefont {Tharp}, \citenamefont {Thompson}, \citenamefont {Thompson}, \citenamefont {Thorpe-Woods}, \citenamefont {Torkzaban}, \citenamefont {Urioni}, \citenamefont {Woosaree},\ and\ \citenamefont {Wurtele}}]{Anderson2023}%
  \BibitemOpen
  \bibfield  {author} {\bibinfo {author} {\bibfnamefont {E.~K.}\ \bibnamefont {Anderson}}, \bibinfo {author} {\bibfnamefont {C.~J.}\ \bibnamefont {Baker}}, \bibinfo {author} {\bibfnamefont {W.}~\bibnamefont {Bertsche}}, \bibinfo {author} {\bibfnamefont {N.~M.}\ \bibnamefont {Bhatt}}, \bibinfo {author} {\bibfnamefont {G.}~\bibnamefont {Bonomi}}, \bibinfo {author} {\bibfnamefont {A.}~\bibnamefont {Capra}}, \bibinfo {author} {\bibfnamefont {I.}~\bibnamefont {Carli}}, \bibinfo {author} {\bibfnamefont {C.~L.}\ \bibnamefont {Cesar}}, \bibinfo {author} {\bibfnamefont {M.}~\bibnamefont {Charlton}}, \bibinfo {author} {\bibfnamefont {A.}~\bibnamefont {Christensen}}, \bibinfo {author} {\bibfnamefont {R.}~\bibnamefont {Collister}}, \bibinfo {author} {\bibfnamefont {A.}~\bibnamefont {Cridland~Mathad}}, \bibinfo {author} {\bibfnamefont {D.}~\bibnamefont {Duque~Quiceno}}, \bibinfo {author} {\bibfnamefont {S.}~\bibnamefont {Eriksson}}, \bibinfo {author} {\bibfnamefont {A.}~\bibnamefont {Evans}}, \bibinfo {author}
  {\bibfnamefont {N.}~\bibnamefont {Evetts}}, \bibinfo {author} {\bibfnamefont {S.}~\bibnamefont {Fabbri}}, \bibinfo {author} {\bibfnamefont {J.}~\bibnamefont {Fajans}}, \bibinfo {author} {\bibfnamefont {A.}~\bibnamefont {Ferwerda}}, \bibinfo {author} {\bibfnamefont {T.}~\bibnamefont {Friesen}}, \bibinfo {author} {\bibfnamefont {M.~C.}\ \bibnamefont {Fujiwara}}, \bibinfo {author} {\bibfnamefont {D.~R.}\ \bibnamefont {Gill}}, \bibinfo {author} {\bibfnamefont {L.~M.}\ \bibnamefont {Golino}}, \bibinfo {author} {\bibfnamefont {M.~B.}\ \bibnamefont {Gomes~Gonçalves}}, \bibinfo {author} {\bibfnamefont {P.}~\bibnamefont {Grandemange}}, \bibinfo {author} {\bibfnamefont {P.}~\bibnamefont {Granum}}, \bibinfo {author} {\bibfnamefont {J.~S.}\ \bibnamefont {Hangst}}, \bibinfo {author} {\bibfnamefont {M.~E.}\ \bibnamefont {Hayden}}, \bibinfo {author} {\bibfnamefont {D.}~\bibnamefont {Hodgkinson}}, \bibinfo {author} {\bibfnamefont {E.~D.}\ \bibnamefont {Hunter}}, \bibinfo {author} {\bibfnamefont {C.~A.}\ \bibnamefont
  {Isaac}}, \bibinfo {author} {\bibfnamefont {A.~J.~U.}\ \bibnamefont {Jimenez}}, \bibinfo {author} {\bibfnamefont {M.~A.}\ \bibnamefont {Johnson}}, \bibinfo {author} {\bibfnamefont {J.~M.}\ \bibnamefont {Jones}}, \bibinfo {author} {\bibfnamefont {S.~A.}\ \bibnamefont {Jones}}, \bibinfo {author} {\bibfnamefont {S.}~\bibnamefont {Jonsell}}, \bibinfo {author} {\bibfnamefont {A.}~\bibnamefont {Khramov}}, \bibinfo {author} {\bibfnamefont {N.}~\bibnamefont {Madsen}}, \bibinfo {author} {\bibfnamefont {L.}~\bibnamefont {Martin}}, \bibinfo {author} {\bibfnamefont {N.}~\bibnamefont {Massacret}}, \bibinfo {author} {\bibfnamefont {D.}~\bibnamefont {Maxwell}}, \bibinfo {author} {\bibfnamefont {J.~T.~K.}\ \bibnamefont {McKenna}}, \bibinfo {author} {\bibfnamefont {S.}~\bibnamefont {Menary}}, \bibinfo {author} {\bibfnamefont {T.}~\bibnamefont {Momose}}, \bibinfo {author} {\bibfnamefont {M.}~\bibnamefont {Mostamand}}, \bibinfo {author} {\bibfnamefont {P.~S.}\ \bibnamefont {Mullan}}, \bibinfo {author} {\bibfnamefont
  {J.}~\bibnamefont {Nauta}}, \bibinfo {author} {\bibfnamefont {K.}~\bibnamefont {Olchanski}}, \bibinfo {author} {\bibfnamefont {A.~N.}\ \bibnamefont {Oliveira}}, \bibinfo {author} {\bibfnamefont {J.}~\bibnamefont {Peszka}}, \bibinfo {author} {\bibfnamefont {A.}~\bibnamefont {Powell}}, \bibinfo {author} {\bibfnamefont {C.~O.}\ \bibnamefont {Rasmussen}}, \bibinfo {author} {\bibfnamefont {F.}~\bibnamefont {Robicheaux}}, \bibinfo {author} {\bibfnamefont {R.~L.}\ \bibnamefont {Sacramento}}, \bibinfo {author} {\bibfnamefont {M.}~\bibnamefont {Sameed}}, \bibinfo {author} {\bibfnamefont {E.}~\bibnamefont {Sarid}}, \bibinfo {author} {\bibfnamefont {J.}~\bibnamefont {Schoonwater}}, \bibinfo {author} {\bibfnamefont {D.~M.}\ \bibnamefont {Silveira}}, \bibinfo {author} {\bibfnamefont {J.}~\bibnamefont {Singh}}, \bibinfo {author} {\bibfnamefont {G.}~\bibnamefont {Smith}}, \bibinfo {author} {\bibfnamefont {C.}~\bibnamefont {So}}, \bibinfo {author} {\bibfnamefont {S.}~\bibnamefont {Stracka}}, \bibinfo {author}
  {\bibfnamefont {G.}~\bibnamefont {Stutter}}, \bibinfo {author} {\bibfnamefont {T.~D.}\ \bibnamefont {Tharp}}, \bibinfo {author} {\bibfnamefont {K.~A.}\ \bibnamefont {Thompson}}, \bibinfo {author} {\bibfnamefont {R.~I.}\ \bibnamefont {Thompson}}, \bibinfo {author} {\bibfnamefont {E.}~\bibnamefont {Thorpe-Woods}}, \bibinfo {author} {\bibfnamefont {C.}~\bibnamefont {Torkzaban}}, \bibinfo {author} {\bibfnamefont {M.}~\bibnamefont {Urioni}}, \bibinfo {author} {\bibfnamefont {P.}~\bibnamefont {Woosaree}},\ and\ \bibinfo {author} {\bibfnamefont {J.~S.}\ \bibnamefont {Wurtele}},\ }\bibfield  {title} {\bibinfo {title} {Observation of the effect of gravity on the motion of antimatter},\ }\href {https://doi.org/10.1038/s41586-023-06527-1} {\bibfield  {journal} {\bibinfo  {journal} {Nature}\ }\textbf {\bibinfo {volume} {621}},\ \bibinfo {pages} {716} (\bibinfo {year} {2023})}\BibitemShut {NoStop}%
\bibitem [{\citenamefont {Teixeira}\ \emph {et~al.}(2022)\citenamefont {Teixeira}, \citenamefont {Aumann}, \citenamefont {Bertulani},\ and\ \citenamefont {Carlson}}]{Teixeira2022}%
  \BibitemOpen
  \bibfield  {author} {\bibinfo {author} {\bibfnamefont {E.~A.}\ \bibnamefont {Teixeira}}, \bibinfo {author} {\bibfnamefont {T.}~\bibnamefont {Aumann}}, \bibinfo {author} {\bibfnamefont {C.~A.}\ \bibnamefont {Bertulani}},\ and\ \bibinfo {author} {\bibfnamefont {B.~V.}\ \bibnamefont {Carlson}},\ }\bibfield  {title} {\bibinfo {title} {Nuclear fragmentation reactions as a probe of neutron skins in nuclei},\ }\href {https://doi.org/10.1140/epja/s10050-022-00849-w} {\bibfield  {journal} {\bibinfo  {journal} {Eur. Phys. J. A}\ }\textbf {\bibinfo {volume} {58}},\ \bibinfo {pages} {205} (\bibinfo {year} {2022})}\BibitemShut {NoStop}%
\bibitem [{\citenamefont {Trzcińska}\ \emph {et~al.}(2004)\citenamefont {Trzcińska}, \citenamefont {Jastrzebski}, \citenamefont {Lubiński}, \citenamefont {Hartmann}, \citenamefont {Schmidt}, \citenamefont {{von Egidy}},\ and\ \citenamefont {Klos}}]{TRZCINSKA2004157}%
  \BibitemOpen
  \bibfield  {author} {\bibinfo {author} {\bibfnamefont {A.}~\bibnamefont {Trzcińska}}, \bibinfo {author} {\bibfnamefont {J.}~\bibnamefont {Jastrzebski}}, \bibinfo {author} {\bibfnamefont {P.}~\bibnamefont {Lubiński}}, \bibinfo {author} {\bibfnamefont {F.}~\bibnamefont {Hartmann}}, \bibinfo {author} {\bibfnamefont {R.}~\bibnamefont {Schmidt}}, \bibinfo {author} {\bibfnamefont {T.}~\bibnamefont {{von Egidy}}},\ and\ \bibinfo {author} {\bibfnamefont {B.}~\bibnamefont {Klos}},\ }\bibfield  {title} {\bibinfo {title} {Information on the nuclear periphery deduced from the properties of heavy antiprotonic atoms},\ }\href {https://doi.org/https://doi.org/10.1016/j.nimb.2003.08.017} {\bibfield  {journal} {\bibinfo  {journal} {Nucl. Instrum. Methods Phys. Res., B}\ }\textbf {\bibinfo {volume} {214}},\ \bibinfo {pages} {157} (\bibinfo {year} {2004})},\ \bibinfo {note} {low Energy Antiproton Physics (LEAP'03)}\BibitemShut {NoStop}%
\bibitem [{\citenamefont {Zhang}\ and\ \citenamefont {Feng}(2025)}]{Zhang2025}%
  \BibitemOpen
  \bibfield  {author} {\bibinfo {author} {\bibfnamefont {B.}~\bibnamefont {Zhang}}\ and\ \bibinfo {author} {\bibfnamefont {Z.-Q.}\ \bibnamefont {Feng}},\ }\bibfield  {title} {\bibinfo {title} {Correlation of the symmetry energy at subsaturation densities and neutron-skin thickness in low-energy antiproton-induced reactions},\ }\href {https://doi.org/10.1103/PhysRevC.111.014607} {\bibfield  {journal} {\bibinfo  {journal} {Phys. Rev. C}\ }\textbf {\bibinfo {volume} {111}},\ \bibinfo {pages} {014607} (\bibinfo {year} {2025})}\BibitemShut {NoStop}%
\bibitem [{\citenamefont {Aumann}\ \emph {et~al.}(2022)\citenamefont {Aumann}, \citenamefont {Bartmann}, \citenamefont {Boine-Frankenheim}, \citenamefont {Bouvard}, \citenamefont {Broche}, \citenamefont {Butin}, \citenamefont {Calvet}, \citenamefont {Carbonell}, \citenamefont {Chiggiato}, \citenamefont {De~Gersem}, \citenamefont {De~Oliveira}, \citenamefont {Dobers}, \citenamefont {Ehm}, \citenamefont {Somoza}, \citenamefont {Fischer}, \citenamefont {Fraser}, \citenamefont {Friedrich}, \citenamefont {Frotscher}, \citenamefont {Gomez-Ramos}, \citenamefont {Grenard}, \citenamefont {Hobl}, \citenamefont {Hupin}, \citenamefont {Husson}, \citenamefont {Indelicato}, \citenamefont {Johnston}, \citenamefont {Klink}, \citenamefont {Kubota}, \citenamefont {Lazauskas}, \citenamefont {Malbrunot-Ettenauer}, \citenamefont {Marsic}, \citenamefont {O~Müller}, \citenamefont {Naimi}, \citenamefont {Nakatsuka}, \citenamefont {Necca}, \citenamefont {Neidherr}, \citenamefont {Neyens}, \citenamefont {Obertelli}, \citenamefont
  {Ono}, \citenamefont {Pasinelli}, \citenamefont {Paul}, \citenamefont {Pollacco}, \citenamefont {Rossi}, \citenamefont {Scheit}, \citenamefont {Schlaich}, \citenamefont {Schmidt}, \citenamefont {Schweikhard}, \citenamefont {Seki}, \citenamefont {Sels}, \citenamefont {Siesling}, \citenamefont {Uesaka}, \citenamefont {Vilén}, \citenamefont {Wada}, \citenamefont {Wienholtz}, \citenamefont {Wycech},\ and\ \citenamefont {Zacarias}}]{Aumann2022}%
  \BibitemOpen
  \bibfield  {author} {\bibinfo {author} {\bibfnamefont {T.}~\bibnamefont {Aumann}}, \bibinfo {author} {\bibfnamefont {W.}~\bibnamefont {Bartmann}}, \bibinfo {author} {\bibfnamefont {O.}~\bibnamefont {Boine-Frankenheim}}, \bibinfo {author} {\bibfnamefont {A.}~\bibnamefont {Bouvard}}, \bibinfo {author} {\bibfnamefont {A.}~\bibnamefont {Broche}}, \bibinfo {author} {\bibfnamefont {F.}~\bibnamefont {Butin}}, \bibinfo {author} {\bibfnamefont {D.}~\bibnamefont {Calvet}}, \bibinfo {author} {\bibfnamefont {J.}~\bibnamefont {Carbonell}}, \bibinfo {author} {\bibfnamefont {P.}~\bibnamefont {Chiggiato}}, \bibinfo {author} {\bibfnamefont {H.}~\bibnamefont {De~Gersem}}, \bibinfo {author} {\bibfnamefont {R.}~\bibnamefont {De~Oliveira}}, \bibinfo {author} {\bibfnamefont {T.}~\bibnamefont {Dobers}}, \bibinfo {author} {\bibfnamefont {F.}~\bibnamefont {Ehm}}, \bibinfo {author} {\bibfnamefont {J.~F.}\ \bibnamefont {Somoza}}, \bibinfo {author} {\bibfnamefont {J.}~\bibnamefont {Fischer}}, \bibinfo {author} {\bibfnamefont
  {M.}~\bibnamefont {Fraser}}, \bibinfo {author} {\bibfnamefont {E.}~\bibnamefont {Friedrich}}, \bibinfo {author} {\bibfnamefont {A.}~\bibnamefont {Frotscher}}, \bibinfo {author} {\bibfnamefont {M.}~\bibnamefont {Gomez-Ramos}}, \bibinfo {author} {\bibfnamefont {J.-L.}\ \bibnamefont {Grenard}}, \bibinfo {author} {\bibfnamefont {A.}~\bibnamefont {Hobl}}, \bibinfo {author} {\bibfnamefont {G.}~\bibnamefont {Hupin}}, \bibinfo {author} {\bibfnamefont {A.}~\bibnamefont {Husson}}, \bibinfo {author} {\bibfnamefont {P.}~\bibnamefont {Indelicato}}, \bibinfo {author} {\bibfnamefont {K.}~\bibnamefont {Johnston}}, \bibinfo {author} {\bibfnamefont {C.}~\bibnamefont {Klink}}, \bibinfo {author} {\bibfnamefont {Y.}~\bibnamefont {Kubota}}, \bibinfo {author} {\bibfnamefont {R.}~\bibnamefont {Lazauskas}}, \bibinfo {author} {\bibfnamefont {S.}~\bibnamefont {Malbrunot-Ettenauer}}, \bibinfo {author} {\bibfnamefont {N.}~\bibnamefont {Marsic}}, \bibinfo {author} {\bibfnamefont {W.~F.}\ \bibnamefont {O~Müller}}, \bibinfo {author}
  {\bibfnamefont {S.}~\bibnamefont {Naimi}}, \bibinfo {author} {\bibfnamefont {N.}~\bibnamefont {Nakatsuka}}, \bibinfo {author} {\bibfnamefont {R.}~\bibnamefont {Necca}}, \bibinfo {author} {\bibfnamefont {D.}~\bibnamefont {Neidherr}}, \bibinfo {author} {\bibfnamefont {G.}~\bibnamefont {Neyens}}, \bibinfo {author} {\bibfnamefont {A.}~\bibnamefont {Obertelli}}, \bibinfo {author} {\bibfnamefont {Y.}~\bibnamefont {Ono}}, \bibinfo {author} {\bibfnamefont {S.}~\bibnamefont {Pasinelli}}, \bibinfo {author} {\bibfnamefont {N.}~\bibnamefont {Paul}}, \bibinfo {author} {\bibfnamefont {E.~C.}\ \bibnamefont {Pollacco}}, \bibinfo {author} {\bibfnamefont {D.}~\bibnamefont {Rossi}}, \bibinfo {author} {\bibfnamefont {H.}~\bibnamefont {Scheit}}, \bibinfo {author} {\bibfnamefont {M.}~\bibnamefont {Schlaich}}, \bibinfo {author} {\bibfnamefont {A.}~\bibnamefont {Schmidt}}, \bibinfo {author} {\bibfnamefont {L.}~\bibnamefont {Schweikhard}}, \bibinfo {author} {\bibfnamefont {R.}~\bibnamefont {Seki}}, \bibinfo {author} {\bibfnamefont
  {S.}~\bibnamefont {Sels}}, \bibinfo {author} {\bibfnamefont {E.}~\bibnamefont {Siesling}}, \bibinfo {author} {\bibfnamefont {T.}~\bibnamefont {Uesaka}}, \bibinfo {author} {\bibfnamefont {M.}~\bibnamefont {Vilén}}, \bibinfo {author} {\bibfnamefont {M.}~\bibnamefont {Wada}}, \bibinfo {author} {\bibfnamefont {F.}~\bibnamefont {Wienholtz}}, \bibinfo {author} {\bibfnamefont {S.}~\bibnamefont {Wycech}},\ and\ \bibinfo {author} {\bibfnamefont {S.}~\bibnamefont {Zacarias}},\ }\bibfield  {title} {\bibinfo {title} {Puma, antiproton unstable matter annihilation},\ }\href {https://doi.org/10.1140/epja/s10050-022-00713-x} {\bibfield  {journal} {\bibinfo  {journal} {Eur. J. Phys. A}\ }\textbf {\bibinfo {volume} {58}},\ \bibinfo {pages} {88} (\bibinfo {year} {2022})}\BibitemShut {NoStop}%
\bibitem [{\citenamefont {Gibney}(2024)}]{Gibney2024}%
  \BibitemOpen
  \bibfield  {author} {\bibinfo {author} {\bibfnamefont {E.}~\bibnamefont {Gibney}},\ }\bibfield  {title} {\bibinfo {title} {Antimatter to be transported outside a lab for first time — in a van},\ }\href {https://doi.org/10.1038/d41586-024-03841-0} {\bibfield  {journal} {\bibinfo  {journal} {Nature}\ }\textbf {\bibinfo {volume} {636}},\ \bibinfo {pages} {13} (\bibinfo {year} {2024})}\BibitemShut {NoStop}%
\bibitem [{\citenamefont {Leonhardt}\ \emph {et~al.}(2025)\citenamefont {Leonhardt}, \citenamefont {Schweitzer}, \citenamefont {Abbass}, \citenamefont {Anjum}, \citenamefont {Arndt}, \citenamefont {Erlewein}, \citenamefont {Endoh}, \citenamefont {Geissler}, \citenamefont {Imamura}, \citenamefont {Jäger}, \citenamefont {Latacz}, \citenamefont {Micke}, \citenamefont {Voelksen}, \citenamefont {Yildiz}, \citenamefont {Blaum}, \citenamefont {Devlin}, \citenamefont {Matsuda}, \citenamefont {Ospelkaus}, \citenamefont {Quint}, \citenamefont {Soter}, \citenamefont {Walz}, \citenamefont {Yamazaki}, \citenamefont {Ulmer},\ and\ \citenamefont {Smorra}}]{Leonhardt2025}%
  \BibitemOpen
  \bibfield  {author} {\bibinfo {author} {\bibfnamefont {M.}~\bibnamefont {Leonhardt}}, \bibinfo {author} {\bibfnamefont {D.}~\bibnamefont {Schweitzer}}, \bibinfo {author} {\bibfnamefont {F.}~\bibnamefont {Abbass}}, \bibinfo {author} {\bibfnamefont {K.~K.}\ \bibnamefont {Anjum}}, \bibinfo {author} {\bibfnamefont {B.}~\bibnamefont {Arndt}}, \bibinfo {author} {\bibfnamefont {S.}~\bibnamefont {Erlewein}}, \bibinfo {author} {\bibfnamefont {S.}~\bibnamefont {Endoh}}, \bibinfo {author} {\bibfnamefont {P.}~\bibnamefont {Geissler}}, \bibinfo {author} {\bibfnamefont {T.}~\bibnamefont {Imamura}}, \bibinfo {author} {\bibfnamefont {J.~I.}\ \bibnamefont {Jäger}}, \bibinfo {author} {\bibfnamefont {B.~M.}\ \bibnamefont {Latacz}}, \bibinfo {author} {\bibfnamefont {P.}~\bibnamefont {Micke}}, \bibinfo {author} {\bibfnamefont {F.}~\bibnamefont {Voelksen}}, \bibinfo {author} {\bibfnamefont {H.}~\bibnamefont {Yildiz}}, \bibinfo {author} {\bibfnamefont {K.}~\bibnamefont {Blaum}}, \bibinfo {author} {\bibfnamefont {J.~A.}\
  \bibnamefont {Devlin}}, \bibinfo {author} {\bibfnamefont {Y.}~\bibnamefont {Matsuda}}, \bibinfo {author} {\bibfnamefont {C.}~\bibnamefont {Ospelkaus}}, \bibinfo {author} {\bibfnamefont {W.}~\bibnamefont {Quint}}, \bibinfo {author} {\bibfnamefont {A.}~\bibnamefont {Soter}}, \bibinfo {author} {\bibfnamefont {J.}~\bibnamefont {Walz}}, \bibinfo {author} {\bibfnamefont {Y.}~\bibnamefont {Yamazaki}}, \bibinfo {author} {\bibfnamefont {S.}~\bibnamefont {Ulmer}},\ and\ \bibinfo {author} {\bibfnamefont {C.}~\bibnamefont {Smorra}},\ }\bibfield  {title} {\bibinfo {title} {Proton transport from the antimatter factory of cern},\ }\href {https://doi.org/10.1038/s41586-025-08926-y} {\bibfield  {journal} {\bibinfo  {journal} {Nature}\ }\textbf {\bibinfo {volume} {641}},\ \bibinfo {pages} {871} (\bibinfo {year} {2025})}\BibitemShut {NoStop}%
\bibitem [{\citenamefont {Shabaev}\ \emph {et~al.}(2018)\citenamefont {Shabaev}, \citenamefont {Bondarev}, \citenamefont {Glazov}, \citenamefont {Kaygorodov}, \citenamefont {Kozhedub}, \citenamefont {Maltsev}, \citenamefont {Malyshev}, \citenamefont {Popov}, \citenamefont {Tupitsyn},\ and\ \citenamefont {Zubova}}]{Shabaev2018}%
  \BibitemOpen
  \bibfield  {author} {\bibinfo {author} {\bibfnamefont {V.~M.}\ \bibnamefont {Shabaev}}, \bibinfo {author} {\bibfnamefont {A.~I.}\ \bibnamefont {Bondarev}}, \bibinfo {author} {\bibfnamefont {D.~A.}\ \bibnamefont {Glazov}}, \bibinfo {author} {\bibfnamefont {M.~Y.}\ \bibnamefont {Kaygorodov}}, \bibinfo {author} {\bibfnamefont {Y.~S.}\ \bibnamefont {Kozhedub}}, \bibinfo {author} {\bibfnamefont {I.~A.}\ \bibnamefont {Maltsev}}, \bibinfo {author} {\bibfnamefont {A.~V.}\ \bibnamefont {Malyshev}}, \bibinfo {author} {\bibfnamefont {R.~V.}\ \bibnamefont {Popov}}, \bibinfo {author} {\bibfnamefont {I.~I.}\ \bibnamefont {Tupitsyn}},\ and\ \bibinfo {author} {\bibfnamefont {N.~A.}\ \bibnamefont {Zubova}},\ }\bibfield  {title} {\bibinfo {title} {Stringent tests of qed using highly charged ions},\ }\href {https://doi.org/10.1007/s10751-018-1537-8} {\bibfield  {journal} {\bibinfo  {journal} {Hyperfine Interact.}\ }\textbf {\bibinfo {volume} {239}},\ \bibinfo {pages} {60} (\bibinfo {year} {2018})}\BibitemShut {NoStop}%
\bibitem [{\citenamefont {Indelicato}(2019)}]{Indelicato2019}%
  \BibitemOpen
  \bibfield  {author} {\bibinfo {author} {\bibfnamefont {P.}~\bibnamefont {Indelicato}},\ }\bibfield  {title} {\bibinfo {title} {Qed tests with highly charged ions},\ }\href {https://doi.org/10.1088/1361-6455/ab42c9} {\bibfield  {journal} {\bibinfo  {journal} {J. Phys. B}\ }\textbf {\bibinfo {volume} {52}},\ \bibinfo {pages} {232001} (\bibinfo {year} {2019})}\BibitemShut {NoStop}%
\bibitem [{\citenamefont {Morgner}\ \emph {et~al.}(2023)\citenamefont {Morgner}, \citenamefont {Tu}, \citenamefont {König}, \citenamefont {Sailer}, \citenamefont {Heiße}, \citenamefont {Bekker}, \citenamefont {Sikora}, \citenamefont {Lyu}, \citenamefont {Yerokhin}, \citenamefont {Harman}, \citenamefont {Crespo López-Urrutia}, \citenamefont {Keitel}, \citenamefont {Sturm},\ and\ \citenamefont {Blaum}}]{Morgner2023}%
  \BibitemOpen
  \bibfield  {author} {\bibinfo {author} {\bibfnamefont {J.}~\bibnamefont {Morgner}}, \bibinfo {author} {\bibfnamefont {B.}~\bibnamefont {Tu}}, \bibinfo {author} {\bibfnamefont {C.~M.}\ \bibnamefont {König}}, \bibinfo {author} {\bibfnamefont {T.}~\bibnamefont {Sailer}}, \bibinfo {author} {\bibfnamefont {F.}~\bibnamefont {Heiße}}, \bibinfo {author} {\bibfnamefont {H.}~\bibnamefont {Bekker}}, \bibinfo {author} {\bibfnamefont {B.}~\bibnamefont {Sikora}}, \bibinfo {author} {\bibfnamefont {C.}~\bibnamefont {Lyu}}, \bibinfo {author} {\bibfnamefont {V.~A.}\ \bibnamefont {Yerokhin}}, \bibinfo {author} {\bibfnamefont {Z.}~\bibnamefont {Harman}}, \bibinfo {author} {\bibfnamefont {J.~R.}\ \bibnamefont {Crespo López-Urrutia}}, \bibinfo {author} {\bibfnamefont {C.~H.}\ \bibnamefont {Keitel}}, \bibinfo {author} {\bibfnamefont {S.}~\bibnamefont {Sturm}},\ and\ \bibinfo {author} {\bibfnamefont {K.}~\bibnamefont {Blaum}},\ }\bibfield  {title} {\bibinfo {title} {Stringent test of qed with hydrogen-like tin},\ }\href
  {https://doi.org/10.1038/s41586-023-06453-2} {\bibfield  {journal} {\bibinfo  {journal} {Nature}\ }\textbf {\bibinfo {volume} {622}},\ \bibinfo {pages} {53} (\bibinfo {year} {2023})}\BibitemShut {NoStop}%
\bibitem [{\citenamefont {Kozlov}\ \emph {et~al.}(2018)\citenamefont {Kozlov}, \citenamefont {Safronova}, \citenamefont {Crespo L\'opez-Urrutia},\ and\ \citenamefont {Schmidt}}]{Kozlov2018}%
  \BibitemOpen
  \bibfield  {author} {\bibinfo {author} {\bibfnamefont {M.~G.}\ \bibnamefont {Kozlov}}, \bibinfo {author} {\bibfnamefont {M.~S.}\ \bibnamefont {Safronova}}, \bibinfo {author} {\bibfnamefont {J.~R.}\ \bibnamefont {Crespo L\'opez-Urrutia}},\ and\ \bibinfo {author} {\bibfnamefont {P.~O.}\ \bibnamefont {Schmidt}},\ }\bibfield  {title} {\bibinfo {title} {Highly charged ions: Optical clocks and applications in fundamental physics},\ }\href {https://doi.org/10.1103/RevModPhys.90.045005} {\bibfield  {journal} {\bibinfo  {journal} {Rev. Mod. Phys.}\ }\textbf {\bibinfo {volume} {90}},\ \bibinfo {pages} {045005} (\bibinfo {year} {2018})}\BibitemShut {NoStop}%
\bibitem [{\citenamefont {Sun}\ \emph {et~al.}(2024)\citenamefont {Sun}, \citenamefont {Valuev},\ and\ \citenamefont {Oreshkina}}]{Sun2024}%
  \BibitemOpen
  \bibfield  {author} {\bibinfo {author} {\bibfnamefont {Z.}~\bibnamefont {Sun}}, \bibinfo {author} {\bibfnamefont {I.~A.}\ \bibnamefont {Valuev}},\ and\ \bibinfo {author} {\bibfnamefont {N.~S.}\ \bibnamefont {Oreshkina}},\ }\bibfield  {title} {\bibinfo {title} {Nuclear deformation effects in the spectra of highly charged ions},\ }\href {https://doi.org/10.1103/PhysRevResearch.6.023327} {\bibfield  {journal} {\bibinfo  {journal} {Phys. Rev. Res.}\ }\textbf {\bibinfo {volume} {6}},\ \bibinfo {pages} {023327} (\bibinfo {year} {2024})}\BibitemShut {NoStop}%
\bibitem [{\citenamefont {Campbell}\ \emph {et~al.}(2016)\citenamefont {Campbell}, \citenamefont {Moore},\ and\ \citenamefont {Pearson}}]{Campbell2016}%
  \BibitemOpen
  \bibfield  {author} {\bibinfo {author} {\bibfnamefont {P.}~\bibnamefont {Campbell}}, \bibinfo {author} {\bibfnamefont {I.}~\bibnamefont {Moore}},\ and\ \bibinfo {author} {\bibfnamefont {M.}~\bibnamefont {Pearson}},\ }\bibfield  {title} {\bibinfo {title} {Laser spectroscopy for nuclear structure physics},\ }\href {https://doi.org/https://doi.org/10.1016/j.ppnp.2015.09.003} {\bibfield  {journal} {\bibinfo  {journal} {Progress in Particle and Nuclear Physics}\ }\textbf {\bibinfo {volume} {86}},\ \bibinfo {pages} {127} (\bibinfo {year} {2016})}\BibitemShut {NoStop}%
\bibitem [{\citenamefont {López-Urrutia}(2016)}]{LopezUrrutia2016}%
  \BibitemOpen
  \bibfield  {author} {\bibinfo {author} {\bibfnamefont {J.~R.~C.}\ \bibnamefont {López-Urrutia}},\ }\bibfield  {title} {\bibinfo {title} {Frequency metrology using highly charged ions},\ }\href {https://doi.org/10.1088/1742-6596/723/1/012052} {\bibfield  {journal} {\bibinfo  {journal} {J. Phys. Conf. Ser.}\ }\textbf {\bibinfo {volume} {723}},\ \bibinfo {pages} {012052} (\bibinfo {year} {2016})}\BibitemShut {NoStop}%
\bibitem [{\citenamefont {Levine}\ \emph {et~al.}(1989)\citenamefont {Levine}, \citenamefont {Marrs}, \citenamefont {Bardsley}, \citenamefont {Beiersdorfer}, \citenamefont {Bennett}, \citenamefont {Chen}, \citenamefont {Cowan}, \citenamefont {Dietrich}, \citenamefont {Henderson}, \citenamefont {Knapp}, \citenamefont {Osterheld}, \citenamefont {Penetrante}, \citenamefont {Schneider},\ and\ \citenamefont {Scofield}}]{Levine1989}%
  \BibitemOpen
  \bibfield  {author} {\bibinfo {author} {\bibfnamefont {M.}~\bibnamefont {Levine}}, \bibinfo {author} {\bibfnamefont {R.}~\bibnamefont {Marrs}}, \bibinfo {author} {\bibfnamefont {J.}~\bibnamefont {Bardsley}}, \bibinfo {author} {\bibfnamefont {P.}~\bibnamefont {Beiersdorfer}}, \bibinfo {author} {\bibfnamefont {C.}~\bibnamefont {Bennett}}, \bibinfo {author} {\bibfnamefont {M.}~\bibnamefont {Chen}}, \bibinfo {author} {\bibfnamefont {T.}~\bibnamefont {Cowan}}, \bibinfo {author} {\bibfnamefont {D.}~\bibnamefont {Dietrich}}, \bibinfo {author} {\bibfnamefont {J.}~\bibnamefont {Henderson}}, \bibinfo {author} {\bibfnamefont {D.}~\bibnamefont {Knapp}}, \bibinfo {author} {\bibfnamefont {A.}~\bibnamefont {Osterheld}}, \bibinfo {author} {\bibfnamefont {B.}~\bibnamefont {Penetrante}}, \bibinfo {author} {\bibfnamefont {M.}~\bibnamefont {Schneider}},\ and\ \bibinfo {author} {\bibfnamefont {J.}~\bibnamefont {Scofield}},\ }\bibfield  {title} {\bibinfo {title} {The use of an electron beam ion trap in the study of highly
  charged ions},\ }\href {https://doi.org/https://doi.org/10.1016/0168-583X(89)90386-8} {\bibfield  {journal} {\bibinfo  {journal} {Nucl. Instrum. Methods Phys. Res. B}\ }\textbf {\bibinfo {volume} {43}},\ \bibinfo {pages} {431} (\bibinfo {year} {1989})}\BibitemShut {NoStop}%
\bibitem [{\citenamefont {Steck}\ and\ \citenamefont {Litvinov}(2020)}]{Steck2020}%
  \BibitemOpen
  \bibfield  {author} {\bibinfo {author} {\bibfnamefont {M.}~\bibnamefont {Steck}}\ and\ \bibinfo {author} {\bibfnamefont {Y.~A.}\ \bibnamefont {Litvinov}},\ }\bibfield  {title} {\bibinfo {title} {Heavy-ion storage rings and their use in precision experiments with highly charged ions},\ }\href {https://doi.org/https://doi.org/10.1016/j.ppnp.2020.103811} {\bibfield  {journal} {\bibinfo  {journal} {Prog. Part. Nucl. Phys.}\ }\textbf {\bibinfo {volume} {115}},\ \bibinfo {pages} {103811} (\bibinfo {year} {2020})}\BibitemShut {NoStop}%
\bibitem [{\citenamefont {Tanabashi}\ \emph {et~al.}(2018)\citenamefont {Tanabashi}, \citenamefont {Hagiwara}, \citenamefont {Hikasa}, \citenamefont {Nakamura}, \citenamefont {Sumino}, \citenamefont {Takahashi}, \citenamefont {Tanaka}, \citenamefont {Agashe}, \citenamefont {Aielli}, \citenamefont {Amsler}, \citenamefont {Antonelli}, \citenamefont {Asner}, \citenamefont {Baer}, \citenamefont {Banerjee}, \citenamefont {Barnett}, \citenamefont {Basaglia}, \citenamefont {Bauer}, \citenamefont {Beatty}, \citenamefont {Belousov}, \citenamefont {Beringer}, \citenamefont {Bethke}, \citenamefont {Bettini}, \citenamefont {Bichsel}, \citenamefont {Biebel}, \citenamefont {Black}, \citenamefont {Blucher}, \citenamefont {Buchmuller}, \citenamefont {Burkert}, \citenamefont {Bychkov}, \citenamefont {Cahn}, \citenamefont {Carena}, \citenamefont {Ceccucci}, \citenamefont {Cerri}, \citenamefont {Chakraborty}, \citenamefont {Chen}, \citenamefont {Chivukula}, \citenamefont {Cowan}, \citenamefont {Dahl}, \citenamefont
  {D'Ambrosio}, \citenamefont {Damour}, \citenamefont {de~Florian}, \citenamefont {de~Gouv\^ea}, \citenamefont {DeGrand}, \citenamefont {de~Jong}, \citenamefont {Dissertori}, \citenamefont {Dobrescu}, \citenamefont {D'Onofrio}, \citenamefont {Doser}, \citenamefont {Drees}, \citenamefont {Dreiner}, \citenamefont {Dwyer}, \citenamefont {Eerola}, \citenamefont {Eidelman}, \citenamefont {Ellis}, \citenamefont {Erler}, \citenamefont {Ezhela}, \citenamefont {Fetscher}, \citenamefont {Fields}, \citenamefont {Firestone}, \citenamefont {Foster}, \citenamefont {Freitas}, \citenamefont {Gallagher}, \citenamefont {Garren}, \citenamefont {Gerber}, \citenamefont {Gerbier}, \citenamefont {Gershon}, \citenamefont {Gershtein}, \citenamefont {Gherghetta}, \citenamefont {Godizov}, \citenamefont {Goodman}, \citenamefont {Grab}, \citenamefont {Gritsan}, \citenamefont {Grojean}, \citenamefont {Groom}, \citenamefont {Gr\"unewald}, \citenamefont {Gurtu}, \citenamefont {Gutsche}, \citenamefont {Haber}, \citenamefont {Hanhart},
  \citenamefont {Hashimoto}, \citenamefont {Hayato}, \citenamefont {Hayes}, \citenamefont {Hebecker}, \citenamefont {Heinemeyer}, \citenamefont {Heltsley}, \citenamefont {Hern\'andez-Rey}, \citenamefont {Hisano}, \citenamefont {H\"ocker}, \citenamefont {Holder}, \citenamefont {Holtkamp}, \citenamefont {Hyodo}, \citenamefont {Irwin}, \citenamefont {Johnson}, \citenamefont {Kado}, \citenamefont {Karliner}, \citenamefont {Katz}, \citenamefont {Klein}, \citenamefont {Klempt}, \citenamefont {Kowalewski}, \citenamefont {Krauss}, \citenamefont {Kreps}, \citenamefont {Krusche}, \citenamefont {Kuyanov}, \citenamefont {Kwon}, \citenamefont {Lahav}, \citenamefont {Laiho}, \citenamefont {Lesgourgues}, \citenamefont {Liddle}, \citenamefont {Ligeti}, \citenamefont {Lin}, \citenamefont {Lippmann}, \citenamefont {Liss}, \citenamefont {Littenberg}, \citenamefont {Lugovsky}, \citenamefont {Lugovsky}, \citenamefont {Lusiani}, \citenamefont {Makida}, \citenamefont {Maltoni}, \citenamefont {Mannel}, \citenamefont {Manohar},
  \citenamefont {Marciano}, \citenamefont {Martin}, \citenamefont {Masoni}, \citenamefont {Matthews}, \citenamefont {Mei\ss{}ner}, \citenamefont {Milstead}, \citenamefont {Mitchell}, \citenamefont {M\"onig}, \citenamefont {Molaro}, \citenamefont {Moortgat}, \citenamefont {Moskovic}, \citenamefont {Murayama}, \citenamefont {Narain}, \citenamefont {Nason}, \citenamefont {Navas}, \citenamefont {Neubert}, \citenamefont {Nevski}, \citenamefont {Nir}, \citenamefont {Olive}, \citenamefont {Pagan~Griso}, \citenamefont {Parsons}, \citenamefont {Patrignani}, \citenamefont {Peacock}, \citenamefont {Pennington}, \citenamefont {Petcov}, \citenamefont {Petrov}, \citenamefont {Pianori}, \citenamefont {Piepke}, \citenamefont {Pomarol}, \citenamefont {Quadt}, \citenamefont {Rademacker}, \citenamefont {Raffelt}, \citenamefont {Ratcliff}, \citenamefont {Richardson}, \citenamefont {Ringwald}, \citenamefont {Roesler}, \citenamefont {Rolli}, \citenamefont {Romaniouk}, \citenamefont {Rosenberg}, \citenamefont {Rosner},
  \citenamefont {Rybka}, \citenamefont {Ryutin}, \citenamefont {Sachrajda}, \citenamefont {Sakai}, \citenamefont {Salam}, \citenamefont {Sarkar}, \citenamefont {Sauli}, \citenamefont {Schneider}, \citenamefont {Scholberg}, \citenamefont {Schwartz}, \citenamefont {Scott}, \citenamefont {Sharma}, \citenamefont {Sharpe}, \citenamefont {Shutt}, \citenamefont {Silari}, \citenamefont {Sj\"ostrand}, \citenamefont {Skands}, \citenamefont {Skwarnicki}, \citenamefont {Smith}, \citenamefont {Smoot}, \citenamefont {Spanier}, \citenamefont {Spieler}, \citenamefont {Spiering}, \citenamefont {Stahl}, \citenamefont {Stone}, \citenamefont {Sumiyoshi}, \citenamefont {Syphers}, \citenamefont {Terashi}, \citenamefont {Terning}, \citenamefont {Thoma}, \citenamefont {Thorne}, \citenamefont {Tiator}, \citenamefont {Titov}, \citenamefont {Tkachenko}, \citenamefont {T\"ornqvist}, \citenamefont {Tovey}, \citenamefont {Valencia}, \citenamefont {Van~de Water}, \citenamefont {Varelas}, \citenamefont {Venanzoni}, \citenamefont {Verde},
  \citenamefont {Vincter}, \citenamefont {Vogel}, \citenamefont {Vogt}, \citenamefont {Wakely}, \citenamefont {Walkowiak}, \citenamefont {Walter}, \citenamefont {Wands}, \citenamefont {Ward}, \citenamefont {Wascko}, \citenamefont {Weiglein}, \citenamefont {Weinberg}, \citenamefont {Weinberg}, \citenamefont {White}, \citenamefont {Wiencke}, \citenamefont {Willocq}, \citenamefont {Wohl}, \citenamefont {Womersley}, \citenamefont {Woody}, \citenamefont {Workman}, \citenamefont {Yao}, \citenamefont {Zeller}, \citenamefont {Zenin}, \citenamefont {Zhu}, \citenamefont {Zhu}, \citenamefont {Zimmermann}, \citenamefont {Zyla}, \citenamefont {Anderson}, \citenamefont {Fuller}, \citenamefont {Lugovsky},\ and\ \citenamefont {Schaffner}}]{PDG}%
  \BibitemOpen
  \bibfield  {author} {\bibinfo {author} {\bibfnamefont {M.}~\bibnamefont {Tanabashi}}, \bibinfo {author} {\bibfnamefont {K.}~\bibnamefont {Hagiwara}}, \bibinfo {author} {\bibfnamefont {K.}~\bibnamefont {Hikasa}}, \bibinfo {author} {\bibfnamefont {K.}~\bibnamefont {Nakamura}}, \bibinfo {author} {\bibfnamefont {Y.}~\bibnamefont {Sumino}}, \bibinfo {author} {\bibfnamefont {F.}~\bibnamefont {Takahashi}}, \bibinfo {author} {\bibfnamefont {J.}~\bibnamefont {Tanaka}}, \bibinfo {author} {\bibfnamefont {K.}~\bibnamefont {Agashe}}, \bibinfo {author} {\bibfnamefont {G.}~\bibnamefont {Aielli}}, \bibinfo {author} {\bibfnamefont {C.}~\bibnamefont {Amsler}}, \bibinfo {author} {\bibfnamefont {M.}~\bibnamefont {Antonelli}}, \bibinfo {author} {\bibfnamefont {D.~M.}\ \bibnamefont {Asner}}, \bibinfo {author} {\bibfnamefont {H.}~\bibnamefont {Baer}}, \bibinfo {author} {\bibfnamefont {S.}~\bibnamefont {Banerjee}}, \bibinfo {author} {\bibfnamefont {R.~M.}\ \bibnamefont {Barnett}}, \bibinfo {author} {\bibfnamefont {T.}~\bibnamefont
  {Basaglia}}, \bibinfo {author} {\bibfnamefont {C.~W.}\ \bibnamefont {Bauer}}, \bibinfo {author} {\bibfnamefont {J.~J.}\ \bibnamefont {Beatty}}, \bibinfo {author} {\bibfnamefont {V.~I.}\ \bibnamefont {Belousov}}, \bibinfo {author} {\bibfnamefont {J.}~\bibnamefont {Beringer}}, \bibinfo {author} {\bibfnamefont {S.}~\bibnamefont {Bethke}}, \bibinfo {author} {\bibfnamefont {A.}~\bibnamefont {Bettini}}, \bibinfo {author} {\bibfnamefont {H.}~\bibnamefont {Bichsel}}, \bibinfo {author} {\bibfnamefont {O.}~\bibnamefont {Biebel}}, \bibinfo {author} {\bibfnamefont {K.~M.}\ \bibnamefont {Black}}, \bibinfo {author} {\bibfnamefont {E.}~\bibnamefont {Blucher}}, \bibinfo {author} {\bibfnamefont {O.}~\bibnamefont {Buchmuller}}, \bibinfo {author} {\bibfnamefont {V.}~\bibnamefont {Burkert}}, \bibinfo {author} {\bibfnamefont {M.~A.}\ \bibnamefont {Bychkov}}, \bibinfo {author} {\bibfnamefont {R.~N.}\ \bibnamefont {Cahn}}, \bibinfo {author} {\bibfnamefont {M.}~\bibnamefont {Carena}}, \bibinfo {author} {\bibfnamefont
  {A.}~\bibnamefont {Ceccucci}}, \bibinfo {author} {\bibfnamefont {A.}~\bibnamefont {Cerri}}, \bibinfo {author} {\bibfnamefont {D.}~\bibnamefont {Chakraborty}}, \bibinfo {author} {\bibfnamefont {M.-C.}\ \bibnamefont {Chen}}, \bibinfo {author} {\bibfnamefont {R.~S.}\ \bibnamefont {Chivukula}}, \bibinfo {author} {\bibfnamefont {G.}~\bibnamefont {Cowan}}, \bibinfo {author} {\bibfnamefont {O.}~\bibnamefont {Dahl}}, \bibinfo {author} {\bibfnamefont {G.}~\bibnamefont {D'Ambrosio}}, \bibinfo {author} {\bibfnamefont {T.}~\bibnamefont {Damour}}, \bibinfo {author} {\bibfnamefont {D.}~\bibnamefont {de~Florian}}, \bibinfo {author} {\bibfnamefont {A.}~\bibnamefont {de~Gouv\^ea}}, \bibinfo {author} {\bibfnamefont {T.}~\bibnamefont {DeGrand}}, \bibinfo {author} {\bibfnamefont {P.}~\bibnamefont {de~Jong}}, \bibinfo {author} {\bibfnamefont {G.}~\bibnamefont {Dissertori}}, \bibinfo {author} {\bibfnamefont {B.~A.}\ \bibnamefont {Dobrescu}}, \bibinfo {author} {\bibfnamefont {M.}~\bibnamefont {D'Onofrio}}, \bibinfo {author}
  {\bibfnamefont {M.}~\bibnamefont {Doser}}, \bibinfo {author} {\bibfnamefont {M.}~\bibnamefont {Drees}}, \bibinfo {author} {\bibfnamefont {H.~K.}\ \bibnamefont {Dreiner}}, \bibinfo {author} {\bibfnamefont {D.~A.}\ \bibnamefont {Dwyer}}, \bibinfo {author} {\bibfnamefont {P.}~\bibnamefont {Eerola}}, \bibinfo {author} {\bibfnamefont {S.}~\bibnamefont {Eidelman}}, \bibinfo {author} {\bibfnamefont {J.}~\bibnamefont {Ellis}}, \bibinfo {author} {\bibfnamefont {J.}~\bibnamefont {Erler}}, \bibinfo {author} {\bibfnamefont {V.~V.}\ \bibnamefont {Ezhela}}, \bibinfo {author} {\bibfnamefont {W.}~\bibnamefont {Fetscher}}, \bibinfo {author} {\bibfnamefont {B.~D.}\ \bibnamefont {Fields}}, \bibinfo {author} {\bibfnamefont {R.}~\bibnamefont {Firestone}}, \bibinfo {author} {\bibfnamefont {B.}~\bibnamefont {Foster}}, \bibinfo {author} {\bibfnamefont {A.}~\bibnamefont {Freitas}}, \bibinfo {author} {\bibfnamefont {H.}~\bibnamefont {Gallagher}}, \bibinfo {author} {\bibfnamefont {L.}~\bibnamefont {Garren}}, \bibinfo {author}
  {\bibfnamefont {H.-J.}\ \bibnamefont {Gerber}}, \bibinfo {author} {\bibfnamefont {G.}~\bibnamefont {Gerbier}}, \bibinfo {author} {\bibfnamefont {T.}~\bibnamefont {Gershon}}, \bibinfo {author} {\bibfnamefont {Y.}~\bibnamefont {Gershtein}}, \bibinfo {author} {\bibfnamefont {T.}~\bibnamefont {Gherghetta}}, \bibinfo {author} {\bibfnamefont {A.~A.}\ \bibnamefont {Godizov}}, \bibinfo {author} {\bibfnamefont {M.}~\bibnamefont {Goodman}}, \bibinfo {author} {\bibfnamefont {C.}~\bibnamefont {Grab}}, \bibinfo {author} {\bibfnamefont {A.~V.}\ \bibnamefont {Gritsan}}, \bibinfo {author} {\bibfnamefont {C.}~\bibnamefont {Grojean}}, \bibinfo {author} {\bibfnamefont {D.~E.}\ \bibnamefont {Groom}}, \bibinfo {author} {\bibfnamefont {M.}~\bibnamefont {Gr\"unewald}}, \bibinfo {author} {\bibfnamefont {A.}~\bibnamefont {Gurtu}}, \bibinfo {author} {\bibfnamefont {T.}~\bibnamefont {Gutsche}}, \bibinfo {author} {\bibfnamefont {H.~E.}\ \bibnamefont {Haber}}, \bibinfo {author} {\bibfnamefont {C.}~\bibnamefont {Hanhart}}, \bibinfo
  {author} {\bibfnamefont {S.}~\bibnamefont {Hashimoto}}, \bibinfo {author} {\bibfnamefont {Y.}~\bibnamefont {Hayato}}, \bibinfo {author} {\bibfnamefont {K.~G.}\ \bibnamefont {Hayes}}, \bibinfo {author} {\bibfnamefont {A.}~\bibnamefont {Hebecker}}, \bibinfo {author} {\bibfnamefont {S.}~\bibnamefont {Heinemeyer}}, \bibinfo {author} {\bibfnamefont {B.}~\bibnamefont {Heltsley}}, \bibinfo {author} {\bibfnamefont {J.~J.}\ \bibnamefont {Hern\'andez-Rey}}, \bibinfo {author} {\bibfnamefont {J.}~\bibnamefont {Hisano}}, \bibinfo {author} {\bibfnamefont {A.}~\bibnamefont {H\"ocker}}, \bibinfo {author} {\bibfnamefont {J.}~\bibnamefont {Holder}}, \bibinfo {author} {\bibfnamefont {A.}~\bibnamefont {Holtkamp}}, \bibinfo {author} {\bibfnamefont {T.}~\bibnamefont {Hyodo}}, \bibinfo {author} {\bibfnamefont {K.~D.}\ \bibnamefont {Irwin}}, \bibinfo {author} {\bibfnamefont {K.~F.}\ \bibnamefont {Johnson}}, \bibinfo {author} {\bibfnamefont {M.}~\bibnamefont {Kado}}, \bibinfo {author} {\bibfnamefont {M.}~\bibnamefont {Karliner}},
  \bibinfo {author} {\bibfnamefont {U.~F.}\ \bibnamefont {Katz}}, \bibinfo {author} {\bibfnamefont {S.~R.}\ \bibnamefont {Klein}}, \bibinfo {author} {\bibfnamefont {E.}~\bibnamefont {Klempt}}, \bibinfo {author} {\bibfnamefont {R.~V.}\ \bibnamefont {Kowalewski}}, \bibinfo {author} {\bibfnamefont {F.}~\bibnamefont {Krauss}}, \bibinfo {author} {\bibfnamefont {M.}~\bibnamefont {Kreps}}, \bibinfo {author} {\bibfnamefont {B.}~\bibnamefont {Krusche}}, \bibinfo {author} {\bibfnamefont {Y.~V.}\ \bibnamefont {Kuyanov}}, \bibinfo {author} {\bibfnamefont {Y.}~\bibnamefont {Kwon}}, \bibinfo {author} {\bibfnamefont {O.}~\bibnamefont {Lahav}}, \bibinfo {author} {\bibfnamefont {J.}~\bibnamefont {Laiho}}, \bibinfo {author} {\bibfnamefont {J.}~\bibnamefont {Lesgourgues}}, \bibinfo {author} {\bibfnamefont {A.}~\bibnamefont {Liddle}}, \bibinfo {author} {\bibfnamefont {Z.}~\bibnamefont {Ligeti}}, \bibinfo {author} {\bibfnamefont {C.-J.}\ \bibnamefont {Lin}}, \bibinfo {author} {\bibfnamefont {C.}~\bibnamefont {Lippmann}}, \bibinfo
  {author} {\bibfnamefont {T.~M.}\ \bibnamefont {Liss}}, \bibinfo {author} {\bibfnamefont {L.}~\bibnamefont {Littenberg}}, \bibinfo {author} {\bibfnamefont {K.~S.}\ \bibnamefont {Lugovsky}}, \bibinfo {author} {\bibfnamefont {S.~B.}\ \bibnamefont {Lugovsky}}, \bibinfo {author} {\bibfnamefont {A.}~\bibnamefont {Lusiani}}, \bibinfo {author} {\bibfnamefont {Y.}~\bibnamefont {Makida}}, \bibinfo {author} {\bibfnamefont {F.}~\bibnamefont {Maltoni}}, \bibinfo {author} {\bibfnamefont {T.}~\bibnamefont {Mannel}}, \bibinfo {author} {\bibfnamefont {A.~V.}\ \bibnamefont {Manohar}}, \bibinfo {author} {\bibfnamefont {W.~J.}\ \bibnamefont {Marciano}}, \bibinfo {author} {\bibfnamefont {A.~D.}\ \bibnamefont {Martin}}, \bibinfo {author} {\bibfnamefont {A.}~\bibnamefont {Masoni}}, \bibinfo {author} {\bibfnamefont {J.}~\bibnamefont {Matthews}}, \bibinfo {author} {\bibfnamefont {U.-G.}\ \bibnamefont {Mei\ss{}ner}}, \bibinfo {author} {\bibfnamefont {D.}~\bibnamefont {Milstead}}, \bibinfo {author} {\bibfnamefont {R.~E.}\
  \bibnamefont {Mitchell}}, \bibinfo {author} {\bibfnamefont {K.}~\bibnamefont {M\"onig}}, \bibinfo {author} {\bibfnamefont {P.}~\bibnamefont {Molaro}}, \bibinfo {author} {\bibfnamefont {F.}~\bibnamefont {Moortgat}}, \bibinfo {author} {\bibfnamefont {M.}~\bibnamefont {Moskovic}}, \bibinfo {author} {\bibfnamefont {H.}~\bibnamefont {Murayama}}, \bibinfo {author} {\bibfnamefont {M.}~\bibnamefont {Narain}}, \bibinfo {author} {\bibfnamefont {P.}~\bibnamefont {Nason}}, \bibinfo {author} {\bibfnamefont {S.}~\bibnamefont {Navas}}, \bibinfo {author} {\bibfnamefont {M.}~\bibnamefont {Neubert}}, \bibinfo {author} {\bibfnamefont {P.}~\bibnamefont {Nevski}}, \bibinfo {author} {\bibfnamefont {Y.}~\bibnamefont {Nir}}, \bibinfo {author} {\bibfnamefont {K.~A.}\ \bibnamefont {Olive}}, \bibinfo {author} {\bibfnamefont {S.}~\bibnamefont {Pagan~Griso}}, \bibinfo {author} {\bibfnamefont {J.}~\bibnamefont {Parsons}}, \bibinfo {author} {\bibfnamefont {C.}~\bibnamefont {Patrignani}}, \bibinfo {author} {\bibfnamefont {J.~A.}\
  \bibnamefont {Peacock}}, \bibinfo {author} {\bibfnamefont {M.}~\bibnamefont {Pennington}}, \bibinfo {author} {\bibfnamefont {S.~T.}\ \bibnamefont {Petcov}}, \bibinfo {author} {\bibfnamefont {V.~A.}\ \bibnamefont {Petrov}}, \bibinfo {author} {\bibfnamefont {E.}~\bibnamefont {Pianori}}, \bibinfo {author} {\bibfnamefont {A.}~\bibnamefont {Piepke}}, \bibinfo {author} {\bibfnamefont {A.}~\bibnamefont {Pomarol}}, \bibinfo {author} {\bibfnamefont {A.}~\bibnamefont {Quadt}}, \bibinfo {author} {\bibfnamefont {J.}~\bibnamefont {Rademacker}}, \bibinfo {author} {\bibfnamefont {G.}~\bibnamefont {Raffelt}}, \bibinfo {author} {\bibfnamefont {B.~N.}\ \bibnamefont {Ratcliff}}, \bibinfo {author} {\bibfnamefont {P.}~\bibnamefont {Richardson}}, \bibinfo {author} {\bibfnamefont {A.}~\bibnamefont {Ringwald}}, \bibinfo {author} {\bibfnamefont {S.}~\bibnamefont {Roesler}}, \bibinfo {author} {\bibfnamefont {S.}~\bibnamefont {Rolli}}, \bibinfo {author} {\bibfnamefont {A.}~\bibnamefont {Romaniouk}}, \bibinfo {author} {\bibfnamefont
  {L.~J.}\ \bibnamefont {Rosenberg}}, \bibinfo {author} {\bibfnamefont {J.~L.}\ \bibnamefont {Rosner}}, \bibinfo {author} {\bibfnamefont {G.}~\bibnamefont {Rybka}}, \bibinfo {author} {\bibfnamefont {R.~A.}\ \bibnamefont {Ryutin}}, \bibinfo {author} {\bibfnamefont {C.~T.}\ \bibnamefont {Sachrajda}}, \bibinfo {author} {\bibfnamefont {Y.}~\bibnamefont {Sakai}}, \bibinfo {author} {\bibfnamefont {G.~P.}\ \bibnamefont {Salam}}, \bibinfo {author} {\bibfnamefont {S.}~\bibnamefont {Sarkar}}, \bibinfo {author} {\bibfnamefont {F.}~\bibnamefont {Sauli}}, \bibinfo {author} {\bibfnamefont {O.}~\bibnamefont {Schneider}}, \bibinfo {author} {\bibfnamefont {K.}~\bibnamefont {Scholberg}}, \bibinfo {author} {\bibfnamefont {A.~J.}\ \bibnamefont {Schwartz}}, \bibinfo {author} {\bibfnamefont {D.}~\bibnamefont {Scott}}, \bibinfo {author} {\bibfnamefont {V.}~\bibnamefont {Sharma}}, \bibinfo {author} {\bibfnamefont {S.~R.}\ \bibnamefont {Sharpe}}, \bibinfo {author} {\bibfnamefont {T.}~\bibnamefont {Shutt}}, \bibinfo {author}
  {\bibfnamefont {M.}~\bibnamefont {Silari}}, \bibinfo {author} {\bibfnamefont {T.}~\bibnamefont {Sj\"ostrand}}, \bibinfo {author} {\bibfnamefont {P.}~\bibnamefont {Skands}}, \bibinfo {author} {\bibfnamefont {T.}~\bibnamefont {Skwarnicki}}, \bibinfo {author} {\bibfnamefont {J.~G.}\ \bibnamefont {Smith}}, \bibinfo {author} {\bibfnamefont {G.~F.}\ \bibnamefont {Smoot}}, \bibinfo {author} {\bibfnamefont {S.}~\bibnamefont {Spanier}}, \bibinfo {author} {\bibfnamefont {H.}~\bibnamefont {Spieler}}, \bibinfo {author} {\bibfnamefont {C.}~\bibnamefont {Spiering}}, \bibinfo {author} {\bibfnamefont {A.}~\bibnamefont {Stahl}}, \bibinfo {author} {\bibfnamefont {S.~L.}\ \bibnamefont {Stone}}, \bibinfo {author} {\bibfnamefont {T.}~\bibnamefont {Sumiyoshi}}, \bibinfo {author} {\bibfnamefont {M.~J.}\ \bibnamefont {Syphers}}, \bibinfo {author} {\bibfnamefont {K.}~\bibnamefont {Terashi}}, \bibinfo {author} {\bibfnamefont {J.}~\bibnamefont {Terning}}, \bibinfo {author} {\bibfnamefont {U.}~\bibnamefont {Thoma}}, \bibinfo {author}
  {\bibfnamefont {R.~S.}\ \bibnamefont {Thorne}}, \bibinfo {author} {\bibfnamefont {L.}~\bibnamefont {Tiator}}, \bibinfo {author} {\bibfnamefont {M.}~\bibnamefont {Titov}}, \bibinfo {author} {\bibfnamefont {N.~P.}\ \bibnamefont {Tkachenko}}, \bibinfo {author} {\bibfnamefont {N.~A.}\ \bibnamefont {T\"ornqvist}}, \bibinfo {author} {\bibfnamefont {D.~R.}\ \bibnamefont {Tovey}}, \bibinfo {author} {\bibfnamefont {G.}~\bibnamefont {Valencia}}, \bibinfo {author} {\bibfnamefont {R.}~\bibnamefont {Van~de Water}}, \bibinfo {author} {\bibfnamefont {N.}~\bibnamefont {Varelas}}, \bibinfo {author} {\bibfnamefont {G.}~\bibnamefont {Venanzoni}}, \bibinfo {author} {\bibfnamefont {L.}~\bibnamefont {Verde}}, \bibinfo {author} {\bibfnamefont {M.~G.}\ \bibnamefont {Vincter}}, \bibinfo {author} {\bibfnamefont {P.}~\bibnamefont {Vogel}}, \bibinfo {author} {\bibfnamefont {A.}~\bibnamefont {Vogt}}, \bibinfo {author} {\bibfnamefont {S.~P.}\ \bibnamefont {Wakely}}, \bibinfo {author} {\bibfnamefont {W.}~\bibnamefont {Walkowiak}},
  \bibinfo {author} {\bibfnamefont {C.~W.}\ \bibnamefont {Walter}}, \bibinfo {author} {\bibfnamefont {D.}~\bibnamefont {Wands}}, \bibinfo {author} {\bibfnamefont {D.~R.}\ \bibnamefont {Ward}}, \bibinfo {author} {\bibfnamefont {M.~O.}\ \bibnamefont {Wascko}}, \bibinfo {author} {\bibfnamefont {G.}~\bibnamefont {Weiglein}}, \bibinfo {author} {\bibfnamefont {D.~H.}\ \bibnamefont {Weinberg}}, \bibinfo {author} {\bibfnamefont {E.~J.}\ \bibnamefont {Weinberg}}, \bibinfo {author} {\bibfnamefont {M.}~\bibnamefont {White}}, \bibinfo {author} {\bibfnamefont {L.~R.}\ \bibnamefont {Wiencke}}, \bibinfo {author} {\bibfnamefont {S.}~\bibnamefont {Willocq}}, \bibinfo {author} {\bibfnamefont {C.~G.}\ \bibnamefont {Wohl}}, \bibinfo {author} {\bibfnamefont {J.}~\bibnamefont {Womersley}}, \bibinfo {author} {\bibfnamefont {C.~L.}\ \bibnamefont {Woody}}, \bibinfo {author} {\bibfnamefont {R.~L.}\ \bibnamefont {Workman}}, \bibinfo {author} {\bibfnamefont {W.-M.}\ \bibnamefont {Yao}}, \bibinfo {author} {\bibfnamefont {G.~P.}\
  \bibnamefont {Zeller}}, \bibinfo {author} {\bibfnamefont {O.~V.}\ \bibnamefont {Zenin}}, \bibinfo {author} {\bibfnamefont {R.-Y.}\ \bibnamefont {Zhu}}, \bibinfo {author} {\bibfnamefont {S.-L.}\ \bibnamefont {Zhu}}, \bibinfo {author} {\bibfnamefont {F.}~\bibnamefont {Zimmermann}}, \bibinfo {author} {\bibfnamefont {P.~A.}\ \bibnamefont {Zyla}}, \bibinfo {author} {\bibfnamefont {J.}~\bibnamefont {Anderson}}, \bibinfo {author} {\bibfnamefont {L.}~\bibnamefont {Fuller}}, \bibinfo {author} {\bibfnamefont {V.~S.}\ \bibnamefont {Lugovsky}},\ and\ \bibinfo {author} {\bibfnamefont {P.}~\bibnamefont {Schaffner}} (\bibinfo {collaboration} {Particle Data Group}),\ }\bibfield  {title} {\bibinfo {title} {Phys. rev. d},\ }\href {https://doi.org/10.1103/PhysRevD.98.030001} {\bibfield  {journal} {\bibinfo  {journal} {Phys. Rev. D}\ }\textbf {\bibinfo {volume} {98}},\ \bibinfo {pages} {030001} (\bibinfo {year} {2018})}\BibitemShut {NoStop}%
\bibitem [{\citenamefont {Kornakov}\ \emph {et~al.}(2023)\citenamefont {Kornakov}, \citenamefont {Cerchiari}, \citenamefont {Zieli\'{n}ski}, \citenamefont {Lappo}, \citenamefont {Sadowski},\ and\ \citenamefont {Doser}}]{Kornakov2022}%
  \BibitemOpen
  \bibfield  {author} {\bibinfo {author} {\bibfnamefont {G.}~\bibnamefont {Kornakov}}, \bibinfo {author} {\bibfnamefont {G.}~\bibnamefont {Cerchiari}}, \bibinfo {author} {\bibfnamefont {J.}~\bibnamefont {Zieli\'{n}ski}}, \bibinfo {author} {\bibfnamefont {L.}~\bibnamefont {Lappo}}, \bibinfo {author} {\bibfnamefont {G.}~\bibnamefont {Sadowski}},\ and\ \bibinfo {author} {\bibfnamefont {M.}~\bibnamefont {Doser}},\ }\bibfield  {title} {\bibinfo {title} {Synthesis of cold and trappable fully stripped highly charged ions via antiproton-induced nuclear fragmentation in traps},\ }\href {https://doi.org/10.1103/PhysRevC.107.034314} {\bibfield  {journal} {\bibinfo  {journal} {Phys. Rev. C}\ }\textbf {\bibinfo {volume} {107}},\ \bibinfo {pages} {034314} (\bibinfo {year} {2023})}\BibitemShut {NoStop}%
\bibitem [{\citenamefont {Gotta}\ \emph {et~al.}(2008)\citenamefont {Gotta}, \citenamefont {Rashid}, \citenamefont {Fricke}, \citenamefont {Indelicato},\ and\ \citenamefont {Simons}}]{Gotta2008}%
  \BibitemOpen
  \bibfield  {author} {\bibinfo {author} {\bibfnamefont {D.}~\bibnamefont {Gotta}}, \bibinfo {author} {\bibfnamefont {K.}~\bibnamefont {Rashid}}, \bibinfo {author} {\bibfnamefont {B.}~\bibnamefont {Fricke}}, \bibinfo {author} {\bibfnamefont {P.}~\bibnamefont {Indelicato}},\ and\ \bibinfo {author} {\bibfnamefont {L.~M.}\ \bibnamefont {Simons}},\ }\bibfield  {title} {\bibinfo {title} {X-ray transitions from antiprotonic noble gases},\ }\href {https://doi.org/10.1140/epjd/e2008-00025-3} {\bibfield  {journal} {\bibinfo  {journal} {Eur. J. Phys. D}\ }\textbf {\bibinfo {volume} {47}},\ \bibinfo {pages} {11} (\bibinfo {year} {2008})}\BibitemShut {NoStop}%
\bibitem [{\citenamefont {Bacher}\ \emph {et~al.}(1988)\citenamefont {Bacher}, \citenamefont {Bl\"um}, \citenamefont {Gotta}, \citenamefont {Heitlinger}, \citenamefont {Schneider}, \citenamefont {Missimer}, \citenamefont {Simons},\ and\ \citenamefont {Elsener}}]{Bacher1988}%
  \BibitemOpen
  \bibfield  {author} {\bibinfo {author} {\bibfnamefont {R.}~\bibnamefont {Bacher}}, \bibinfo {author} {\bibfnamefont {P.}~\bibnamefont {Bl\"um}}, \bibinfo {author} {\bibfnamefont {D.}~\bibnamefont {Gotta}}, \bibinfo {author} {\bibfnamefont {K.}~\bibnamefont {Heitlinger}}, \bibinfo {author} {\bibfnamefont {M.}~\bibnamefont {Schneider}}, \bibinfo {author} {\bibfnamefont {J.}~\bibnamefont {Missimer}}, \bibinfo {author} {\bibfnamefont {L.~M.}\ \bibnamefont {Simons}},\ and\ \bibinfo {author} {\bibfnamefont {K.}~\bibnamefont {Elsener}},\ }\bibfield  {title} {\bibinfo {title} {Degree of ionization in antiprotonic noble gases},\ }\href {https://doi.org/10.1103/PhysRevA.38.4395} {\bibfield  {journal} {\bibinfo  {journal} {Phys. Rev. A}\ }\textbf {\bibinfo {volume} {38}},\ \bibinfo {pages} {4395} (\bibinfo {year} {1988})}\BibitemShut {NoStop}%
\bibitem [{\citenamefont {Amoretti}\ \emph {et~al.}(2002)\citenamefont {Amoretti}, \citenamefont {Amsler}, \citenamefont {Bonomi}, \citenamefont {Bouchta}, \citenamefont {Bowe}, \citenamefont {Carraro}, \citenamefont {Cesar}, \citenamefont {Charlton}, \citenamefont {Collier}, \citenamefont {Doser}, \citenamefont {Filippini}, \citenamefont {Fine}, \citenamefont {Fontana}, \citenamefont {Fujiwara}, \citenamefont {Funakoshi}, \citenamefont {Genova}, \citenamefont {Hangst}, \citenamefont {Hayano}, \citenamefont {Holzscheiter}, \citenamefont {Jørgensen}, \citenamefont {Lagomarsino}, \citenamefont {Landua}, \citenamefont {Lindelöf}, \citenamefont {Rizzini}, \citenamefont {Macrì}, \citenamefont {Madsen}, \citenamefont {Manuzio}, \citenamefont {Marchesotti}, \citenamefont {Montagna}, \citenamefont {Pruys}, \citenamefont {Regenfus}, \citenamefont {Riedler}, \citenamefont {Rochet}, \citenamefont {Rotondi}, \citenamefont {Rouleau}, \citenamefont {Testera}, \citenamefont {Variola}, \citenamefont {Watson},\ and\
  \citenamefont {van~der Werf}}]{Amoretti2002}%
  \BibitemOpen
  \bibfield  {author} {\bibinfo {author} {\bibfnamefont {M.}~\bibnamefont {Amoretti}}, \bibinfo {author} {\bibfnamefont {C.}~\bibnamefont {Amsler}}, \bibinfo {author} {\bibfnamefont {G.}~\bibnamefont {Bonomi}}, \bibinfo {author} {\bibfnamefont {A.}~\bibnamefont {Bouchta}}, \bibinfo {author} {\bibfnamefont {P.}~\bibnamefont {Bowe}}, \bibinfo {author} {\bibfnamefont {C.}~\bibnamefont {Carraro}}, \bibinfo {author} {\bibfnamefont {C.~L.}\ \bibnamefont {Cesar}}, \bibinfo {author} {\bibfnamefont {M.}~\bibnamefont {Charlton}}, \bibinfo {author} {\bibfnamefont {M.~J.~T.}\ \bibnamefont {Collier}}, \bibinfo {author} {\bibfnamefont {M.}~\bibnamefont {Doser}}, \bibinfo {author} {\bibfnamefont {V.}~\bibnamefont {Filippini}}, \bibinfo {author} {\bibfnamefont {K.~S.}\ \bibnamefont {Fine}}, \bibinfo {author} {\bibfnamefont {A.}~\bibnamefont {Fontana}}, \bibinfo {author} {\bibfnamefont {M.~C.}\ \bibnamefont {Fujiwara}}, \bibinfo {author} {\bibfnamefont {R.}~\bibnamefont {Funakoshi}}, \bibinfo {author} {\bibfnamefont
  {P.}~\bibnamefont {Genova}}, \bibinfo {author} {\bibfnamefont {J.~S.}\ \bibnamefont {Hangst}}, \bibinfo {author} {\bibfnamefont {R.~S.}\ \bibnamefont {Hayano}}, \bibinfo {author} {\bibfnamefont {M.~H.}\ \bibnamefont {Holzscheiter}}, \bibinfo {author} {\bibfnamefont {L.~V.}\ \bibnamefont {Jørgensen}}, \bibinfo {author} {\bibfnamefont {V.}~\bibnamefont {Lagomarsino}}, \bibinfo {author} {\bibfnamefont {R.}~\bibnamefont {Landua}}, \bibinfo {author} {\bibfnamefont {D.}~\bibnamefont {Lindelöf}}, \bibinfo {author} {\bibfnamefont {E.~L.}\ \bibnamefont {Rizzini}}, \bibinfo {author} {\bibfnamefont {M.}~\bibnamefont {Macrì}}, \bibinfo {author} {\bibfnamefont {N.}~\bibnamefont {Madsen}}, \bibinfo {author} {\bibfnamefont {G.}~\bibnamefont {Manuzio}}, \bibinfo {author} {\bibfnamefont {M.}~\bibnamefont {Marchesotti}}, \bibinfo {author} {\bibfnamefont {P.}~\bibnamefont {Montagna}}, \bibinfo {author} {\bibfnamefont {H.}~\bibnamefont {Pruys}}, \bibinfo {author} {\bibfnamefont {C.}~\bibnamefont {Regenfus}}, \bibinfo
  {author} {\bibfnamefont {P.}~\bibnamefont {Riedler}}, \bibinfo {author} {\bibfnamefont {J.}~\bibnamefont {Rochet}}, \bibinfo {author} {\bibfnamefont {A.}~\bibnamefont {Rotondi}}, \bibinfo {author} {\bibfnamefont {G.}~\bibnamefont {Rouleau}}, \bibinfo {author} {\bibfnamefont {G.}~\bibnamefont {Testera}}, \bibinfo {author} {\bibfnamefont {A.}~\bibnamefont {Variola}}, \bibinfo {author} {\bibfnamefont {T.~L.}\ \bibnamefont {Watson}},\ and\ \bibinfo {author} {\bibfnamefont {D.~P.}\ \bibnamefont {van~der Werf}},\ }\bibfield  {title} {\bibinfo {title} {Production and detection of cold antihydrogen atoms},\ }\href {https://doi.org/10.1038/nature01096} {\bibfield  {journal} {\bibinfo  {journal} {Nature}\ }\textbf {\bibinfo {volume} {419}},\ \bibinfo {pages} {456} (\bibinfo {year} {2002})}\BibitemShut {NoStop}%
\bibitem [{\citenamefont {Gerber}\ \emph {et~al.}(2019)\citenamefont {Gerber}, \citenamefont {Doser},\ and\ \citenamefont {Comparat}}]{Gerber2019}%
  \BibitemOpen
  \bibfield  {author} {\bibinfo {author} {\bibfnamefont {S.}~\bibnamefont {Gerber}}, \bibinfo {author} {\bibfnamefont {M.}~\bibnamefont {Doser}},\ and\ \bibinfo {author} {\bibfnamefont {D.}~\bibnamefont {Comparat}},\ }\bibfield  {title} {\bibinfo {title} {Pulsed production of cold protonium in penning traps},\ }\href {https://doi.org/10.1103/PhysRevA.100.063418} {\bibfield  {journal} {\bibinfo  {journal} {Phys. Rev. A}\ }\textbf {\bibinfo {volume} {100}},\ \bibinfo {pages} {063418} (\bibinfo {year} {2019})}\BibitemShut {NoStop}%
\bibitem [{\citenamefont {Ludlow}\ \emph {et~al.}(2015)\citenamefont {Ludlow}, \citenamefont {Boyd}, \citenamefont {Ye}, \citenamefont {Peik},\ and\ \citenamefont {Schmidt}}]{Ludlow2015}%
  \BibitemOpen
  \bibfield  {author} {\bibinfo {author} {\bibfnamefont {A.~D.}\ \bibnamefont {Ludlow}}, \bibinfo {author} {\bibfnamefont {M.~M.}\ \bibnamefont {Boyd}}, \bibinfo {author} {\bibfnamefont {J.}~\bibnamefont {Ye}}, \bibinfo {author} {\bibfnamefont {E.}~\bibnamefont {Peik}},\ and\ \bibinfo {author} {\bibfnamefont {P.~O.}\ \bibnamefont {Schmidt}},\ }\bibfield  {title} {\bibinfo {title} {Optical atomic clocks},\ }\href {https://doi.org/10.1103/RevModPhys.87.637} {\bibfield  {journal} {\bibinfo  {journal} {Rev. Mod. Phys.}\ }\textbf {\bibinfo {volume} {87}},\ \bibinfo {pages} {637} (\bibinfo {year} {2015})}\BibitemShut {NoStop}%
\bibitem [{\citenamefont {Sellner}\ \emph {et~al.}(2017)\citenamefont {Sellner}, \citenamefont {Besirli}, \citenamefont {Bohman}, \citenamefont {Borchert}, \citenamefont {Harrington}, \citenamefont {Higuchi}, \citenamefont {Mooser}, \citenamefont {Nagahama}, \citenamefont {Schneider}, \citenamefont {Smorra}, \citenamefont {Tanaka}, \citenamefont {Blaum}, \citenamefont {Matsuda}, \citenamefont {Ospelkaus}, \citenamefont {Quint}, \citenamefont {Walz}, \citenamefont {Yamazaki},\ and\ \citenamefont {Ulmer}}]{Sellner2017}%
  \BibitemOpen
  \bibfield  {author} {\bibinfo {author} {\bibfnamefont {S.}~\bibnamefont {Sellner}}, \bibinfo {author} {\bibfnamefont {M.}~\bibnamefont {Besirli}}, \bibinfo {author} {\bibfnamefont {M.}~\bibnamefont {Bohman}}, \bibinfo {author} {\bibfnamefont {M.~J.}\ \bibnamefont {Borchert}}, \bibinfo {author} {\bibfnamefont {J.}~\bibnamefont {Harrington}}, \bibinfo {author} {\bibfnamefont {T.}~\bibnamefont {Higuchi}}, \bibinfo {author} {\bibfnamefont {A.}~\bibnamefont {Mooser}}, \bibinfo {author} {\bibfnamefont {H.}~\bibnamefont {Nagahama}}, \bibinfo {author} {\bibfnamefont {G.}~\bibnamefont {Schneider}}, \bibinfo {author} {\bibfnamefont {C.}~\bibnamefont {Smorra}}, \bibinfo {author} {\bibfnamefont {T.}~\bibnamefont {Tanaka}}, \bibinfo {author} {\bibfnamefont {K.}~\bibnamefont {Blaum}}, \bibinfo {author} {\bibfnamefont {Y.}~\bibnamefont {Matsuda}}, \bibinfo {author} {\bibfnamefont {C.}~\bibnamefont {Ospelkaus}}, \bibinfo {author} {\bibfnamefont {W.}~\bibnamefont {Quint}}, \bibinfo {author} {\bibfnamefont {J.}~\bibnamefont
  {Walz}}, \bibinfo {author} {\bibfnamefont {Y.}~\bibnamefont {Yamazaki}},\ and\ \bibinfo {author} {\bibfnamefont {S.}~\bibnamefont {Ulmer}},\ }\bibfield  {title} {\bibinfo {title} {Improved limit on the directly measured antiproton lifetime},\ }\href {https://doi.org/10.1088/1367-2630/aa7e73} {\bibfield  {journal} {\bibinfo  {journal} {New J. Phys.}\ }\textbf {\bibinfo {volume} {19}},\ \bibinfo {pages} {083023} (\bibinfo {year} {2017})}\BibitemShut {NoStop}%
\bibitem [{\citenamefont {Micke}\ \emph {et~al.}(2018)\citenamefont {Micke}, \citenamefont {Kühn}, \citenamefont {Buchauer}, \citenamefont {Harries}, \citenamefont {Bücking}, \citenamefont {Blaum}, \citenamefont {Cieluch}, \citenamefont {Egl}, \citenamefont {Hollain}, \citenamefont {Kraemer}, \citenamefont {Pfeifer}, \citenamefont {Schmidt}, \citenamefont {Schüssler}, \citenamefont {Schweiger}, \citenamefont {Stöhlker}, \citenamefont {Sturm}, \citenamefont {Wolf}, \citenamefont {Bernitt},\ and\ \citenamefont {Crespo López-Urrutia}}]{Micke2018}%
  \BibitemOpen
  \bibfield  {author} {\bibinfo {author} {\bibfnamefont {P.}~\bibnamefont {Micke}}, \bibinfo {author} {\bibfnamefont {S.}~\bibnamefont {Kühn}}, \bibinfo {author} {\bibfnamefont {L.}~\bibnamefont {Buchauer}}, \bibinfo {author} {\bibfnamefont {J.~R.}\ \bibnamefont {Harries}}, \bibinfo {author} {\bibfnamefont {T.~M.}\ \bibnamefont {Bücking}}, \bibinfo {author} {\bibfnamefont {K.}~\bibnamefont {Blaum}}, \bibinfo {author} {\bibfnamefont {A.}~\bibnamefont {Cieluch}}, \bibinfo {author} {\bibfnamefont {A.}~\bibnamefont {Egl}}, \bibinfo {author} {\bibfnamefont {D.}~\bibnamefont {Hollain}}, \bibinfo {author} {\bibfnamefont {S.}~\bibnamefont {Kraemer}}, \bibinfo {author} {\bibfnamefont {T.}~\bibnamefont {Pfeifer}}, \bibinfo {author} {\bibfnamefont {P.~O.}\ \bibnamefont {Schmidt}}, \bibinfo {author} {\bibfnamefont {R.~X.}\ \bibnamefont {Schüssler}}, \bibinfo {author} {\bibfnamefont {C.}~\bibnamefont {Schweiger}}, \bibinfo {author} {\bibfnamefont {T.}~\bibnamefont {Stöhlker}}, \bibinfo {author} {\bibfnamefont
  {S.}~\bibnamefont {Sturm}}, \bibinfo {author} {\bibfnamefont {R.~N.}\ \bibnamefont {Wolf}}, \bibinfo {author} {\bibfnamefont {S.}~\bibnamefont {Bernitt}},\ and\ \bibinfo {author} {\bibfnamefont {J.~R.}\ \bibnamefont {Crespo López-Urrutia}},\ }\bibfield  {title} {\bibinfo {title} {{The Heidelberg compact electron beam ion traps}},\ }\href {https://doi.org/10.1063/1.5026961} {\bibfield  {journal} {\bibinfo  {journal} {Rev. Sci. Instrum.}\ }\textbf {\bibinfo {volume} {89}},\ \bibinfo {pages} {063109} (\bibinfo {year} {2018})}\BibitemShut {NoStop}%
\bibitem [{\citenamefont {Shabaev}(1994)}]{Shabaev1994}%
  \BibitemOpen
  \bibfield  {author} {\bibinfo {author} {\bibfnamefont {V.~M.}\ \bibnamefont {Shabaev}},\ }\bibfield  {title} {\bibinfo {title} {Hyperfine structure of hydrogen-like ions},\ }\href {https://doi.org/10.1088/0953-4075/27/24/006} {\bibfield  {journal} {\bibinfo  {journal} {J. Phys. B}\ }\textbf {\bibinfo {volume} {27}},\ \bibinfo {pages} {5825} (\bibinfo {year} {1994})}\BibitemShut {NoStop}%
\bibitem [{\citenamefont {Mertzimekis}\ \emph {et~al.}(2016)\citenamefont {Mertzimekis}, \citenamefont {Stamou},\ and\ \citenamefont {Psaltis}}]{Mertzimekis2016}%
  \BibitemOpen
  \bibfield  {author} {\bibinfo {author} {\bibfnamefont {T.}~\bibnamefont {Mertzimekis}}, \bibinfo {author} {\bibfnamefont {K.}~\bibnamefont {Stamou}},\ and\ \bibinfo {author} {\bibfnamefont {A.}~\bibnamefont {Psaltis}},\ }\bibfield  {title} {\bibinfo {title} {An online database of nuclear electromagnetic moments},\ }\href {https://doi.org/https://doi.org/10.1016/j.nima.2015.10.096} {\bibfield  {journal} {\bibinfo  {journal} {Nucl. Instrum. Methods Phys. Res. A}\ }\textbf {\bibinfo {volume} {807}},\ \bibinfo {pages} {56} (\bibinfo {year} {2016})}\BibitemShut {NoStop}%
\bibitem [{\citenamefont {Jönsson}\ \emph {et~al.}(2013)\citenamefont {Jönsson}, \citenamefont {Gaigalas}, \citenamefont {Bieroń}, \citenamefont {Fischer},\ and\ \citenamefont {Grant}}]{Jonsson2013}%
  \BibitemOpen
  \bibfield  {author} {\bibinfo {author} {\bibfnamefont {P.}~\bibnamefont {Jönsson}}, \bibinfo {author} {\bibfnamefont {G.}~\bibnamefont {Gaigalas}}, \bibinfo {author} {\bibfnamefont {J.}~\bibnamefont {Bieroń}}, \bibinfo {author} {\bibfnamefont {C.~F.}\ \bibnamefont {Fischer}},\ and\ \bibinfo {author} {\bibfnamefont {I.}~\bibnamefont {Grant}},\ }\bibfield  {title} {\bibinfo {title} {New version: Grasp2k relativistic atomic structure package},\ }\href {https://doi.org/https://doi.org/10.1016/j.cpc.2013.02.016} {\bibfield  {journal} {\bibinfo  {journal} {Comput. Phys. Commun.}\ }\textbf {\bibinfo {volume} {184}},\ \bibinfo {pages} {2197} (\bibinfo {year} {2013})}\BibitemShut {NoStop}%
\bibitem [{\citenamefont {Jönsson}\ \emph {et~al.}(2007)\citenamefont {Jönsson}, \citenamefont {He}, \citenamefont {{Froese Fischer}},\ and\ \citenamefont {Grant}}]{Jonnsson2007}%
  \BibitemOpen
  \bibfield  {author} {\bibinfo {author} {\bibfnamefont {P.}~\bibnamefont {Jönsson}}, \bibinfo {author} {\bibfnamefont {X.}~\bibnamefont {He}}, \bibinfo {author} {\bibfnamefont {C.}~\bibnamefont {{Froese Fischer}}},\ and\ \bibinfo {author} {\bibfnamefont {I.}~\bibnamefont {Grant}},\ }\bibfield  {title} {\bibinfo {title} {The grasp2k relativistic atomic structure package},\ }\href {https://doi.org/https://doi.org/10.1016/j.cpc.2007.06.002} {\bibfield  {journal} {\bibinfo  {journal} {Comput. Phys. Commun.}\ }\textbf {\bibinfo {volume} {177}},\ \bibinfo {pages} {597} (\bibinfo {year} {2007})}\BibitemShut {NoStop}%
\bibitem [{\citenamefont {Bieroń}\ \emph {et~al.}(2023)\citenamefont {Bieroń}, \citenamefont {Fischer},\ and\ \citenamefont {Jönsson}}]{Bieron2023}%
  \BibitemOpen
  \bibfield  {author} {\bibinfo {author} {\bibfnamefont {J.}~\bibnamefont {Bieroń}}, \bibinfo {author} {\bibfnamefont {C.~F.}\ \bibnamefont {Fischer}},\ and\ \bibinfo {author} {\bibfnamefont {P.}~\bibnamefont {Jönsson}},\ }\bibfield  {title} {\bibinfo {title} {Editorial of the special issue “general relativistic atomic structure program—grasp”},\ }\bibfield  {journal} {\bibinfo  {journal} {Atoms}\ }\textbf {\bibinfo {volume} {11}},\ \href {https://doi.org/10.3390/atoms11060093} {10.3390/atoms11060093} (\bibinfo {year} {2023})\BibitemShut {NoStop}%
\bibitem [{\citenamefont {Shabaev}\ \emph {et~al.}(1998)\citenamefont {Shabaev}, \citenamefont {Shabaeva}, \citenamefont {Tupitsyn}, \citenamefont {Yerokhin}, \citenamefont {Artemyev}, \citenamefont {K\"uhl}, \citenamefont {Tomaselli},\ and\ \citenamefont {Zherebtsov}}]{shabaev1998}%
  \BibitemOpen
  \bibfield  {author} {\bibinfo {author} {\bibfnamefont {V.~M.}\ \bibnamefont {Shabaev}}, \bibinfo {author} {\bibfnamefont {M.~B.}\ \bibnamefont {Shabaeva}}, \bibinfo {author} {\bibfnamefont {I.~I.}\ \bibnamefont {Tupitsyn}}, \bibinfo {author} {\bibfnamefont {V.~A.}\ \bibnamefont {Yerokhin}}, \bibinfo {author} {\bibfnamefont {A.~N.}\ \bibnamefont {Artemyev}}, \bibinfo {author} {\bibfnamefont {T.}~\bibnamefont {K\"uhl}}, \bibinfo {author} {\bibfnamefont {M.}~\bibnamefont {Tomaselli}},\ and\ \bibinfo {author} {\bibfnamefont {O.~M.}\ \bibnamefont {Zherebtsov}},\ }\bibfield  {title} {\bibinfo {title} {Transition energy and lifetime for the ground-state hyperfine splitting of high-$z$ lithiumlike ions},\ }\href {https://doi.org/10.1103/PhysRevA.57.149} {\bibfield  {journal} {\bibinfo  {journal} {Phys. Rev. A}\ }\textbf {\bibinfo {volume} {57}},\ \bibinfo {pages} {149} (\bibinfo {year} {1998})}\BibitemShut {NoStop}%
\bibitem [{\citenamefont {Agostinelli}\ \emph {et~al.}(2003)\citenamefont {Agostinelli}, \citenamefont {Allison}, \citenamefont {Amako}, \citenamefont {Apostolakis}, \citenamefont {Araujo}, \citenamefont {Arce}, \citenamefont {Asai}, \citenamefont {Axen}, \citenamefont {Banerjee}, \citenamefont {Barrand}, \citenamefont {Behner}, \citenamefont {Bellagamba}, \citenamefont {Boudreau}, \citenamefont {Broglia}, \citenamefont {Brunengo}, \citenamefont {Burkhardt}, \citenamefont {Chauvie}, \citenamefont {Chuma}, \citenamefont {Chytracek}, \citenamefont {Cooperman}, \citenamefont {Cosmo}, \citenamefont {Degtyarenko}, \citenamefont {Dell'Acqua}, \citenamefont {Depaola}, \citenamefont {Dietrich}, \citenamefont {Enami}, \citenamefont {Feliciello}, \citenamefont {Ferguson}, \citenamefont {Fesefeldt}, \citenamefont {Folger}, \citenamefont {Foppiano}, \citenamefont {Forti}, \citenamefont {Garelli}, \citenamefont {Giani}, \citenamefont {Giannitrapani}, \citenamefont {Gibin}, \citenamefont {Gómez~Cadenas}, \citenamefont
  {González}, \citenamefont {Gracia~Abril}, \citenamefont {Greeniaus}, \citenamefont {Greiner}, \citenamefont {Grichine}, \citenamefont {Grossheim}, \citenamefont {Guatelli}, \citenamefont {Gumplinger}, \citenamefont {Hamatsu}, \citenamefont {Hashimoto}, \citenamefont {Hasui}, \citenamefont {Heikkinen}, \citenamefont {Howard}, \citenamefont {Ivanchenko}, \citenamefont {Johnson}, \citenamefont {Jones}, \citenamefont {Kallenbach}, \citenamefont {Kanaya}, \citenamefont {Kawabata}, \citenamefont {Kawabata}, \citenamefont {Kawaguti}, \citenamefont {Kelner}, \citenamefont {Kent}, \citenamefont {Kimura}, \citenamefont {Kodama}, \citenamefont {Kokoulin}, \citenamefont {Kossov}, \citenamefont {Kurashige}, \citenamefont {Lamanna}, \citenamefont {Lampén}, \citenamefont {Lara}, \citenamefont {Lefebure}, \citenamefont {Lei}, \citenamefont {Liendl}, \citenamefont {Lockman}, \citenamefont {Longo}, \citenamefont {Magni}, \citenamefont {Maire}, \citenamefont {Medernach}, \citenamefont {Minamimoto}, \citenamefont {Mora~de
  Freitas}, \citenamefont {Morita}, \citenamefont {Murakami}, \citenamefont {Nagamatu}, \citenamefont {Nartallo}, \citenamefont {Nieminen}, \citenamefont {Nishimura}, \citenamefont {Ohtsubo}, \citenamefont {Okamura}, \citenamefont {O'Neale}, \citenamefont {Oohata}, \citenamefont {Paech}, \citenamefont {Perl}, \citenamefont {Pfeiffer}, \citenamefont {Pia}, \citenamefont {Ranjard}, \citenamefont {Rybin}, \citenamefont {Sadilov}, \citenamefont {Di~Salvo}, \citenamefont {Santin}, \citenamefont {Sasaki}, \citenamefont {Savvas}, \citenamefont {Sawada}, \citenamefont {Scherer}, \citenamefont {Sei}, \citenamefont {Sirotenko}, \citenamefont {Smith}, \citenamefont {Starkov}, \citenamefont {Stoecker}, \citenamefont {Sulkimo}, \citenamefont {Takahata}, \citenamefont {Tanaka}, \citenamefont {Tcherniaev}, \citenamefont {Safai~Tehrani}, \citenamefont {Tropeano}, \citenamefont {Truscott}, \citenamefont {Uno}, \citenamefont {Urban}, \citenamefont {Urban}, \citenamefont {Verderi}, \citenamefont {Walkden}, \citenamefont
  {Wander}, \citenamefont {Weber}, \citenamefont {Wellisch}, \citenamefont {Wenaus}, \citenamefont {Williams}, \citenamefont {Wright}, \citenamefont {Yamada}, \citenamefont {Yoshida},\ and\ \citenamefont {Zschiesche}}]{agostinelli_geant4simulation_2003}%
  \BibitemOpen
  \bibfield  {author} {\bibinfo {author} {\bibfnamefont {S.}~\bibnamefont {Agostinelli}}, \bibinfo {author} {\bibfnamefont {J.}~\bibnamefont {Allison}}, \bibinfo {author} {\bibfnamefont {K.}~\bibnamefont {Amako}}, \bibinfo {author} {\bibfnamefont {J.}~\bibnamefont {Apostolakis}}, \bibinfo {author} {\bibfnamefont {H.}~\bibnamefont {Araujo}}, \bibinfo {author} {\bibfnamefont {P.}~\bibnamefont {Arce}}, \bibinfo {author} {\bibfnamefont {M.}~\bibnamefont {Asai}}, \bibinfo {author} {\bibfnamefont {D.}~\bibnamefont {Axen}}, \bibinfo {author} {\bibfnamefont {S.}~\bibnamefont {Banerjee}}, \bibinfo {author} {\bibfnamefont {G.}~\bibnamefont {Barrand}}, \bibinfo {author} {\bibfnamefont {F.}~\bibnamefont {Behner}}, \bibinfo {author} {\bibfnamefont {L.}~\bibnamefont {Bellagamba}}, \bibinfo {author} {\bibfnamefont {J.}~\bibnamefont {Boudreau}}, \bibinfo {author} {\bibfnamefont {L.}~\bibnamefont {Broglia}}, \bibinfo {author} {\bibfnamefont {A.}~\bibnamefont {Brunengo}}, \bibinfo {author} {\bibfnamefont {H.}~\bibnamefont
  {Burkhardt}}, \bibinfo {author} {\bibfnamefont {S.}~\bibnamefont {Chauvie}}, \bibinfo {author} {\bibfnamefont {J.}~\bibnamefont {Chuma}}, \bibinfo {author} {\bibfnamefont {R.}~\bibnamefont {Chytracek}}, \bibinfo {author} {\bibfnamefont {G.}~\bibnamefont {Cooperman}}, \bibinfo {author} {\bibfnamefont {G.}~\bibnamefont {Cosmo}}, \bibinfo {author} {\bibfnamefont {P.}~\bibnamefont {Degtyarenko}}, \bibinfo {author} {\bibfnamefont {A.}~\bibnamefont {Dell'Acqua}}, \bibinfo {author} {\bibfnamefont {G.}~\bibnamefont {Depaola}}, \bibinfo {author} {\bibfnamefont {D.}~\bibnamefont {Dietrich}}, \bibinfo {author} {\bibfnamefont {R.}~\bibnamefont {Enami}}, \bibinfo {author} {\bibfnamefont {A.}~\bibnamefont {Feliciello}}, \bibinfo {author} {\bibfnamefont {C.}~\bibnamefont {Ferguson}}, \bibinfo {author} {\bibfnamefont {H.}~\bibnamefont {Fesefeldt}}, \bibinfo {author} {\bibfnamefont {G.}~\bibnamefont {Folger}}, \bibinfo {author} {\bibfnamefont {F.}~\bibnamefont {Foppiano}}, \bibinfo {author} {\bibfnamefont {A.}~\bibnamefont
  {Forti}}, \bibinfo {author} {\bibfnamefont {S.}~\bibnamefont {Garelli}}, \bibinfo {author} {\bibfnamefont {S.}~\bibnamefont {Giani}}, \bibinfo {author} {\bibfnamefont {R.}~\bibnamefont {Giannitrapani}}, \bibinfo {author} {\bibfnamefont {D.}~\bibnamefont {Gibin}}, \bibinfo {author} {\bibfnamefont {J.}~\bibnamefont {Gómez~Cadenas}}, \bibinfo {author} {\bibfnamefont {I.}~\bibnamefont {González}}, \bibinfo {author} {\bibfnamefont {G.}~\bibnamefont {Gracia~Abril}}, \bibinfo {author} {\bibfnamefont {G.}~\bibnamefont {Greeniaus}}, \bibinfo {author} {\bibfnamefont {W.}~\bibnamefont {Greiner}}, \bibinfo {author} {\bibfnamefont {V.}~\bibnamefont {Grichine}}, \bibinfo {author} {\bibfnamefont {A.}~\bibnamefont {Grossheim}}, \bibinfo {author} {\bibfnamefont {S.}~\bibnamefont {Guatelli}}, \bibinfo {author} {\bibfnamefont {P.}~\bibnamefont {Gumplinger}}, \bibinfo {author} {\bibfnamefont {R.}~\bibnamefont {Hamatsu}}, \bibinfo {author} {\bibfnamefont {K.}~\bibnamefont {Hashimoto}}, \bibinfo {author} {\bibfnamefont
  {H.}~\bibnamefont {Hasui}}, \bibinfo {author} {\bibfnamefont {A.}~\bibnamefont {Heikkinen}}, \bibinfo {author} {\bibfnamefont {A.}~\bibnamefont {Howard}}, \bibinfo {author} {\bibfnamefont {V.}~\bibnamefont {Ivanchenko}}, \bibinfo {author} {\bibfnamefont {A.}~\bibnamefont {Johnson}}, \bibinfo {author} {\bibfnamefont {F.}~\bibnamefont {Jones}}, \bibinfo {author} {\bibfnamefont {J.}~\bibnamefont {Kallenbach}}, \bibinfo {author} {\bibfnamefont {N.}~\bibnamefont {Kanaya}}, \bibinfo {author} {\bibfnamefont {M.}~\bibnamefont {Kawabata}}, \bibinfo {author} {\bibfnamefont {Y.}~\bibnamefont {Kawabata}}, \bibinfo {author} {\bibfnamefont {M.}~\bibnamefont {Kawaguti}}, \bibinfo {author} {\bibfnamefont {S.}~\bibnamefont {Kelner}}, \bibinfo {author} {\bibfnamefont {P.}~\bibnamefont {Kent}}, \bibinfo {author} {\bibfnamefont {A.}~\bibnamefont {Kimura}}, \bibinfo {author} {\bibfnamefont {T.}~\bibnamefont {Kodama}}, \bibinfo {author} {\bibfnamefont {R.}~\bibnamefont {Kokoulin}}, \bibinfo {author} {\bibfnamefont
  {M.}~\bibnamefont {Kossov}}, \bibinfo {author} {\bibfnamefont {H.}~\bibnamefont {Kurashige}}, \bibinfo {author} {\bibfnamefont {E.}~\bibnamefont {Lamanna}}, \bibinfo {author} {\bibfnamefont {T.}~\bibnamefont {Lampén}}, \bibinfo {author} {\bibfnamefont {V.}~\bibnamefont {Lara}}, \bibinfo {author} {\bibfnamefont {V.}~\bibnamefont {Lefebure}}, \bibinfo {author} {\bibfnamefont {F.}~\bibnamefont {Lei}}, \bibinfo {author} {\bibfnamefont {M.}~\bibnamefont {Liendl}}, \bibinfo {author} {\bibfnamefont {W.}~\bibnamefont {Lockman}}, \bibinfo {author} {\bibfnamefont {F.}~\bibnamefont {Longo}}, \bibinfo {author} {\bibfnamefont {S.}~\bibnamefont {Magni}}, \bibinfo {author} {\bibfnamefont {M.}~\bibnamefont {Maire}}, \bibinfo {author} {\bibfnamefont {E.}~\bibnamefont {Medernach}}, \bibinfo {author} {\bibfnamefont {K.}~\bibnamefont {Minamimoto}}, \bibinfo {author} {\bibfnamefont {P.}~\bibnamefont {Mora~de Freitas}}, \bibinfo {author} {\bibfnamefont {Y.}~\bibnamefont {Morita}}, \bibinfo {author} {\bibfnamefont
  {K.}~\bibnamefont {Murakami}}, \bibinfo {author} {\bibfnamefont {M.}~\bibnamefont {Nagamatu}}, \bibinfo {author} {\bibfnamefont {R.}~\bibnamefont {Nartallo}}, \bibinfo {author} {\bibfnamefont {P.}~\bibnamefont {Nieminen}}, \bibinfo {author} {\bibfnamefont {T.}~\bibnamefont {Nishimura}}, \bibinfo {author} {\bibfnamefont {K.}~\bibnamefont {Ohtsubo}}, \bibinfo {author} {\bibfnamefont {M.}~\bibnamefont {Okamura}}, \bibinfo {author} {\bibfnamefont {S.}~\bibnamefont {O'Neale}}, \bibinfo {author} {\bibfnamefont {Y.}~\bibnamefont {Oohata}}, \bibinfo {author} {\bibfnamefont {K.}~\bibnamefont {Paech}}, \bibinfo {author} {\bibfnamefont {J.}~\bibnamefont {Perl}}, \bibinfo {author} {\bibfnamefont {A.}~\bibnamefont {Pfeiffer}}, \bibinfo {author} {\bibfnamefont {M.}~\bibnamefont {Pia}}, \bibinfo {author} {\bibfnamefont {F.}~\bibnamefont {Ranjard}}, \bibinfo {author} {\bibfnamefont {A.}~\bibnamefont {Rybin}}, \bibinfo {author} {\bibfnamefont {S.}~\bibnamefont {Sadilov}}, \bibinfo {author} {\bibfnamefont {E.}~\bibnamefont
  {Di~Salvo}}, \bibinfo {author} {\bibfnamefont {G.}~\bibnamefont {Santin}}, \bibinfo {author} {\bibfnamefont {T.}~\bibnamefont {Sasaki}}, \bibinfo {author} {\bibfnamefont {N.}~\bibnamefont {Savvas}}, \bibinfo {author} {\bibfnamefont {Y.}~\bibnamefont {Sawada}}, \bibinfo {author} {\bibfnamefont {S.}~\bibnamefont {Scherer}}, \bibinfo {author} {\bibfnamefont {S.}~\bibnamefont {Sei}}, \bibinfo {author} {\bibfnamefont {V.}~\bibnamefont {Sirotenko}}, \bibinfo {author} {\bibfnamefont {D.}~\bibnamefont {Smith}}, \bibinfo {author} {\bibfnamefont {N.}~\bibnamefont {Starkov}}, \bibinfo {author} {\bibfnamefont {H.}~\bibnamefont {Stoecker}}, \bibinfo {author} {\bibfnamefont {J.}~\bibnamefont {Sulkimo}}, \bibinfo {author} {\bibfnamefont {M.}~\bibnamefont {Takahata}}, \bibinfo {author} {\bibfnamefont {S.}~\bibnamefont {Tanaka}}, \bibinfo {author} {\bibfnamefont {E.}~\bibnamefont {Tcherniaev}}, \bibinfo {author} {\bibfnamefont {E.}~\bibnamefont {Safai~Tehrani}}, \bibinfo {author} {\bibfnamefont {M.}~\bibnamefont
  {Tropeano}}, \bibinfo {author} {\bibfnamefont {P.}~\bibnamefont {Truscott}}, \bibinfo {author} {\bibfnamefont {H.}~\bibnamefont {Uno}}, \bibinfo {author} {\bibfnamefont {L.}~\bibnamefont {Urban}}, \bibinfo {author} {\bibfnamefont {P.}~\bibnamefont {Urban}}, \bibinfo {author} {\bibfnamefont {M.}~\bibnamefont {Verderi}}, \bibinfo {author} {\bibfnamefont {A.}~\bibnamefont {Walkden}}, \bibinfo {author} {\bibfnamefont {W.}~\bibnamefont {Wander}}, \bibinfo {author} {\bibfnamefont {H.}~\bibnamefont {Weber}}, \bibinfo {author} {\bibfnamefont {J.}~\bibnamefont {Wellisch}}, \bibinfo {author} {\bibfnamefont {T.}~\bibnamefont {Wenaus}}, \bibinfo {author} {\bibfnamefont {D.}~\bibnamefont {Williams}}, \bibinfo {author} {\bibfnamefont {D.}~\bibnamefont {Wright}}, \bibinfo {author} {\bibfnamefont {T.}~\bibnamefont {Yamada}}, \bibinfo {author} {\bibfnamefont {H.}~\bibnamefont {Yoshida}},\ and\ \bibinfo {author} {\bibfnamefont {D.}~\bibnamefont {Zschiesche}},\ }\bibfield  {title} {\bibinfo {title} {Geant4—a simulation
  toolkit},\ }\href {https://doi.org/10.1016/S0168-9002(03)01368-8} {\bibfield  {journal} {\bibinfo  {journal} {Nucl. Instrum. Methods Phys. Res. A}\ }\textbf {\bibinfo {volume} {506}},\ \bibinfo {pages} {250} (\bibinfo {year} {2003})}\BibitemShut {NoStop}%
\bibitem [{\citenamefont {{J. Allison}}\ \emph {et~al.}(2006)\citenamefont {{J. Allison}}, \citenamefont {{K. Amako}}, \citenamefont {{J. Apostolakis}}, \citenamefont {{H. Araujo}}, \citenamefont {{P. Arce Dubois}}, \citenamefont {{M. Asai}}, \citenamefont {{G. Barrand}}, \citenamefont {{R. Capra}}, \citenamefont {{S. Chauvie}}, \citenamefont {{R. Chytracek}}, \citenamefont {{G. A. P. Cirrone}}, \citenamefont {{G. Cooperman}}, \citenamefont {{G. Cosmo}}, \citenamefont {{G. Cuttone}}, \citenamefont {{G. G. Daquino}}, \citenamefont {{M. Donszelmann}}, \citenamefont {{M. Dressel}}, \citenamefont {{G. Folger}}, \citenamefont {{F. Foppiano}}, \citenamefont {{J. Generowicz}}, \citenamefont {{V. Grichine}}, \citenamefont {{S. Guatelli}}, \citenamefont {{P. Gumplinger}}, \citenamefont {{A. Heikkinen}}, \citenamefont {{I. Hrivnacova}}, \citenamefont {{A. Howard}}, \citenamefont {{S. Incerti}}, \citenamefont {{V. Ivanchenko}}, \citenamefont {{T. Johnson}}, \citenamefont {{F. Jones}}, \citenamefont {{T. Koi}},
  \citenamefont {{R. Kokoulin}}, \citenamefont {{M. Kossov}}, \citenamefont {{H. Kurashige}}, \citenamefont {{V. Lara}}, \citenamefont {{S. Larsson}}, \citenamefont {{F. Lei}}, \citenamefont {{O. Link}}, \citenamefont {{F. Longo}}, \citenamefont {{M. Maire}}, \citenamefont {{A. Mantero}}, \citenamefont {{B. Mascialino}}, \citenamefont {{I. McLaren}}, \citenamefont {{P. Mendez Lorenzo}}, \citenamefont {{K. Minamimoto}}, \citenamefont {{K. Murakami}}, \citenamefont {{P. Nieminen}}, \citenamefont {{L. Pandola}}, \citenamefont {{S. Parlati}}, \citenamefont {{L. Peralta}}, \citenamefont {{J. Perl}}, \citenamefont {{A. Pfeiffer}}, \citenamefont {{M. G. Pia}}, \citenamefont {{A. Ribon}}, \citenamefont {{P. Rodrigues}}, \citenamefont {{G. Russo}}, \citenamefont {{S. Sadilov}}, \citenamefont {{G. Santin}}, \citenamefont {{T. Sasaki}}, \citenamefont {{D. Smith}}, \citenamefont {{N. Starkov}}, \citenamefont {{S. Tanaka}}, \citenamefont {{E. Tcherniaev}}, \citenamefont {{B. Tome}}, \citenamefont {{A. Trindade}},
  \citenamefont {{P. Truscott}}, \citenamefont {{L. Urban}}, \citenamefont {{M. Verderi}}, \citenamefont {{A. Walkden}}, \citenamefont {{J. P. Wellisch}}, \citenamefont {{D. C. Williams}}, \citenamefont {{D. Wright}},\ and\ \citenamefont {{H. Yoshida}}}]{j_allison_geant4_2006}%
  \BibitemOpen
  \bibfield  {author} {\bibinfo {author} {\bibnamefont {{J. Allison}}}, \bibinfo {author} {\bibnamefont {{K. Amako}}}, \bibinfo {author} {\bibnamefont {{J. Apostolakis}}}, \bibinfo {author} {\bibnamefont {{H. Araujo}}}, \bibinfo {author} {\bibnamefont {{P. Arce Dubois}}}, \bibinfo {author} {\bibnamefont {{M. Asai}}}, \bibinfo {author} {\bibnamefont {{G. Barrand}}}, \bibinfo {author} {\bibnamefont {{R. Capra}}}, \bibinfo {author} {\bibnamefont {{S. Chauvie}}}, \bibinfo {author} {\bibnamefont {{R. Chytracek}}}, \bibinfo {author} {\bibnamefont {{G. A. P. Cirrone}}}, \bibinfo {author} {\bibnamefont {{G. Cooperman}}}, \bibinfo {author} {\bibnamefont {{G. Cosmo}}}, \bibinfo {author} {\bibnamefont {{G. Cuttone}}}, \bibinfo {author} {\bibnamefont {{G. G. Daquino}}}, \bibinfo {author} {\bibnamefont {{M. Donszelmann}}}, \bibinfo {author} {\bibnamefont {{M. Dressel}}}, \bibinfo {author} {\bibnamefont {{G. Folger}}}, \bibinfo {author} {\bibnamefont {{F. Foppiano}}}, \bibinfo {author} {\bibnamefont {{J. Generowicz}}},
  \bibinfo {author} {\bibnamefont {{V. Grichine}}}, \bibinfo {author} {\bibnamefont {{S. Guatelli}}}, \bibinfo {author} {\bibnamefont {{P. Gumplinger}}}, \bibinfo {author} {\bibnamefont {{A. Heikkinen}}}, \bibinfo {author} {\bibnamefont {{I. Hrivnacova}}}, \bibinfo {author} {\bibnamefont {{A. Howard}}}, \bibinfo {author} {\bibnamefont {{S. Incerti}}}, \bibinfo {author} {\bibnamefont {{V. Ivanchenko}}}, \bibinfo {author} {\bibnamefont {{T. Johnson}}}, \bibinfo {author} {\bibnamefont {{F. Jones}}}, \bibinfo {author} {\bibnamefont {{T. Koi}}}, \bibinfo {author} {\bibnamefont {{R. Kokoulin}}}, \bibinfo {author} {\bibnamefont {{M. Kossov}}}, \bibinfo {author} {\bibnamefont {{H. Kurashige}}}, \bibinfo {author} {\bibnamefont {{V. Lara}}}, \bibinfo {author} {\bibnamefont {{S. Larsson}}}, \bibinfo {author} {\bibnamefont {{F. Lei}}}, \bibinfo {author} {\bibnamefont {{O. Link}}}, \bibinfo {author} {\bibnamefont {{F. Longo}}}, \bibinfo {author} {\bibnamefont {{M. Maire}}}, \bibinfo {author} {\bibnamefont {{A. Mantero}}},
  \bibinfo {author} {\bibnamefont {{B. Mascialino}}}, \bibinfo {author} {\bibnamefont {{I. McLaren}}}, \bibinfo {author} {\bibnamefont {{P. Mendez Lorenzo}}}, \bibinfo {author} {\bibnamefont {{K. Minamimoto}}}, \bibinfo {author} {\bibnamefont {{K. Murakami}}}, \bibinfo {author} {\bibnamefont {{P. Nieminen}}}, \bibinfo {author} {\bibnamefont {{L. Pandola}}}, \bibinfo {author} {\bibnamefont {{S. Parlati}}}, \bibinfo {author} {\bibnamefont {{L. Peralta}}}, \bibinfo {author} {\bibnamefont {{J. Perl}}}, \bibinfo {author} {\bibnamefont {{A. Pfeiffer}}}, \bibinfo {author} {\bibnamefont {{M. G. Pia}}}, \bibinfo {author} {\bibnamefont {{A. Ribon}}}, \bibinfo {author} {\bibnamefont {{P. Rodrigues}}}, \bibinfo {author} {\bibnamefont {{G. Russo}}}, \bibinfo {author} {\bibnamefont {{S. Sadilov}}}, \bibinfo {author} {\bibnamefont {{G. Santin}}}, \bibinfo {author} {\bibnamefont {{T. Sasaki}}}, \bibinfo {author} {\bibnamefont {{D. Smith}}}, \bibinfo {author} {\bibnamefont {{N. Starkov}}}, \bibinfo {author} {\bibnamefont {{S.
  Tanaka}}}, \bibinfo {author} {\bibnamefont {{E. Tcherniaev}}}, \bibinfo {author} {\bibnamefont {{B. Tome}}}, \bibinfo {author} {\bibnamefont {{A. Trindade}}}, \bibinfo {author} {\bibnamefont {{P. Truscott}}}, \bibinfo {author} {\bibnamefont {{L. Urban}}}, \bibinfo {author} {\bibnamefont {{M. Verderi}}}, \bibinfo {author} {\bibnamefont {{A. Walkden}}}, \bibinfo {author} {\bibnamefont {{J. P. Wellisch}}}, \bibinfo {author} {\bibnamefont {{D. C. Williams}}}, \bibinfo {author} {\bibnamefont {{D. Wright}}},\ and\ \bibinfo {author} {\bibnamefont {{H. Yoshida}}},\ }\bibfield  {title} {\bibinfo {title} {Geant4 developments and applications},\ }\href {https://doi.org/10.1109/TNS.2006.869826} {\bibfield  {journal} {\bibinfo  {journal} {IEEE Trans. Nucl. Sci.}\ }\textbf {\bibinfo {volume} {53}},\ \bibinfo {pages} {270} (\bibinfo {year} {2006})}\BibitemShut {NoStop}%
\bibitem [{\citenamefont {Allison}\ \emph {et~al.}(2016)\citenamefont {Allison}, \citenamefont {Amako}, \citenamefont {Apostolakis}, \citenamefont {Arce}, \citenamefont {Asai}, \citenamefont {Aso}, \citenamefont {Bagli}, \citenamefont {Bagulya}, \citenamefont {Banerjee}, \citenamefont {Barrand}, \citenamefont {Beck}, \citenamefont {Bogdanov}, \citenamefont {Brandt}, \citenamefont {Brown}, \citenamefont {Burkhardt}, \citenamefont {Canal}, \citenamefont {Cano-Ott}, \citenamefont {Chauvie}, \citenamefont {Cho}, \citenamefont {Cirrone}, \citenamefont {Cooperman}, \citenamefont {Cortés-Giraldo}, \citenamefont {Cosmo}, \citenamefont {Cuttone}, \citenamefont {Depaola}, \citenamefont {Desorgher}, \citenamefont {Dong}, \citenamefont {Dotti}, \citenamefont {Elvira}, \citenamefont {Folger}, \citenamefont {Francis}, \citenamefont {Galoyan}, \citenamefont {Garnier}, \citenamefont {Gayer}, \citenamefont {Genser}, \citenamefont {Grichine}, \citenamefont {Guatelli}, \citenamefont {Guèye}, \citenamefont {Gumplinger},
  \citenamefont {Howard}, \citenamefont {Hřivnáčová}, \citenamefont {Hwang}, \citenamefont {Incerti}, \citenamefont {Ivanchenko}, \citenamefont {Ivanchenko}, \citenamefont {Jones}, \citenamefont {Jun}, \citenamefont {Kaitaniemi}, \citenamefont {Karakatsanis}, \citenamefont {Karamitros}, \citenamefont {Kelsey}, \citenamefont {Kimura}, \citenamefont {Koi}, \citenamefont {Kurashige}, \citenamefont {Lechner}, \citenamefont {Lee}, \citenamefont {Longo}, \citenamefont {Maire}, \citenamefont {Mancusi}, \citenamefont {Mantero}, \citenamefont {Mendoza}, \citenamefont {Morgan}, \citenamefont {Murakami}, \citenamefont {Nikitina}, \citenamefont {Pandola}, \citenamefont {Paprocki}, \citenamefont {Perl}, \citenamefont {Petrović}, \citenamefont {Pia}, \citenamefont {Pokorski}, \citenamefont {Quesada}, \citenamefont {Raine}, \citenamefont {Reis}, \citenamefont {Ribon}, \citenamefont {Ristić~Fira}, \citenamefont {Romano}, \citenamefont {Russo}, \citenamefont {Santin}, \citenamefont {Sasaki}, \citenamefont {Sawkey},
  \citenamefont {Shin}, \citenamefont {Strakovsky}, \citenamefont {Taborda}, \citenamefont {Tanaka}, \citenamefont {Tomé}, \citenamefont {Toshito}, \citenamefont {Tran}, \citenamefont {Truscott}, \citenamefont {Urban}, \citenamefont {Uzhinsky}, \citenamefont {Verbeke}, \citenamefont {Verderi}, \citenamefont {Wendt}, \citenamefont {Wenzel}, \citenamefont {Wright}, \citenamefont {Wright}, \citenamefont {Yamashita}, \citenamefont {Yarba},\ and\ \citenamefont {Yoshida}}]{allison_recent_2016}%
  \BibitemOpen
  \bibfield  {author} {\bibinfo {author} {\bibfnamefont {J.}~\bibnamefont {Allison}}, \bibinfo {author} {\bibfnamefont {K.}~\bibnamefont {Amako}}, \bibinfo {author} {\bibfnamefont {J.}~\bibnamefont {Apostolakis}}, \bibinfo {author} {\bibfnamefont {P.}~\bibnamefont {Arce}}, \bibinfo {author} {\bibfnamefont {M.}~\bibnamefont {Asai}}, \bibinfo {author} {\bibfnamefont {T.}~\bibnamefont {Aso}}, \bibinfo {author} {\bibfnamefont {E.}~\bibnamefont {Bagli}}, \bibinfo {author} {\bibfnamefont {A.}~\bibnamefont {Bagulya}}, \bibinfo {author} {\bibfnamefont {S.}~\bibnamefont {Banerjee}}, \bibinfo {author} {\bibfnamefont {G.}~\bibnamefont {Barrand}}, \bibinfo {author} {\bibfnamefont {B.}~\bibnamefont {Beck}}, \bibinfo {author} {\bibfnamefont {A.}~\bibnamefont {Bogdanov}}, \bibinfo {author} {\bibfnamefont {D.}~\bibnamefont {Brandt}}, \bibinfo {author} {\bibfnamefont {J.}~\bibnamefont {Brown}}, \bibinfo {author} {\bibfnamefont {H.}~\bibnamefont {Burkhardt}}, \bibinfo {author} {\bibfnamefont {P.}~\bibnamefont {Canal}},
  \bibinfo {author} {\bibfnamefont {D.}~\bibnamefont {Cano-Ott}}, \bibinfo {author} {\bibfnamefont {S.}~\bibnamefont {Chauvie}}, \bibinfo {author} {\bibfnamefont {K.}~\bibnamefont {Cho}}, \bibinfo {author} {\bibfnamefont {G.}~\bibnamefont {Cirrone}}, \bibinfo {author} {\bibfnamefont {G.}~\bibnamefont {Cooperman}}, \bibinfo {author} {\bibfnamefont {M.}~\bibnamefont {Cortés-Giraldo}}, \bibinfo {author} {\bibfnamefont {G.}~\bibnamefont {Cosmo}}, \bibinfo {author} {\bibfnamefont {G.}~\bibnamefont {Cuttone}}, \bibinfo {author} {\bibfnamefont {G.}~\bibnamefont {Depaola}}, \bibinfo {author} {\bibfnamefont {L.}~\bibnamefont {Desorgher}}, \bibinfo {author} {\bibfnamefont {X.}~\bibnamefont {Dong}}, \bibinfo {author} {\bibfnamefont {A.}~\bibnamefont {Dotti}}, \bibinfo {author} {\bibfnamefont {V.}~\bibnamefont {Elvira}}, \bibinfo {author} {\bibfnamefont {G.}~\bibnamefont {Folger}}, \bibinfo {author} {\bibfnamefont {Z.}~\bibnamefont {Francis}}, \bibinfo {author} {\bibfnamefont {A.}~\bibnamefont {Galoyan}}, \bibinfo
  {author} {\bibfnamefont {L.}~\bibnamefont {Garnier}}, \bibinfo {author} {\bibfnamefont {M.}~\bibnamefont {Gayer}}, \bibinfo {author} {\bibfnamefont {K.}~\bibnamefont {Genser}}, \bibinfo {author} {\bibfnamefont {V.}~\bibnamefont {Grichine}}, \bibinfo {author} {\bibfnamefont {S.}~\bibnamefont {Guatelli}}, \bibinfo {author} {\bibfnamefont {P.}~\bibnamefont {Guèye}}, \bibinfo {author} {\bibfnamefont {P.}~\bibnamefont {Gumplinger}}, \bibinfo {author} {\bibfnamefont {A.}~\bibnamefont {Howard}}, \bibinfo {author} {\bibfnamefont {I.}~\bibnamefont {Hřivnáčová}}, \bibinfo {author} {\bibfnamefont {S.}~\bibnamefont {Hwang}}, \bibinfo {author} {\bibfnamefont {S.}~\bibnamefont {Incerti}}, \bibinfo {author} {\bibfnamefont {A.}~\bibnamefont {Ivanchenko}}, \bibinfo {author} {\bibfnamefont {V.}~\bibnamefont {Ivanchenko}}, \bibinfo {author} {\bibfnamefont {F.}~\bibnamefont {Jones}}, \bibinfo {author} {\bibfnamefont {S.}~\bibnamefont {Jun}}, \bibinfo {author} {\bibfnamefont {P.}~\bibnamefont {Kaitaniemi}}, \bibinfo
  {author} {\bibfnamefont {N.}~\bibnamefont {Karakatsanis}}, \bibinfo {author} {\bibfnamefont {M.}~\bibnamefont {Karamitros}}, \bibinfo {author} {\bibfnamefont {M.}~\bibnamefont {Kelsey}}, \bibinfo {author} {\bibfnamefont {A.}~\bibnamefont {Kimura}}, \bibinfo {author} {\bibfnamefont {T.}~\bibnamefont {Koi}}, \bibinfo {author} {\bibfnamefont {H.}~\bibnamefont {Kurashige}}, \bibinfo {author} {\bibfnamefont {A.}~\bibnamefont {Lechner}}, \bibinfo {author} {\bibfnamefont {S.}~\bibnamefont {Lee}}, \bibinfo {author} {\bibfnamefont {F.}~\bibnamefont {Longo}}, \bibinfo {author} {\bibfnamefont {M.}~\bibnamefont {Maire}}, \bibinfo {author} {\bibfnamefont {D.}~\bibnamefont {Mancusi}}, \bibinfo {author} {\bibfnamefont {A.}~\bibnamefont {Mantero}}, \bibinfo {author} {\bibfnamefont {E.}~\bibnamefont {Mendoza}}, \bibinfo {author} {\bibfnamefont {B.}~\bibnamefont {Morgan}}, \bibinfo {author} {\bibfnamefont {K.}~\bibnamefont {Murakami}}, \bibinfo {author} {\bibfnamefont {T.}~\bibnamefont {Nikitina}}, \bibinfo {author}
  {\bibfnamefont {L.}~\bibnamefont {Pandola}}, \bibinfo {author} {\bibfnamefont {P.}~\bibnamefont {Paprocki}}, \bibinfo {author} {\bibfnamefont {J.}~\bibnamefont {Perl}}, \bibinfo {author} {\bibfnamefont {I.}~\bibnamefont {Petrović}}, \bibinfo {author} {\bibfnamefont {M.}~\bibnamefont {Pia}}, \bibinfo {author} {\bibfnamefont {W.}~\bibnamefont {Pokorski}}, \bibinfo {author} {\bibfnamefont {J.}~\bibnamefont {Quesada}}, \bibinfo {author} {\bibfnamefont {M.}~\bibnamefont {Raine}}, \bibinfo {author} {\bibfnamefont {M.}~\bibnamefont {Reis}}, \bibinfo {author} {\bibfnamefont {A.}~\bibnamefont {Ribon}}, \bibinfo {author} {\bibfnamefont {A.}~\bibnamefont {Ristić~Fira}}, \bibinfo {author} {\bibfnamefont {F.}~\bibnamefont {Romano}}, \bibinfo {author} {\bibfnamefont {G.}~\bibnamefont {Russo}}, \bibinfo {author} {\bibfnamefont {G.}~\bibnamefont {Santin}}, \bibinfo {author} {\bibfnamefont {T.}~\bibnamefont {Sasaki}}, \bibinfo {author} {\bibfnamefont {D.}~\bibnamefont {Sawkey}}, \bibinfo {author} {\bibfnamefont
  {J.}~\bibnamefont {Shin}}, \bibinfo {author} {\bibfnamefont {I.}~\bibnamefont {Strakovsky}}, \bibinfo {author} {\bibfnamefont {A.}~\bibnamefont {Taborda}}, \bibinfo {author} {\bibfnamefont {S.}~\bibnamefont {Tanaka}}, \bibinfo {author} {\bibfnamefont {B.}~\bibnamefont {Tomé}}, \bibinfo {author} {\bibfnamefont {T.}~\bibnamefont {Toshito}}, \bibinfo {author} {\bibfnamefont {H.}~\bibnamefont {Tran}}, \bibinfo {author} {\bibfnamefont {P.}~\bibnamefont {Truscott}}, \bibinfo {author} {\bibfnamefont {L.}~\bibnamefont {Urban}}, \bibinfo {author} {\bibfnamefont {V.}~\bibnamefont {Uzhinsky}}, \bibinfo {author} {\bibfnamefont {J.}~\bibnamefont {Verbeke}}, \bibinfo {author} {\bibfnamefont {M.}~\bibnamefont {Verderi}}, \bibinfo {author} {\bibfnamefont {B.}~\bibnamefont {Wendt}}, \bibinfo {author} {\bibfnamefont {H.}~\bibnamefont {Wenzel}}, \bibinfo {author} {\bibfnamefont {D.}~\bibnamefont {Wright}}, \bibinfo {author} {\bibfnamefont {D.}~\bibnamefont {Wright}}, \bibinfo {author} {\bibfnamefont {T.}~\bibnamefont
  {Yamashita}}, \bibinfo {author} {\bibfnamefont {J.}~\bibnamefont {Yarba}},\ and\ \bibinfo {author} {\bibfnamefont {H.}~\bibnamefont {Yoshida}},\ }\bibfield  {title} {\bibinfo {title} {Recent developments in {Geant4}},\ }\href {https://doi.org/10.1016/j.nima.2016.06.125} {\bibfield  {journal} {\bibinfo  {journal} {Nucl. Instrum. Methods Phys. Res. A}\ }\textbf {\bibinfo {volume} {835}},\ \bibinfo {pages} {186} (\bibinfo {year} {2016})}\BibitemShut {NoStop}%
\bibitem [{\citenamefont {Battistoni}\ \emph {et~al.}(2015)\citenamefont {Battistoni}, \citenamefont {Boehlen}, \citenamefont {Cerutti}, \citenamefont {Chin}, \citenamefont {Esposito}, \citenamefont {Fassò}, \citenamefont {Ferrari}, \citenamefont {Lechner}, \citenamefont {Empl}, \citenamefont {Mairani}, \citenamefont {Mereghetti}, \citenamefont {Ortega}, \citenamefont {Ranft}, \citenamefont {Roesler}, \citenamefont {Sala}, \citenamefont {Vlachoudis},\ and\ \citenamefont {Smirnov}}]{fluka2015}%
  \BibitemOpen
  \bibfield  {author} {\bibinfo {author} {\bibfnamefont {G.}~\bibnamefont {Battistoni}}, \bibinfo {author} {\bibfnamefont {T.}~\bibnamefont {Boehlen}}, \bibinfo {author} {\bibfnamefont {F.}~\bibnamefont {Cerutti}}, \bibinfo {author} {\bibfnamefont {P.~W.}\ \bibnamefont {Chin}}, \bibinfo {author} {\bibfnamefont {L.~S.}\ \bibnamefont {Esposito}}, \bibinfo {author} {\bibfnamefont {A.}~\bibnamefont {Fassò}}, \bibinfo {author} {\bibfnamefont {A.}~\bibnamefont {Ferrari}}, \bibinfo {author} {\bibfnamefont {A.}~\bibnamefont {Lechner}}, \bibinfo {author} {\bibfnamefont {A.}~\bibnamefont {Empl}}, \bibinfo {author} {\bibfnamefont {A.}~\bibnamefont {Mairani}}, \bibinfo {author} {\bibfnamefont {A.}~\bibnamefont {Mereghetti}}, \bibinfo {author} {\bibfnamefont {P.~G.}\ \bibnamefont {Ortega}}, \bibinfo {author} {\bibfnamefont {J.}~\bibnamefont {Ranft}}, \bibinfo {author} {\bibfnamefont {S.}~\bibnamefont {Roesler}}, \bibinfo {author} {\bibfnamefont {P.~R.}\ \bibnamefont {Sala}}, \bibinfo {author} {\bibfnamefont
  {V.}~\bibnamefont {Vlachoudis}},\ and\ \bibinfo {author} {\bibfnamefont {G.}~\bibnamefont {Smirnov}},\ }\bibfield  {title} {\bibinfo {title} {Overview of the {FLUKA} code},\ }\href {https://doi.org/10.1016/j.anucene.2014.11.007} {\bibfield  {journal} {\bibinfo  {journal} {Joint International Conference on Supercomputing in Nuclear Applications and Monte Carlo 2013, SNA + MC 2013. Pluri- and Trans-disciplinarity, Towards New Modeling and Numerical Simulation Paradigms}\ }\textbf {\bibinfo {volume} {82}},\ \bibinfo {pages} {10} (\bibinfo {year} {2015})}\BibitemShut {NoStop}%
\bibitem [{\citenamefont {Ahdida}\ \emph {et~al.}(2022)\citenamefont {Ahdida}, \citenamefont {Bozzato}, \citenamefont {Calzolari}, \citenamefont {Cerutti}, \citenamefont {Charitonidis}, \citenamefont {Cimmino}, \citenamefont {Coronetti}, \citenamefont {D’Alessandro}, \citenamefont {Donadon~Servelle},\ and\ \citenamefont {Esposito}}]{fluka2022}%
  \BibitemOpen
  \bibfield  {author} {\bibinfo {author} {\bibfnamefont {C.}~\bibnamefont {Ahdida}}, \bibinfo {author} {\bibfnamefont {D.}~\bibnamefont {Bozzato}}, \bibinfo {author} {\bibfnamefont {D.}~\bibnamefont {Calzolari}}, \bibinfo {author} {\bibfnamefont {F.}~\bibnamefont {Cerutti}}, \bibinfo {author} {\bibfnamefont {N.}~\bibnamefont {Charitonidis}}, \bibinfo {author} {\bibfnamefont {A.}~\bibnamefont {Cimmino}}, \bibinfo {author} {\bibfnamefont {A.}~\bibnamefont {Coronetti}}, \bibinfo {author} {\bibfnamefont {G.}~\bibnamefont {D’Alessandro}}, \bibinfo {author} {\bibfnamefont {A.}~\bibnamefont {Donadon~Servelle}},\ and\ \bibinfo {author} {\bibfnamefont {L.}~\bibnamefont {Esposito}},\ }\bibfield  {title} {\bibinfo {title} {New capabilities of the {FLUKA} multi-purpose code},\ }\href {https://www.frontiersin.org/articles/10.3389/fphy.2021.788253/full} {\bibfield  {journal} {\bibinfo  {journal} {Frontiers in Physics}\ }\textbf {\bibinfo {volume} {9}},\ \bibinfo {pages} {788253} (\bibinfo {year} {2022})},\ \bibinfo
  {note} {publisher: Frontiers}\BibitemShut {NoStop}%
\bibitem [{\citenamefont {Drobychev}\ \emph {et~al.}(2007)\citenamefont {Drobychev}, \citenamefont {Nédélec}, \citenamefont {Sillou}, \citenamefont {Gribakin}, \citenamefont {Walters}, \citenamefont {Ferrari}, \citenamefont {Prevedelli}, \citenamefont {Tino}, \citenamefont {Doser}, \citenamefont {Canali}, \citenamefont {Carraro}, \citenamefont {Lagomarsino}, \citenamefont {Manuzio}, \citenamefont {Testera}, \citenamefont {Zavatarelli}, \citenamefont {Amoretti}, \citenamefont {Kellerbauer}, \citenamefont {Meier}, \citenamefont {Warring}, \citenamefont {Oberthaler}, \citenamefont {Boscolo}, \citenamefont {Castelli}, \citenamefont {Cialdi}, \citenamefont {Formaro}, \citenamefont {Gervasini}, \citenamefont {Giammarchi}, \citenamefont {Vairo}, \citenamefont {Consolati}, \citenamefont {Dupasquier}, \citenamefont {Quasso}, \citenamefont {Stroke}, \citenamefont {Belov}, \citenamefont {Gninenko}, \citenamefont {Matveev}, \citenamefont {Byakov}, \citenamefont {Stepanov}, \citenamefont {Zvezhinskij}, \citenamefont
  {De~Combarieu}, \citenamefont {Forget}, \citenamefont {Pari}, \citenamefont {Cabaret}, \citenamefont {Comparat}, \citenamefont {Bonomi}, \citenamefont {Rotondi}, \citenamefont {Djourelov}, \citenamefont {Jacquey}, \citenamefont {Büchner}, \citenamefont {Trénec}, \citenamefont {Vigué}, \citenamefont {Brusa}, \citenamefont {Mariazzi}, \citenamefont {Hogan}, \citenamefont {Merkt}, \citenamefont {Badertscher}, \citenamefont {Crivelli}, \citenamefont {Gendotti},\ and\ \citenamefont {Rubbia}}]{drobychev_proposal_2007}%
  \BibitemOpen
  \bibfield  {author} {\bibinfo {author} {\bibfnamefont {G.~Y.}\ \bibnamefont {Drobychev}}, \bibinfo {author} {\bibfnamefont {P.}~\bibnamefont {Nédélec}}, \bibinfo {author} {\bibfnamefont {D.}~\bibnamefont {Sillou}}, \bibinfo {author} {\bibfnamefont {G.}~\bibnamefont {Gribakin}}, \bibinfo {author} {\bibfnamefont {H.}~\bibnamefont {Walters}}, \bibinfo {author} {\bibfnamefont {G.}~\bibnamefont {Ferrari}}, \bibinfo {author} {\bibfnamefont {M.}~\bibnamefont {Prevedelli}}, \bibinfo {author} {\bibfnamefont {G.~M.}\ \bibnamefont {Tino}}, \bibinfo {author} {\bibfnamefont {M.}~\bibnamefont {Doser}}, \bibinfo {author} {\bibfnamefont {C.}~\bibnamefont {Canali}}, \bibinfo {author} {\bibfnamefont {C.}~\bibnamefont {Carraro}}, \bibinfo {author} {\bibfnamefont {V.}~\bibnamefont {Lagomarsino}}, \bibinfo {author} {\bibfnamefont {G.}~\bibnamefont {Manuzio}}, \bibinfo {author} {\bibfnamefont {G.}~\bibnamefont {Testera}}, \bibinfo {author} {\bibfnamefont {S.}~\bibnamefont {Zavatarelli}}, \bibinfo {author} {\bibfnamefont
  {M.}~\bibnamefont {Amoretti}}, \bibinfo {author} {\bibfnamefont {A.~G.}\ \bibnamefont {Kellerbauer}}, \bibinfo {author} {\bibfnamefont {J.}~\bibnamefont {Meier}}, \bibinfo {author} {\bibfnamefont {U.}~\bibnamefont {Warring}}, \bibinfo {author} {\bibfnamefont {M.~K.}\ \bibnamefont {Oberthaler}}, \bibinfo {author} {\bibfnamefont {I.}~\bibnamefont {Boscolo}}, \bibinfo {author} {\bibfnamefont {F.}~\bibnamefont {Castelli}}, \bibinfo {author} {\bibfnamefont {S.}~\bibnamefont {Cialdi}}, \bibinfo {author} {\bibfnamefont {L.}~\bibnamefont {Formaro}}, \bibinfo {author} {\bibfnamefont {A.}~\bibnamefont {Gervasini}}, \bibinfo {author} {\bibfnamefont {G.}~\bibnamefont {Giammarchi}}, \bibinfo {author} {\bibfnamefont {A.}~\bibnamefont {Vairo}}, \bibinfo {author} {\bibfnamefont {G.}~\bibnamefont {Consolati}}, \bibinfo {author} {\bibfnamefont {A.}~\bibnamefont {Dupasquier}}, \bibinfo {author} {\bibfnamefont {F.}~\bibnamefont {Quasso}}, \bibinfo {author} {\bibfnamefont {H.~H.}\ \bibnamefont {Stroke}}, \bibinfo {author}
  {\bibfnamefont {A.~S.}\ \bibnamefont {Belov}}, \bibinfo {author} {\bibfnamefont {S.~N.}\ \bibnamefont {Gninenko}}, \bibinfo {author} {\bibfnamefont {V.~A.}\ \bibnamefont {Matveev}}, \bibinfo {author} {\bibfnamefont {V.~M.}\ \bibnamefont {Byakov}}, \bibinfo {author} {\bibfnamefont {S.~V.}\ \bibnamefont {Stepanov}}, \bibinfo {author} {\bibfnamefont {D.~S.}\ \bibnamefont {Zvezhinskij}}, \bibinfo {author} {\bibfnamefont {M.}~\bibnamefont {De~Combarieu}}, \bibinfo {author} {\bibfnamefont {P.}~\bibnamefont {Forget}}, \bibinfo {author} {\bibfnamefont {P.}~\bibnamefont {Pari}}, \bibinfo {author} {\bibfnamefont {L.}~\bibnamefont {Cabaret}}, \bibinfo {author} {\bibfnamefont {D.}~\bibnamefont {Comparat}}, \bibinfo {author} {\bibfnamefont {G.}~\bibnamefont {Bonomi}}, \bibinfo {author} {\bibfnamefont {A.}~\bibnamefont {Rotondi}}, \bibinfo {author} {\bibfnamefont {N.}~\bibnamefont {Djourelov}}, \bibinfo {author} {\bibfnamefont {M.}~\bibnamefont {Jacquey}}, \bibinfo {author} {\bibfnamefont {M.}~\bibnamefont {Büchner}},
  \bibinfo {author} {\bibfnamefont {G.}~\bibnamefont {Trénec}}, \bibinfo {author} {\bibfnamefont {J.}~\bibnamefont {Vigué}}, \bibinfo {author} {\bibfnamefont {R.~S.}\ \bibnamefont {Brusa}}, \bibinfo {author} {\bibfnamefont {S.}~\bibnamefont {Mariazzi}}, \bibinfo {author} {\bibfnamefont {S.}~\bibnamefont {Hogan}}, \bibinfo {author} {\bibfnamefont {F.}~\bibnamefont {Merkt}}, \bibinfo {author} {\bibfnamefont {A.}~\bibnamefont {Badertscher}}, \bibinfo {author} {\bibfnamefont {P.}~\bibnamefont {Crivelli}}, \bibinfo {author} {\bibfnamefont {U.}~\bibnamefont {Gendotti}},\ and\ \bibinfo {author} {\bibfnamefont {A.}~\bibnamefont {Rubbia}},\ }\href {https://cds.cern.ch/record/1037532} {\emph {\bibinfo {title} {Proposal for the {AEGIS} experiment at the {CERN} antiproton decelerator ({Antimatter} {Experiment}: {Gravity}, {Interferometry}, {Spectroscopy})}}},\ \bibinfo {type} {Tech. Rep.}\ (\bibinfo  {institution} {CERN},\ \bibinfo {address} {Geneva},\ \bibinfo {year} {2007})\BibitemShut {NoStop}%
\bibitem [{\citenamefont {Doser}(2019)}]{doser_aegis_2019}%
  \BibitemOpen
  \bibfield  {author} {\bibinfo {author} {\bibfnamefont {M.}~\bibnamefont {Doser}},\ }\href@noop {} {\emph {\bibinfo {title} {Aegis program and physics prospects from 2020 up to and beyond {LS3}}}},\ \bibinfo {type} {Tech. Rep.}\ (\bibinfo {year} {2019})\BibitemShut {NoStop}%
\bibitem [{\citenamefont {Mohr}(2016)}]{PhysRevC.93.065804}%
  \BibitemOpen
  \bibfield  {author} {\bibinfo {author} {\bibfnamefont {P.}~\bibnamefont {Mohr}},\ }\bibfield  {title} {\bibinfo {title} {Nucleosynthesis of $^{92}\text{Nb}$ and the relevance of the low-lying isomer at 135.5 kev},\ }\href {https://doi.org/10.1103/PhysRevC.93.065804} {\bibfield  {journal} {\bibinfo  {journal} {Phys. Rev. C}\ }\textbf {\bibinfo {volume} {93}},\ \bibinfo {pages} {065804} (\bibinfo {year} {2016})}\BibitemShut {NoStop}%
\bibitem [{\citenamefont {Li}\ \emph {et~al.}(2020)\citenamefont {Li}, \citenamefont {Grumer}, \citenamefont {Brage},\ and\ \citenamefont {Jönsson}}]{Li2020}%
  \BibitemOpen
  \bibfield  {author} {\bibinfo {author} {\bibfnamefont {W.}~\bibnamefont {Li}}, \bibinfo {author} {\bibfnamefont {J.}~\bibnamefont {Grumer}}, \bibinfo {author} {\bibfnamefont {T.}~\bibnamefont {Brage}},\ and\ \bibinfo {author} {\bibfnamefont {P.}~\bibnamefont {Jönsson}},\ }\bibfield  {title} {\bibinfo {title} {Hfszeeman95—a program for computing weak and intermediate magnetic-field- and hyperfine-induced transition rates},\ }\href {https://doi.org/10.1016/j.cpc.2020.107211} {\bibfield  {journal} {\bibinfo  {journal} {Comput. Phys. Commun.}\ }\textbf {\bibinfo {volume} {253}},\ \bibinfo {pages} {107211} (\bibinfo {year} {2020})}\BibitemShut {NoStop}%
\bibitem [{\citenamefont {Schüssler}\ \emph {et~al.}(2020)\citenamefont {Schüssler}, \citenamefont {Bekker}, \citenamefont {Braß}, \citenamefont {Cakir}, \citenamefont {Crespo López-Urrutia}, \citenamefont {Door}, \citenamefont {Filianin}, \citenamefont {Harman}, \citenamefont {Haverkort}, \citenamefont {Huang}, \citenamefont {Indelicato}, \citenamefont {Keitel}, \citenamefont {König}, \citenamefont {Kromer}, \citenamefont {Müller}, \citenamefont {Novikov}, \citenamefont {Rischka}, \citenamefont {Schweiger}, \citenamefont {Sturm}, \citenamefont {Ulmer}, \citenamefont {Eliseev},\ and\ \citenamefont {Blaum}}]{Schuessler2020}%
  \BibitemOpen
  \bibfield  {author} {\bibinfo {author} {\bibfnamefont {R.~X.}\ \bibnamefont {Schüssler}}, \bibinfo {author} {\bibfnamefont {H.}~\bibnamefont {Bekker}}, \bibinfo {author} {\bibfnamefont {M.}~\bibnamefont {Braß}}, \bibinfo {author} {\bibfnamefont {H.}~\bibnamefont {Cakir}}, \bibinfo {author} {\bibfnamefont {J.~R.}\ \bibnamefont {Crespo López-Urrutia}}, \bibinfo {author} {\bibfnamefont {M.}~\bibnamefont {Door}}, \bibinfo {author} {\bibfnamefont {P.}~\bibnamefont {Filianin}}, \bibinfo {author} {\bibfnamefont {Z.}~\bibnamefont {Harman}}, \bibinfo {author} {\bibfnamefont {M.~W.}\ \bibnamefont {Haverkort}}, \bibinfo {author} {\bibfnamefont {W.~J.}\ \bibnamefont {Huang}}, \bibinfo {author} {\bibfnamefont {P.}~\bibnamefont {Indelicato}}, \bibinfo {author} {\bibfnamefont {C.~H.}\ \bibnamefont {Keitel}}, \bibinfo {author} {\bibfnamefont {C.~M.}\ \bibnamefont {König}}, \bibinfo {author} {\bibfnamefont {K.}~\bibnamefont {Kromer}}, \bibinfo {author} {\bibfnamefont {M.}~\bibnamefont {Müller}}, \bibinfo {author}
  {\bibfnamefont {Y.~N.}\ \bibnamefont {Novikov}}, \bibinfo {author} {\bibfnamefont {A.}~\bibnamefont {Rischka}}, \bibinfo {author} {\bibfnamefont {C.}~\bibnamefont {Schweiger}}, \bibinfo {author} {\bibfnamefont {S.}~\bibnamefont {Sturm}}, \bibinfo {author} {\bibfnamefont {S.}~\bibnamefont {Ulmer}}, \bibinfo {author} {\bibfnamefont {S.}~\bibnamefont {Eliseev}},\ and\ \bibinfo {author} {\bibfnamefont {K.}~\bibnamefont {Blaum}},\ }\bibfield  {title} {\bibinfo {title} {Detection of metastable electronic states by penning trap mass spectrometry},\ }\href {https://doi.org/10.1038/s41586-020-2221-0} {\bibfield  {journal} {\bibinfo  {journal} {Nature}\ }\textbf {\bibinfo {volume} {581}},\ \bibinfo {pages} {42} (\bibinfo {year} {2020})}\BibitemShut {NoStop}%
\bibitem [{\citenamefont {Tu}\ \emph {et~al.}(2023)\citenamefont {Tu}, \citenamefont {Si}, \citenamefont {Shen}, \citenamefont {Wang}, \citenamefont {Wei}, \citenamefont {Chen}, \citenamefont {Yao},\ and\ \citenamefont {Zou}}]{Tu2023}%
  \BibitemOpen
  \bibfield  {author} {\bibinfo {author} {\bibfnamefont {B.}~\bibnamefont {Tu}}, \bibinfo {author} {\bibfnamefont {R.}~\bibnamefont {Si}}, \bibinfo {author} {\bibfnamefont {Y.}~\bibnamefont {Shen}}, \bibinfo {author} {\bibfnamefont {J.}~\bibnamefont {Wang}}, \bibinfo {author} {\bibfnamefont {B.}~\bibnamefont {Wei}}, \bibinfo {author} {\bibfnamefont {C.}~\bibnamefont {Chen}}, \bibinfo {author} {\bibfnamefont {K.}~\bibnamefont {Yao}},\ and\ \bibinfo {author} {\bibfnamefont {Y.}~\bibnamefont {Zou}},\ }\bibfield  {title} {\bibinfo {title} {Experimental access to observing decay from extremely long-lived metastable electronic states via penning trap spectrometry},\ }\href {https://doi.org/10.1103/PhysRevResearch.5.043014} {\bibfield  {journal} {\bibinfo  {journal} {Phys. Rev. Res.}\ }\textbf {\bibinfo {volume} {5}},\ \bibinfo {pages} {043014} (\bibinfo {year} {2023})}\BibitemShut {NoStop}%
\bibitem [{\citenamefont {Flambaum}\ \emph {et~al.}(2018)\citenamefont {Flambaum}, \citenamefont {Geddes},\ and\ \citenamefont {Viatkina}}]{Flambaum2018}%
  \BibitemOpen
  \bibfield  {author} {\bibinfo {author} {\bibfnamefont {V.~V.}\ \bibnamefont {Flambaum}}, \bibinfo {author} {\bibfnamefont {A.~J.}\ \bibnamefont {Geddes}},\ and\ \bibinfo {author} {\bibfnamefont {A.~V.}\ \bibnamefont {Viatkina}},\ }\bibfield  {title} {\bibinfo {title} {Isotope shift, nonlinearity of king plots, and the search for new particles},\ }\href {https://doi.org/10.1103/PhysRevA.97.032510} {\bibfield  {journal} {\bibinfo  {journal} {Phys. Rev. A}\ }\textbf {\bibinfo {volume} {97}},\ \bibinfo {pages} {032510} (\bibinfo {year} {2018})}\BibitemShut {NoStop}%
\bibitem [{\citenamefont {Tiedau}\ \emph {et~al.}(2024)\citenamefont {Tiedau}, \citenamefont {Okhapkin}, \citenamefont {Zhang}, \citenamefont {Thielking}, \citenamefont {Zitzer}, \citenamefont {Peik}, \citenamefont {Schaden}, \citenamefont {Pronebner}, \citenamefont {Morawetz}, \citenamefont {De~Col}, \citenamefont {Schneider}, \citenamefont {Leitner}, \citenamefont {Pressler}, \citenamefont {Kazakov}, \citenamefont {Beeks}, \citenamefont {Sikorsky},\ and\ \citenamefont {Schumm}}]{Tiedau2024}%
  \BibitemOpen
  \bibfield  {author} {\bibinfo {author} {\bibfnamefont {J.}~\bibnamefont {Tiedau}}, \bibinfo {author} {\bibfnamefont {M.~V.}\ \bibnamefont {Okhapkin}}, \bibinfo {author} {\bibfnamefont {K.}~\bibnamefont {Zhang}}, \bibinfo {author} {\bibfnamefont {J.}~\bibnamefont {Thielking}}, \bibinfo {author} {\bibfnamefont {G.}~\bibnamefont {Zitzer}}, \bibinfo {author} {\bibfnamefont {E.}~\bibnamefont {Peik}}, \bibinfo {author} {\bibfnamefont {F.}~\bibnamefont {Schaden}}, \bibinfo {author} {\bibfnamefont {T.}~\bibnamefont {Pronebner}}, \bibinfo {author} {\bibfnamefont {I.}~\bibnamefont {Morawetz}}, \bibinfo {author} {\bibfnamefont {L.~T.}\ \bibnamefont {De~Col}}, \bibinfo {author} {\bibfnamefont {F.}~\bibnamefont {Schneider}}, \bibinfo {author} {\bibfnamefont {A.}~\bibnamefont {Leitner}}, \bibinfo {author} {\bibfnamefont {M.}~\bibnamefont {Pressler}}, \bibinfo {author} {\bibfnamefont {G.~A.}\ \bibnamefont {Kazakov}}, \bibinfo {author} {\bibfnamefont {K.}~\bibnamefont {Beeks}}, \bibinfo {author} {\bibfnamefont
  {T.}~\bibnamefont {Sikorsky}},\ and\ \bibinfo {author} {\bibfnamefont {T.}~\bibnamefont {Schumm}},\ }\bibfield  {title} {\bibinfo {title} {Laser excitation of the th-229 nucleus},\ }\href {https://doi.org/10.1103/PhysRevLett.132.182501} {\bibfield  {journal} {\bibinfo  {journal} {Phys. Rev. Lett.}\ }\textbf {\bibinfo {volume} {132}},\ \bibinfo {pages} {182501} (\bibinfo {year} {2024})}\BibitemShut {NoStop}%
\bibitem [{\citenamefont {Shabaev}\ \emph {et~al.}(1995)\citenamefont {Shabaev}, \citenamefont {Shabaeva},\ and\ \citenamefont {Tupitsyn}}]{Shabaev_1995}%
  \BibitemOpen
  \bibfield  {author} {\bibinfo {author} {\bibfnamefont {V.~M.}\ \bibnamefont {Shabaev}}, \bibinfo {author} {\bibfnamefont {M.~B.}\ \bibnamefont {Shabaeva}},\ and\ \bibinfo {author} {\bibfnamefont {I.~I.}\ \bibnamefont {Tupitsyn}},\ }\bibfield  {title} {\bibinfo {title} {Hyperfine structure of hydrogenlike and lithiumlike atoms},\ }\href {https://doi.org/10.1103/PhysRevA.52.3686} {\bibfield  {journal} {\bibinfo  {journal} {Phys. Rev. A}\ }\textbf {\bibinfo {volume} {52}},\ \bibinfo {pages} {3686} (\bibinfo {year} {1995})}\BibitemShut {NoStop}%
\bibitem [{\citenamefont {Schmöger}\ \emph {et~al.}(2015)\citenamefont {Schmöger}, \citenamefont {Versolato}, \citenamefont {Schwarz}, \citenamefont {Kohnen}, \citenamefont {Windberger}, \citenamefont {Piest}, \citenamefont {Feuchtenbeiner}, \citenamefont {Pedregosa-Gutierrez}, \citenamefont {Leopold}, \citenamefont {Micke}, \citenamefont {Hansen}, \citenamefont {Baumann}, \citenamefont {Drewsen}, \citenamefont {Ullrich}, \citenamefont {Schmidt},\ and\ \citenamefont {López-Urrutia}}]{Schmoeger2015}%
  \BibitemOpen
  \bibfield  {author} {\bibinfo {author} {\bibfnamefont {L.}~\bibnamefont {Schmöger}}, \bibinfo {author} {\bibfnamefont {O.~O.}\ \bibnamefont {Versolato}}, \bibinfo {author} {\bibfnamefont {M.}~\bibnamefont {Schwarz}}, \bibinfo {author} {\bibfnamefont {M.}~\bibnamefont {Kohnen}}, \bibinfo {author} {\bibfnamefont {A.}~\bibnamefont {Windberger}}, \bibinfo {author} {\bibfnamefont {B.}~\bibnamefont {Piest}}, \bibinfo {author} {\bibfnamefont {S.}~\bibnamefont {Feuchtenbeiner}}, \bibinfo {author} {\bibfnamefont {J.}~\bibnamefont {Pedregosa-Gutierrez}}, \bibinfo {author} {\bibfnamefont {T.}~\bibnamefont {Leopold}}, \bibinfo {author} {\bibfnamefont {P.}~\bibnamefont {Micke}}, \bibinfo {author} {\bibfnamefont {A.~K.}\ \bibnamefont {Hansen}}, \bibinfo {author} {\bibfnamefont {T.~M.}\ \bibnamefont {Baumann}}, \bibinfo {author} {\bibfnamefont {M.}~\bibnamefont {Drewsen}}, \bibinfo {author} {\bibfnamefont {J.}~\bibnamefont {Ullrich}}, \bibinfo {author} {\bibfnamefont {P.~O.}\ \bibnamefont {Schmidt}},\ and\ \bibinfo
  {author} {\bibfnamefont {J.~R.~C.}\ \bibnamefont {López-Urrutia}},\ }\bibfield  {title} {\bibinfo {title} {Coulomb crystallization of highly charged ions},\ }\href {https://doi.org/10.1126/science.aaa2960} {\bibfield  {journal} {\bibinfo  {journal} {Science}\ }\textbf {\bibinfo {volume} {347}},\ \bibinfo {pages} {1233} (\bibinfo {year} {2015})}\BibitemShut {NoStop}%
\bibitem [{\citenamefont {Ebrahimi}\ \emph {et~al.}(2018)\citenamefont {Ebrahimi}, \citenamefont {Guo}, \citenamefont {Vogel}, \citenamefont {Wiesel}, \citenamefont {Birkl},\ and\ \citenamefont {Quint}}]{Ebrahimi2018}%
  \BibitemOpen
  \bibfield  {author} {\bibinfo {author} {\bibfnamefont {M.~S.}\ \bibnamefont {Ebrahimi}}, \bibinfo {author} {\bibfnamefont {Z.}~\bibnamefont {Guo}}, \bibinfo {author} {\bibfnamefont {M.}~\bibnamefont {Vogel}}, \bibinfo {author} {\bibfnamefont {M.}~\bibnamefont {Wiesel}}, \bibinfo {author} {\bibfnamefont {G.}~\bibnamefont {Birkl}},\ and\ \bibinfo {author} {\bibfnamefont {W.}~\bibnamefont {Quint}},\ }\bibfield  {title} {\bibinfo {title} {Resistive cooling of highly charged ions in a penning trap to a fluidlike state},\ }\href {https://doi.org/10.1103/PhysRevA.98.023423} {\bibfield  {journal} {\bibinfo  {journal} {Phys. Rev. A}\ }\textbf {\bibinfo {volume} {98}},\ \bibinfo {pages} {023423} (\bibinfo {year} {2018})}\BibitemShut {NoStop}%
\bibitem [{\citenamefont {Oshima}\ \emph {et~al.}(2005)\citenamefont {Oshima}, \citenamefont {Niigaki}, \citenamefont {Lebois}, \citenamefont {Mohri}, \citenamefont {Komaki},\ and\ \citenamefont {Yamazaki}}]{OSHIMA2005504}%
  \BibitemOpen
  \bibfield  {author} {\bibinfo {author} {\bibfnamefont {N.}~\bibnamefont {Oshima}}, \bibinfo {author} {\bibfnamefont {M.}~\bibnamefont {Niigaki}}, \bibinfo {author} {\bibfnamefont {M.}~\bibnamefont {Lebois}}, \bibinfo {author} {\bibfnamefont {A.}~\bibnamefont {Mohri}}, \bibinfo {author} {\bibfnamefont {K.}~\bibnamefont {Komaki}},\ and\ \bibinfo {author} {\bibfnamefont {Y.}~\bibnamefont {Yamazaki}},\ }\bibfield  {title} {\bibinfo {title} {Simultaneous cooling of highly charged ions with electrons and positrons},\ }\href {https://doi.org/https://doi.org/10.1016/j.nimb.2005.03.233} {\bibfield  {journal} {\bibinfo  {journal} {Nuclear Instruments and Methods in Physics Research Section B: Beam Interactions with Materials and Atoms}\ }\textbf {\bibinfo {volume} {235}},\ \bibinfo {pages} {504} (\bibinfo {year} {2005})},\ \bibinfo {note} {the Physics of Highly Charged Ions}\BibitemShut {NoStop}%
\bibitem [{\citenamefont {Tomita}\ \emph {et~al.}(2016)\citenamefont {Tomita}, \citenamefont {Takatsuka}, \citenamefont {Takamatsu}, \citenamefont {Adachi}, \citenamefont {Furuta}, \citenamefont {Noto}, \citenamefont {Iguchi}, \citenamefont {Sonnenschein}, \citenamefont {Wendt}, \citenamefont {Ito},\ and\ \citenamefont {Maeda}}]{tomita2016development}%
  \BibitemOpen
  \bibfield  {author} {\bibinfo {author} {\bibfnamefont {H.}~\bibnamefont {Tomita}}, \bibinfo {author} {\bibfnamefont {T.}~\bibnamefont {Takatsuka}}, \bibinfo {author} {\bibfnamefont {T.}~\bibnamefont {Takamatsu}}, \bibinfo {author} {\bibfnamefont {Y.}~\bibnamefont {Adachi}}, \bibinfo {author} {\bibfnamefont {Y.}~\bibnamefont {Furuta}}, \bibinfo {author} {\bibfnamefont {T.}~\bibnamefont {Noto}}, \bibinfo {author} {\bibfnamefont {T.}~\bibnamefont {Iguchi}}, \bibinfo {author} {\bibfnamefont {V.}~\bibnamefont {Sonnenschein}}, \bibinfo {author} {\bibfnamefont {K.}~\bibnamefont {Wendt}}, \bibinfo {author} {\bibfnamefont {C.}~\bibnamefont {Ito}},\ and\ \bibinfo {author} {\bibfnamefont {S.}~\bibnamefont {Maeda}},\ }\bibfield  {title} {\bibinfo {title} {Development of high resolution resonance ionization mass spectrometry for neutron dosimetry technique with $^{93}\textrm{Nb}$(n, n') $^{93\textrm{m}}\textrm{Nb}$ reaction},\ }\href {https://doi.org/10.1051/epjconf/201610605002} {\bibfield  {journal} {\bibinfo
  {journal} {EPJ Web Conf.}\ }\textbf {\bibinfo {volume} {106}},\ \bibinfo {pages} {05002} (\bibinfo {year} {2016})}\BibitemShut {NoStop}%
\bibitem [{\citenamefont {Lu}\ \emph {et~al.}(2022)\citenamefont {Lu}, \citenamefont {Varvarezos}, \citenamefont {Nicolosi}, \citenamefont {Andrighetto}, \citenamefont {Scarpa}, \citenamefont {Mariotti},\ and\ \citenamefont {Costello}}]{lu2022laser}%
  \BibitemOpen
  \bibfield  {author} {\bibinfo {author} {\bibfnamefont {H.}~\bibnamefont {Lu}}, \bibinfo {author} {\bibfnamefont {L.}~\bibnamefont {Varvarezos}}, \bibinfo {author} {\bibfnamefont {P.}~\bibnamefont {Nicolosi}}, \bibinfo {author} {\bibfnamefont {A.}~\bibnamefont {Andrighetto}}, \bibinfo {author} {\bibfnamefont {D.}~\bibnamefont {Scarpa}}, \bibinfo {author} {\bibfnamefont {E.}~\bibnamefont {Mariotti}},\ and\ \bibinfo {author} {\bibfnamefont {J.~T.}\ \bibnamefont {Costello}},\ }\bibfield  {title} {\bibinfo {title} {Laser double optical resonance excitation-ionization of mo with optogalvanic detection},\ }\href {https://doi.org/10.1088/1402-4896/ac48a7} {\bibfield  {journal} {\bibinfo  {journal} {Phys. Scr.}\ }\textbf {\bibinfo {volume} {97}},\ \bibinfo {pages} {024004} (\bibinfo {year} {2022})}\BibitemShut {NoStop}%
\bibitem [{\citenamefont {Cerchiari}\ \emph {et~al.}(2018)\citenamefont {Cerchiari}, \citenamefont {Erlewein}, \citenamefont {K\"onig},\ and\ \citenamefont {Kellerbauer}}]{Cerchiari_Penning_2018}%
  \BibitemOpen
  \bibfield  {author} {\bibinfo {author} {\bibfnamefont {G.}~\bibnamefont {Cerchiari}}, \bibinfo {author} {\bibfnamefont {S.}~\bibnamefont {Erlewein}}, \bibinfo {author} {\bibfnamefont {C.}~\bibnamefont {K\"onig}},\ and\ \bibinfo {author} {\bibfnamefont {A.}~\bibnamefont {Kellerbauer}},\ }\bibfield  {title} {\bibinfo {title} {Loading of a continuous anion beam into a penning trap with a view to laser cooling},\ }\href {https://doi.org/10.1103/PhysRevA.98.021402} {\bibfield  {journal} {\bibinfo  {journal} {Phys. Rev. A}\ }\textbf {\bibinfo {volume} {98}},\ \bibinfo {pages} {021402} (\bibinfo {year} {2018})}\BibitemShut {NoStop}%
\bibitem [{\citenamefont {Shabaev}(1991)}]{Shabaev1991}%
  \BibitemOpen
  \bibfield  {author} {\bibinfo {author} {\bibfnamefont {V.~M.}\ \bibnamefont {Shabaev}},\ }\bibfield  {title} {\bibinfo {title} {Generalizations of the virial relations for the dirac equation in a central field and their applications to the coulomb field},\ }\href {https://doi.org/10.1088/0953-4075/24/21/004} {\bibfield  {journal} {\bibinfo  {journal} {J. Phys. B}\ }\textbf {\bibinfo {volume} {24}},\ \bibinfo {pages} {4479} (\bibinfo {year} {1991})}\BibitemShut {NoStop}%
\end{thebibliography}%

\clearpage
\newpage
\appendix

\section{Hyperfine splitting calculations}\label{app:calculations}
The article focuses on the hyperfine structure of ions with a single orbiting electron (hydrogen-like ions). We calculated the energy splitting of the hyperfine ground state following the calculations of Ref.~\cite{Shabaev1994}. In this appendix, the key steps that we used to derive the values reported in the tables are reported for completeness. The hyperfine splitting energy $\Delta E_\text{HF}$ is given by

\begin{equation}
    \begin{split}
        \Delta E_\text{HF} =& \alpha(\alpha Z)^3 \frac{m_e}{m_p}\frac{\mu}{\mu_N}\frac{2(2I+1)}{3I}
        \\
        &\cdot m_ec^2\left[A(1-\delta)(1-\epsilon)\right].
    \end{split}
\end{equation}

Here, $\alpha$ is the fine structure constant, $Z$ is the number of protons, $I$ is the nuclear spin, and $\mu/\mu_N$ is the nuclear magnetic moment of the respective nuclei in units of the nuclear magneton. The masses $m_\text{e}$ and $m_\text{p}$ are those of the electron and proton, respectively, and $c$ is the speed of light in a vacuum. Here, $A$ is a relativistic correction factor. Additionally, $\delta$ and $\epsilon$ denote the nuclear charge distribution and nuclear magnetization (Bohr-Weisskopf) distribution correction factors, respectively. However, this equation does not account for radiative corrections, which are negligible in this case.

\begin{table}[]
    \jcaption{The hyperfine splitting wavelength $\lambda_\text{HF}$ for ground state hydrogen-like ion candidates with atomic numbers between Kr and Xe and a half-life above $60$ minutes. Here $A$ and $\delta$ are the relativistic and nuclear charge distribution correction factors given by Ref.~\cite{Shabaev1994}. The nuclear magnetization distribution (Bohr-Weisskopf) correction $\epsilon$ is given by Eq.~\ref{eq: epsilon}. The calculations are conducted using the nuclear data provided in Ref.~\cite{Mertzimekis2016}.}
    \centering
    \small
    \begin{tabular}{cccccccccc}
\toprule
            Ion &     $I$  &    $\mu/\mu_N$ &   $A$ &    $\delta \times 10^2$ &  $\epsilon \times 10^2$ &  $\lambda_\text{HF}$(\si{\micro\meter}) \\
\midrule
    $^{83}$Kr$^{35+}$ &  1/2- & 0.5910 & 1.114708 & 0.8328 &  0.4939 &       19.42455 \\
 $^{85}$Kr$^{35+}$ &  1/2- & 0.6320 & 1.114708 & 0.8328 &  0.4939 &       18.16442 \\
 $^{82}$Rb$^{36+}$ &    5- & 1.5096 & 1.121909 & 0.8769 & -0.6574 &  12.51460 \\
 $^{85}$Sr$^{37+}$ &  1/2- & 0.5990 & 1.129399 & 0.9236 &  0.5423 &  16.10615 \\
 $^{87}$Sr$^{37+}$ &  1/2- & 0.6240 & 1.129399 & 0.9236 &  0.5423 &  15.46087 \\
  $^{85}$Y$^{38+}$ &  9/2+ & 6.2000 & 1.137188 & 0.9742 &  0.2540 &  2.567061 \\
  $^{87}$Y$^{38+}$ &  9/2+ & 6.2400 & 1.137188 & 0.9742 &  0.2550 &  2.550630 \\
  $^{90}$Y$^{38+}$ &    7+ & 5.2800 & 1.137188 & 0.9742 &  0.2389 &  3.125522 \\
 $^{92}$Nb$^{40+}$ &    2+ & 6.1370 & 1.153701 & 1.0917 &  0.8096 &  1.968983 \\
 $^{93}$Mo$^{41+}$ & 21/2+ & 9.9300 & 1.162446 & 1.1587 &  0.4354 &  1.336406 \\
 $^{99}$Rh$^{44+}$ &  9/2+ & 5.6200 & 1.190779 & 1.3800 &  0.3219 &  1.769000 \\
$^{101}$Rh$^{44+}$ &  9/2+ & 5.4300 & 1.190779 & 1.3800 &  0.3141 &  1.830757 \\
$^{102}$Rh$^{44+}$ &    6+ & 4.0100 & 1.190779 & 1.3800 &  0.2122 &  2.540022 \\
$^{106}$Ag$^{46+}$ &    6+ & 3.7040 & 1.211547 & 1.5364 &  0.1453 &  2.374345 \\
$^{108}$Ag$^{46+}$ &    6+ & 3.5800 & 1.211547 & 1.5364 &  0.1055 &  2.455607 \\
$^{110}$Ag$^{46+}$ &    6+ & 3.6020 & 1.211547 & 1.5364 &  0.1127 &  2.440786 \\
$^{110}$In$^{48+}$ &    2+ & 4.3650 & 1.233957 & 1.7065 &  1.0757 &  1.529835 \\
$^{114}$In$^{48+}$ &    5+ & 4.6460 & 1.233957 & 1.7065 &  0.5796 &  1.625155 \\
$^{116}$Sb$^{50+}$ &    8- & 2.5900 & 1.258147 & 1.9005 & -1.0601 &  2.587845 \\
$^{118}$Sb$^{50+}$ &    8- & 2.3200 & 1.258147 & 1.9005 & -1.3627 &  2.880392 \\
$^{120}$Sb$^{50+}$ &    8- & 2.3400 & 1.258147 & 1.9005 & -1.3379 &  2.856473 \\
$^{119}$Te$^{51+}$ & 11/2- & 0.8940 & 1.270956 & 2.0077 &  0.6792 &  6.946341 \\
$^{121}$Te$^{51+}$ & 11/2- & 0.8950 & 1.270956 & 2.0077 &  0.6792 &  6.938580 \\
\bottomrule
\end{tabular}
    
    \label{tab:HF_corrections}
\end{table}

The correction factors $A$ and $\delta$ were calculated as outlined in Ref.~\cite{Shabaev1994}. The correction factor $\epsilon$ is given by~\cite{Shabaev1994} 
\begin{equation}\label{eq: epsilon}
    \epsilon=\alpha_S\left[\langle K_S\rangle+\xi\left(\langle K_S\rangle-\langle K_L\rangle\right)\right]+\alpha_L\langle K_L\rangle ,
\end{equation}

where $\alpha_S$ and $\alpha_L$ are the fractions of the spin and orbital contributions to the magnetic moment, and $K_S(R)$ and $K_L(R)$ correspond to radial overlap integrals of the electron's Dirac wavefunctions up to the nuclear radius. We assume that the total nuclear moment is possessed by the odd nucleon, differentiating between three cases, having an odd number of neutrons, an odd number of protons, or an odd number of both protons and neutrons. 

For nuclei with an even number of protons and an odd number of neutrons, one sets $\alpha_S=1$ and $\alpha_L=0$, whereas for an odd number of protons and an even number of neutrons, these are given by

\begin{equation}
    \alpha_S=\frac{g_s(\mu-I)}{\mu(g_s-1)}\quad,\quad \alpha_L=1-\alpha_S\;.
\end{equation}

Here $g_s$ is the g-factor of the proton when $|I+1/2|$ is odd or the g-factor of the neutron and when it is even. We assign a signed value for $I$ depending on the parity, for example, for a nucleus like $^{83}$Kr$^{35+}$ with $I$ reported in the literature as $1/2-$, the value is assigned as $I=-1/2$. Similarly, the asymmetry parameter $\xi$ is computed for the two cases by
\begin{align}\label{eq: xi}
    \xi=\frac{2I-1}{4(I+1)}\quad \textrm{if}\:|I+1/2|\:\text{odd} \,,\\
    \xi=\frac{2I+3}{4I}\quad \textrm{if}\:|I+1/2|\:\text{even} \,.
\end{align}
The expectation values $\langle K_S\rangle$ and $\langle K_L\rangle$ were computed assuming a homogeneously distributed probability density of the odd nucleon over the nuclear volume, as explained in Ref.~\cite{Shabaev1994}. 

In cases where both the number of protons and neutrons are odd i.e. $I$ is an even number, we compute the nuclear magnetization correction factor $\epsilon$ via Eq.~\ref{eq: epsilon} for both cases of $|I+1/2|$ odd and even separately, and add them to compute the total correction factor~\cite{Shabaev_1995}.

The hyperfine splitting wavelength $\lambda_\text{HF}$ can then be calculated from the resulting energy splitting for each respective ion. The results for each of the ions with their respective $I$ and $\mu/\mu_N$ are given along the correction factors and the resulting wavelength in Tab.~\ref{tab:HF_corrections} for candidate ions with a half-life of more than $60$ minutes.
\section{Extended candidate table}\label{app:extended table}
An extension of Tab.~\ref{tab:candidates} for ions with half-lives above $1$ hour is presented in Tab.~\ref{tab:extra_1min}.

\begin{table}[!htbp]
    \jcaption{Extension of Tab.~\ref{tab:candidates} for potential candidates including masses above the mass of Xe and sorted by ascending proton number. The nuclear spin $I$, magnetic moment $\mu/\mu_N$ , excitement energy $E$, half-life $t_{1/2}$, and calculated ground state hyperfine splitting wavelength $\lambda_\text{HF}$ for each of the respective ion is shown. The calculations are conducted using the nuclear data provided in Ref.~\cite{Mertzimekis2016}.}
    \centering
    \small
    \begin{tabular}{cccccc}
    \toprule
                Ion &     $I$ &     $\mu$ &  $E$ (keV) &  $t_{1/2}$ &  $\lambda_\text{HF}$(\si{\micro\meter}) \\
    \midrule
      $^{44}$Sc$^{20+}$ &    6+ & 3.8310 &          271 &      58.6 h &       29.68554 \\
 $^{69}$Zn$^{29+}$ &  9/2+ & 1.1605 &          439 &      13.7 h &       31.67983 \\
 $^{80}$Br$^{34+}$ &    5- & 1.3165 &           86 &      4.42 h &       17.13157 \\
 $^{83}$Kr$^{35+}$ &  1/2- & 0.5910 &           42 &      1.83 h &       19.42455 \\
 $^{85}$Kr$^{35+}$ &  1/2- & 0.6320 &          305 &      4.48 h &       18.16442 \\
 $^{82}$Rb$^{36+}$ &    5- & 1.5096 &           69 &      6.47 h &       12.51460 \\
 $^{85}$Sr$^{37+}$ &  1/2- & 0.5990 &          239 &      67.6 m &       16.10615 \\
 $^{87}$Sr$^{37+}$ &  1/2- & 0.6240 &          388 &      2.82 h &       15.46087 \\
  $^{85}$Y$^{38+}$ &  9/2+ & 6.2000 &           20 &       4.9 h &        2.567061 \\
  $^{87}$Y$^{38+}$ &  9/2+ & 6.2400 &          381 &      13.4 h &        2.550630 \\
  $^{90}$Y$^{38+}$ &    7+ & 5.2800 &          682 &      3.19 h &        3.125522 \\
 $^{92}$Nb$^{40+}$ &    2+ & 6.1370 &          135 &     10.15 d &        1.968983 \\
 $^{93}$Mo$^{41+}$ & 21/2+ & 9.9300 &         2425 &      6.85 h &        1.336406 \\
 $^{99}$Rh$^{44+}$ &  9/2+ & 5.6200 &           65 &       4.7 h &        1.769000 \\
$^{101}$Rh$^{44+}$ &  9/2+ & 5.4300 &          157 &      4.34 d &        1.830757 \\
$^{102}$Rh$^{44+}$ &    6+ & 4.0100 &          141 &      3.74 y &        2.540022 \\
$^{106}$Ag$^{46+}$ &    6+ & 3.7040 &           90 &       8.3 d &        2.374345 \\
$^{108}$Ag$^{46+}$ &    6+ & 3.5800 &          109 &       438 y &        2.455607 \\
$^{110}$Ag$^{46+}$ &    6+ & 3.6020 &          118 &       250 d &        2.440786 \\
$^{110}$In$^{48+}$ &    2+ & 4.3650 &           62 &      69.1 m &        1.529835 \\
$^{114}$In$^{48+}$ &    5+ & 4.6460 &          190 &      49.5 d &        1.625155 \\
$^{116}$Sb$^{50+}$ &    8- & 2.5900 &          383 &      60.3 m &        2.587845 \\
$^{118}$Sb$^{50+}$ &    8- & 2.3200 &          250 &       5.0 h &        2.880392 \\
$^{120}$Sb$^{50+}$ &    8- & 2.3400 &        0 + x &      5.76 d &        2.856473 \\
$^{119}$Te$^{51+}$ & 11/2- & 0.8940 &          261 &      4.70 d &        6.946341 \\
$^{121}$Te$^{51+}$ & 11/2- & 0.8950 &          294 &       164 d &        6.938580 \\
$^{134}$Cs$^{54+}$ &    8- & 1.0959 &          139 &      2.91 h &        4.497931 \\
$^{137}$Ce$^{57+}$ & 11/2- & 1.0100 &          254 &      34.4 h &        4.184041 \\
$^{148}$Pm$^{60+}$ &    6- & 1.8000 &          138 &      41.3 d &        1.910689 \\
$^{154}$Tb$^{64+}$ &    7- & 0.9000 &        0 + x &      22.7 h &        2.845837 \\
$^{154}$Tb$^{64+}$ &    3- & 1.6000 &        0 + x &       9.4 h &        1.602681 \\
$^{160}$Ho$^{66+}$ &    2- & 2.5100 &           60 &      5.02 h &        0.864546 \\
$^{162}$Ho$^{66+}$ &    6- & 3.5900 &          106 &        67 m &        0.687144 \\
$^{166}$Ho$^{66+}$ &    7- & 3.6200 &            6 &      1200 y &        0.686029 \\
$^{174}$Lu$^{70+}$ &    6- & 1.4870 &          171 &       142 d &        1.261300 \\
$^{176}$Lu$^{70+}$ &    1- & 0.3180 &          123 &      3.66 h &        4.308825 \\
$^{177}$Lu$^{70+}$ & 23/2- & 2.3010 &          970 &       160 d &        0.859488 \\
$^{178}$Hf$^{71+}$ &   16+ & 8.1300 &         2446 &        31 y &        0.247105 \\
$^{179}$Hf$^{71+}$ & 25/2- & 7.4000 &         1106 &      25.1 d &        0.264762 \\
$^{180}$Hf$^{71+}$ &    8- & 8.7000 &         1142 &      5.53 h &        0.224133 \\
$^{180}$Ta$^{72+}$ &    9- & 4.8140 &           75 & $>$ 7.1$\times10^{15}$ y &        0.372471 \\
$^{182}$Re$^{74+}$ &    2+ & 3.2600 &        0 + x &      14.1 h &        0.427852 \\
$^{184}$Re$^{74+}$ &    8+ & 2.8800 &          188 &       169 d &        0.541766 \\
$^{186}$Ir$^{76+}$ &    2- & 0.6380 &        0 + x &      1.90 h &        1.836954 \\
$^{193}$Pt$^{77+}$ & 13/2+ & 0.7530 &          150 &       4.3 d &        1.805178 \\
$^{196}$Au$^{78+}$ &   12- & 5.7000 &          596 &       9.6 h &        0.227025 \\
$^{198}$Au$^{78+}$ &   12- & 5.8300 &          812 &      2.27 d &        0.222241 \\
$^{200}$Au$^{78+}$ &   12- & 5.8800 &          962 &      18.7 h &        0.220454 \\
$^{196}$Tl$^{80+}$ &    7+ & 0.5470 &          394 &      1.41 h &        1.572351 \\
$^{198}$Tl$^{80+}$ &    7+ & 0.6390 &          544 &      1.87 h &        1.403119 \\
$^{210}$Bi$^{82+}$ &    9- & 2.7200 &          271 &   3.0$\times10^6$ y &        0.362775 \\
$^{242}$Am$^{94+}$ &    5- & 1.0000 &           49 &       152 y &        0.441956 \\
$^{254}$Es$^{98+}$ &    2+ & 2.9000 &           84 &      39.3 h &        0.133164 \\
    \bottomrule
    \end{tabular}
    
    \label{tab:extra_1min}
\end{table}

\section{Synthesis from co-trapped negative ions}\label{app:NbandMo}
\begin{figure}[h!]
    \centering
    \includegraphics[width=1.0\columnwidth]{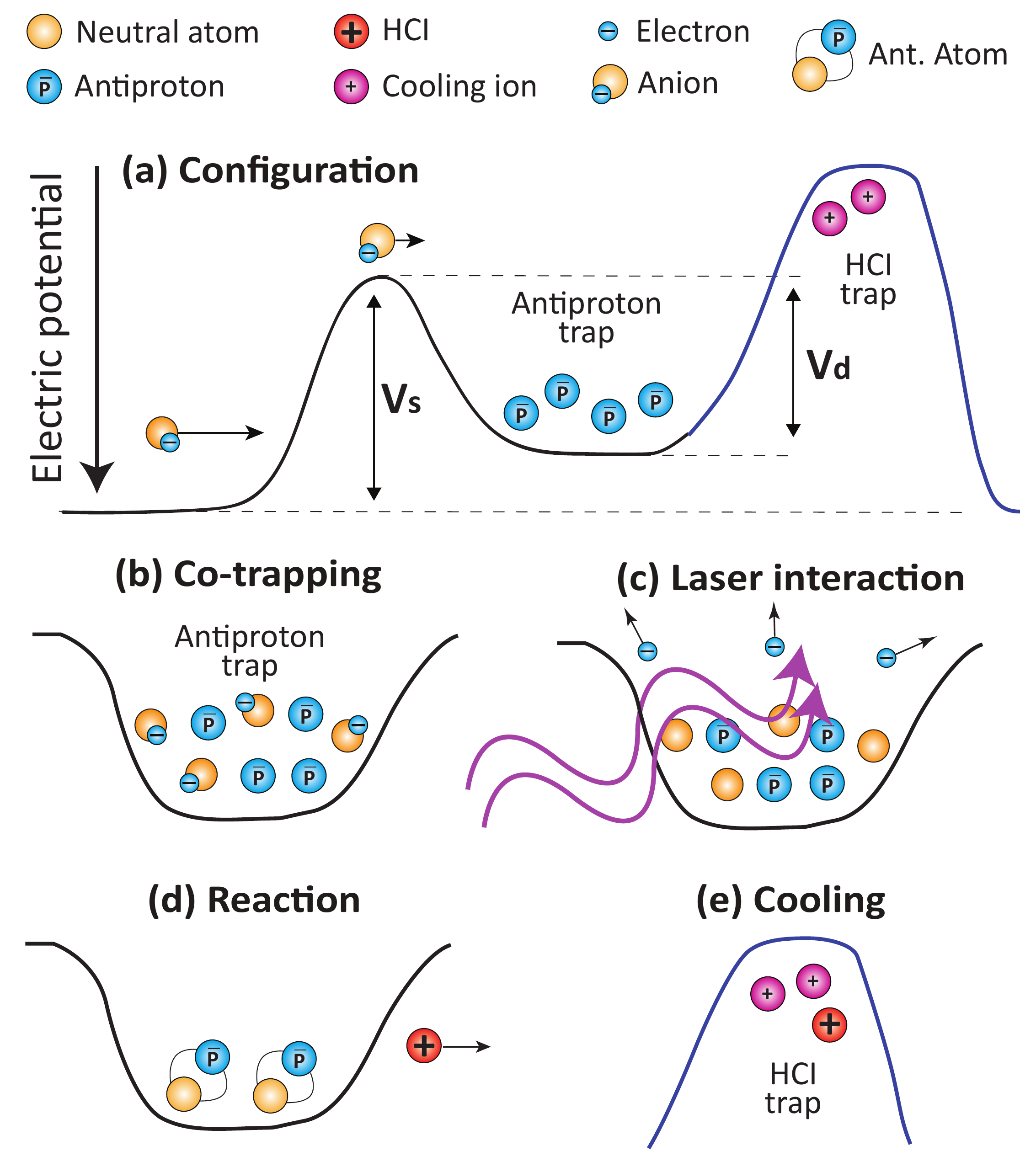}
    \jcaption{Schematic drawing of the formation process of the Highly Charge Ions (HCIs) via negative ions. The figure depicts a possible configuration of the axial electric potential in a Penning trap. (a) Experimental configuration. Negative ions are guided towards the antiproton trap at a kinetic energy of several keV, just a few Volts higher than the stopping potential $V_s$. Under this condition, optimal trapping is obtained with a trap depth $V_d$ of tens of V~\cite{Cerchiari_Penning_2018}. A second area of the trap (HCI trap) is dedicated to the confinement of positive ions. (b) Thanks to the loading procedure, anions are confined with antiprotons in the antiproton trap. Electron cooling (not shown) can be further implemented to reduce the kinetic energy of both species. (c) Laser pulses are used to photodetach the anions and Rydberg-excite the resulting neutral atoms. (d) The excited atoms react with antiprotons forming antiprotonic atoms from which HCIs are synthesized. The HCIs are positive and leave the antiproton trap. (e) The HCIs accumulate in the HCI trap by synpathetic cooling with other positive ions (positrons or laser cooled ions).}
    \label{fig:anions}
\end{figure}

We propose utilizing co-trap negative ions with antiprotons for the synthesis of the isomers. A schematic drawing of the various steps the procedure to control the reaction is presented in Fig.~\ref{fig:anions}. The use of negative ions for the synthesis facilitates the division of the trap into a region where negative ions are trapped, from which the positive HCIs naturally escape. A second trapping region for positive ions shall be prepared where the newly formed HCI can accumulate. There, a buffer of laser-cooled positive ions, positrons, or resistive cooling can be used to reduce the kinetic energy of HCIs in preparation for spectroscopy, as already demonstrated or proposed in other works~\cite{Schmoeger2015, Ebrahimi2018,OSHIMA2005504}. To initiate the reaction between the negative ions and antiprotons, laser excitation is needed to remove the extra electron and trigger the reaction. We further clarify this laser excitation \st{process} step with the example of Mo$^-$ and Nb$^-$ co-trapped with antiprotons. These elements are our proposed reagents to produce $^{92\textrm{m}1}$Nb (see Tab.~\ref{tab:shortcandidates}).
\begin{figure}
    \centering
    \includegraphics[width=0.75\columnwidth]{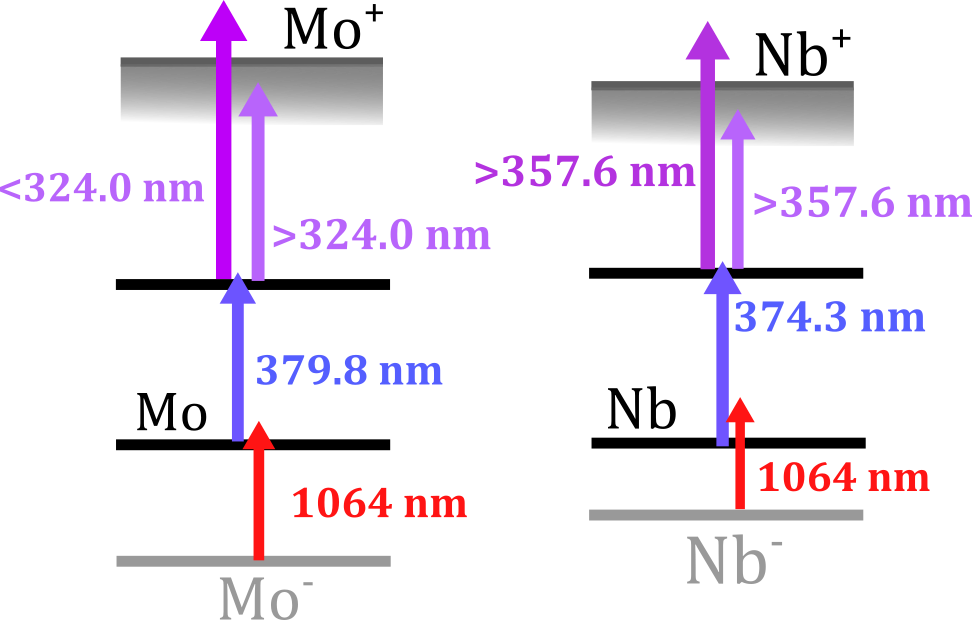}
    \jcaption{Proposed laser photodetachment and excitation schemes for Mo and Nb ~\cite{tomita2016development,lu2022laser}.}
    \label{fig:laser_excitation}
\end{figure}

Figure.~\ref{fig:laser_excitation} graphically summarizes the excitation scheme we propose to utilize. Interaction can be triggered via a three-pulse excitation scheme aimed at creating either singly ionized $^{94}$Mo$^+$ and $^{93}$Nb$^+$, or, preferentially, the neutral Rydberg-excited atom. This option is preferable, as any positive ions formed could escape the trapping fields and thus contaminate the trapping region where the positively charged HCIs are supposed to remain. The first laser pulse will photodetach the electron, forming an atom in the ground state; this can be realized with a wavelength $\qty{1064}{\nano\meter}$ (Nd:YAG fundamental) for both $^{94}$Mo and $^{93}$Nb. From the ground state of Mo, a second laser pulse at \(\qty{379.8}{\nano\meter}\) can excite the atom to an intermediate state, followed by a third step at \(\qty{324.0}{\nano\meter}\) for photoionization \cite{lu2022laser}. Alternatively, using a longer wavelength in the final step allows excitation to a selected Rydberg level for enhancing the antiproton-atom capture cross-section. Similarly, neutral Nb atoms can be excited to an intermediate state with \(\qty{374.3}{\nano\meter}\), followed by a pulse with wavelength around \(\qty{357.6}{\nano\meter}\) for either Rydberg excitation or photoionization \cite{tomita2016development}. The excitation scheme for the other candidates mentioned in Tab.~\ref{tab:shortcandidates} involves an analogous three-step process, and we summarize the necessary wavelengths in Tab.~\ref{tab:pulses_corrected}.

\begin{table}[]
    \jcaption{Laser pulses to trigger the reaction of atoms and antiprotons. The table presents the wavelengths to be used for the three-step excitation scheme from the negative ion to the Rydberg neutral atom. $\lambda_{PD}$: photo-detachment threshold, $\lambda_{Exc.}$: neutral atom excitation, $\lambda_{PI}$: photoionization threshold.}
    \centering
    \begin{tabular}{ccccccc}
    \toprule
        Candidate & $\lambda_{PD}$ (\si{\micro\meter}) & $\lambda_{Exc.}$ (\si{\nano\meter}) & $\lambda_{PI}$ (\si{\nano\meter}) \\
    \midrule
      Y & 3.98 & 404.8  & 393 \\
      Zr & 2.86 & 357.2 & 392 \\
      Nb & 1.35 & 374.3 & 358 \\
      Mo & 1.66 & 379.8 & 324 \\
      Pd & 2.21 & 276 & 322 \\
      Rh & 1.08 & 272 & 428 \\
      In & 3.23 & 304 & 726 \\
      Sb & 1.18 & 231.1 & 382 \\  
    \bottomrule
    \end{tabular}
    
    \label{tab:pulses_corrected}
\end{table}


The table reports the threshold wavelengths for the photodetachment ($\lambda_p$) and photoionization ($\lambda_i$) process from the intermediate state. In the experiments, the photodetachment laser should be tuned at a wavelength lower than $\lambda_p$ to ensure neutralization of the negative ion and the Rydberg excitation laser should be tuned slightly above the value $\lambda_{i}$.

\section{Lifetime calculation}

The lifetimes mentioned in Tab.~\ref{tab:shortcandidates} have been calculated following the guidance in Ref.~\cite{shabaev1998}. We first calculate the transition rates between the states via:

\begin{equation}
\begin{split}
w_{F \to F'} =& \alpha \frac{\omega^3}{c_0^2} \frac{(2F' + 1)(2j + 1)^3}{3j(j + 1)}
\\
& \cdot \begin{Bmatrix}
j & F' & I \\ 
F & j & 1 
\end{Bmatrix}^2 
\left[ \int_0^\infty g(r) f(r) r^3 dr \right]^2 \,,
\end{split}
\label{eq:lifetime}
\end{equation}
which describes a M1 transition for the approximation of a hydrogen-like ion. 
The variable $\mathrm{\omega}$ is the transition frequency, $F$ and $F'$ are the total angular momentum of the initial and final state, $I$ is the nuclear spin, and $g(r)$ and $f(r)$ are the upper and lower radial components of the hydrogen-like Dirac wave function. For a point nucleus the integral in equation \ref{eq:lifetime}, following Ref. \cite{Shabaev1991}, solves as
\begin{equation}
    \int_0^\infty g(r) f(r) r^3 dr = \frac{2\kappa\epsilon - m_e c_0^2}{4m_e c_0^2}\frac{\hbar}{m_e c_0}\,.
\end{equation}
Only considering the s state, we therefore obtain the transition frequency
\begin{equation}
    w_{F \to F'} = \alpha \omega^3 \frac{\hbar^2}{m_e^2 c_0^4} \frac{4}{27} \frac{I}{2I + 1} 
\left[ \frac{2\epsilon}{m_ec_0^2} + 1 \right]^2\,,
\end{equation}
in which the one-electron Dirac energy $\mathrm{\epsilon}$ is given by
\begin{equation}
 \epsilon = \frac{m_e c_0^2}{\sqrt{1 + \left( \frac{\alpha Z}{\gamma + n_r} \right)^2}}\,.
\end{equation}
where the radial quantum number $n_r$ is 0 for a hydrogen-like ion, and $\mathrm{\gamma} = \sqrt{\kappa^2 - (\alpha Z)^2}$, where $\mathrm{\kappa}$ is -1 for $l = 0$ and $j=1/2$. We can therefore rewrite the equation as
\begin{equation} 
    \epsilon = m_ec_0^2\sqrt{1-(\alpha Z)^2}\,.
\end{equation}
The lifetime of the upper level of the ground state hyperfine splitting is calculated by taking the inverse of the transition rate. It is advised to be careful, as a factor of $2\mathrm{\pi}$ within $\mathrm{\omega}$ can be overlooked quite easily.

\section{Hyperfine splitting calculation with GRASP}\label{app:HyperGRASP}
\begin{figure}[h!]
    \centering
    \includegraphics[width=\columnwidth]{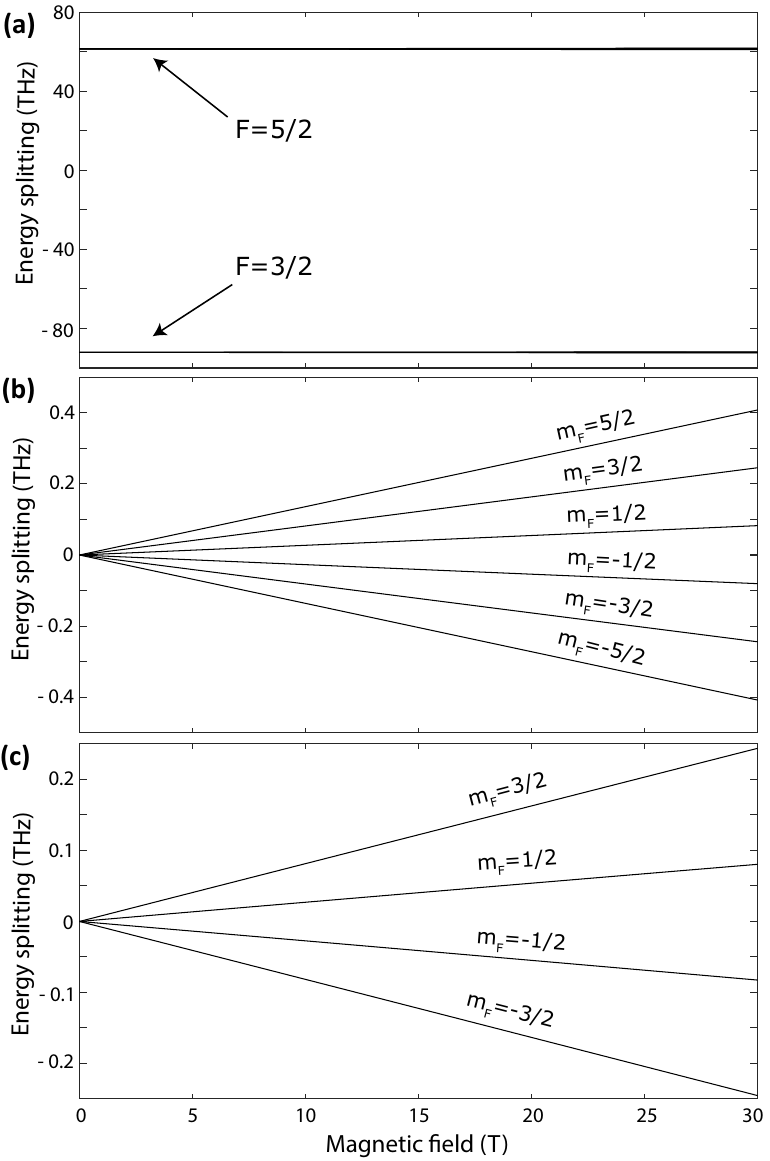}
    \jcaption{Hyperfine and magnetic field splitting of the ground state of $^{92\textrm{m}1}$Nb$^{40+}$ calculated via the GRASP package \cite{Jonsson2013,Jonnsson2007,Bieron2023}. (a) Hyperfine splitting of the states with total angular momentum $\textrm{F}=5/2$ and $\textrm{F}=3/2$. The separation of the two states is \textrm{153,49} THz. (b, c) Zoomed-in view of the magnetic field splitting for the Hyperfine sublevels of the state with $\textrm{F}=5/2$ (b) and the state with $\textrm{F}=3/2$ (c). In both (b) and (c) the offset frequency of the hyperfine splitting was removed. The magnetic field splitting shows a linear behavior as a function of the magnetic field with an energy shift of $0.5434 \times m_F \times \textrm{MHz}/\textrm{G}$ for both ground states, indicating that we are still well within the linear Zeeman regime, even for such large magnetic fields.}
    \label{fig:HFS5}
\end{figure}
The grasp2K relativistic atomic structure package includes different programs that, when combined, allow one to calculate atomic spectra, energy levels, oscillator strength, and transition rates, while using a fully relativistic approach \cite{Jonnsson2007,Jonsson2013}.~\\
As explained in the manual given in Ref. \cite{Jonnsson2007}, the calculation includes multiple steps. One can define whether they want to simulate the nucleus as an extended Fermi distribution or a point charge, which states should be included in the calculations, how many excitations there are, if you want the program to consider relativistic corrections, and so on.~\\
After creating the necessary atomic data, a corresponding Matlab script allows one to calculate and plot the energy splittings and transition rates of chosen energy levels. Figure~\ref{fig:HFS5} presents the simulation results for the example of $^{92\textrm{m}1}$Nb. We plot the two hyperfine ground states once combined, to show the energy splitting, and separately to show the linear behavior of the hyperfine sublevels even up to high magnetic fields of 30 Tesla. However, calculating the transition rates with the GRASP package was not successful, as it requires levels of different principal quantum numbers.


\end{document}